\documentclass[aps,amsmath,amssymb,prfluids,preprint,superscriptaddress,11pt,floatfix]{revtex4-2}

\usepackage{graphicx}% Include figure files
\usepackage{dcolumn}% Align table columns on decimal point
\usepackage{bm}% bold math
\usepackage{hyperref}% add hypertext capabilities
\usepackage{subeqnarray}

\usepackage{todonotes}
\usepackage{caption}

\newcommand{\pd}[2]{\frac{\partial #1}{\partial #2}}
\newcommand{\DIM}[1]{\hat{#1}}

\newcommand{\bnabla}{\boldsymbol{\nabla}}
\newcommand{\pdiff}[2]{\pd{#1}{#2}}

\newcommand{\width}{\Wint}
\newcommand{\Dwidth}{W}
\newcommand{\DLa}{L_a}
\newcommand{\DHa}{H_a}
\newcommand{\Drho}{\rho_0}

\newcommand{\La}{\mathcal{L}_{j}}
\newcommand{\Ha}{\mathcal{H}_{j}}
\newcommand{\Harbitrary}{\mathcal{H}}
\newcommand{\bx}{\mathbf{x}_{j}}
\newcommand{\bv}{\mathbf{v}_{j}}

\newcommand\Sc{\mbox{\textit{Sc}}}  % Schmidt number
\newcommand{\rhojetnd}{\rho'_{j}}
\newcommand{\rhointnd}{\rho'_{i}}

\newcommand{\pjetnd}{p_\textit{j}}
\newcommand{\pintnd}{p_\textit{i}}
\newcommand\Rey{\mbox{\textit{Re}}}

\newcommand{\uint}{u_\textit{i}}
\newcommand{\vint}{v_\textit{i}}

\newcommand{\xint}{x_\textit{i}}
\newcommand{\yint}{y_\textit{i}}
\newcommand{\zint}{z_\textit{i}}
\newcommand{\wint}{w_\textit{i}}
\newcommand{\tint}{t_\textit{i}}
\newcommand{\bvint}{\mathbf{v}_\textit{i}}
\newcommand{\Lint}{\mathcal{L}_\textit{i}}
\newcommand{\Hint}{\mathcal{H}_\textit{i}}
\newcommand{\Hjet}{\mathcal{H}_\textit{j}}
\newcommand{\Wint}{\mathcal{W}}
\newcommand{\xjet}{x_\textit{j}}
\newcommand{\yjet}{y_\textit{j}}
\newcommand{\zjet}{z_\textit{j}}
\newcommand{\wjet}{w_\textit{j}}
\providecommand{\DLz}{L_z}
\newcommand{\Lzjet}{\mathcal{L}_{z,\textit{j}}} 
\newcommand{\Lzint}{\mathcal{L}_{z,\textit{i}}}
\providecommand{\spav}[1]{\langle #1 \rangle}
\newcommand{\tjet}{t_\textit{j}}
\newcommand{\ujet}{u_\textit{j}}
\newcommand{\vjet}{v_\textit{j}}
\newcommand{\DIMbv}{\hat{\mathbf{v}}}
\newcommand{\DIMbx}{\hat{\mathbf{x}}}

\newcommand{\etal}{\textit{et al.}}

\usepackage{titlesec}
\usepackage{booktabs}
\usepackage{tabularx}
\renewcommand\thesubsubsection{\arabic{subsubsection}} % 1, 2, ...

\titleclass{\subsubsubsection}{straight}[\subsubsection]
\newcounter{subsubsubsection}[subsubsection]
\renewcommand\thesubsubsubsection{\thesubsubsection.\arabic{subsubsubsection}}

\titleformat{\subsubsubsection}
  {\normalfont\normalsize\bfseries}{\thesubsubsubsection}{1em}{}
\titlespacing*{\subsubsubsection}{0pt}{3.25ex plus 1ex minus .2ex}{1.5ex plus .2ex}

\begin{document}

% \preprint{APS/123-QED}

\title{\textbf{Planar intrusions from a narrow, jet-like source in a linearly stratified ambient} 
}% 

\author{Hai Duc Vu}
\email{carl.vu@monash.edu}
\affiliation{School of Mathematics, Monash University, Victoria 3800, Australia}
\affiliation{Department of Data Science and Artificial Intelligence, Monash University, Victoria 3800, Australia}

\author{Murray Rudman}
\affiliation{Department of Mechanical and Aerospace Engineering, Monash University, Victoria 3800, Australia}

\author{Anja C. Slim}
\affiliation{School of Mathematics, Monash University, Victoria 3800, Australia}
\affiliation{School of Earth, Atmosphere and Environment, Monash University, Victoria 3800, Australia}

\collaboration{Received xxx; published xxx}%\noaffiliation

\date{\today}% It is always \today, today,

\begin{abstract}
% We present high-resolution numerical simulations of planar intrusions injected from a continuously supplied, narrow source into a linearly stratified ambient at the level of neutral buoyancy. We identify two distinct flow regimes governed by the dimensionless source width, $\Wint$: unstable jets ($\Wint \le 1.5$), which undergo symmetry breaking, and stable jets ($1.5 < \Wint \le 2.2$), which remain laminar. Focusing on the unstable regime, we show that the flow evolution proceeds through three distinct phases. The initial phase is dominated by the formation of a coherent vortex dipole that propagates as a "moving topography," exciting a field of mode-4 internal gravity waves that manifest as phase-locked lee waves. This is followed by a transition phase driven by an antisymmetric (sinuous) instability, which we quantify via spectral analysis to show exponential spatial growth consistent with linear jet theory. In the final phase, the flow relaxes into a self-similar state where the front position and cross-sectional area scale approximately as $t^{0.63}$ and $t^{0.7}$ respectively, recovering the sub-linear power laws observed in laboratory experiments. Furthermore, we found that the data collapse remarkably well under buoyancy-inertia scaling.

We present two and three dimensional direct numerical simulations of planar intrusions injected from a continuously supplied, narrow source into a linearly stratified ambient at the level of neutral buoyancy. The evolution passes through three phases. A laminar leading vortex dipole first advances at constant speed with a varicose wave train behind it; the trailing jet then undergoes a sinuous instability whose initial growth rate and wavelength agree with inviscid unstratified planar-jet theory evaluated on the local jet; the flow finally relaxes into a buoyancy-controlled wedge whose area grows linearly in time. Three-dimensional simulations show that the transition remains organized by the planar instability, while three-dimensionality reshapes the near-source region into a slowly evolving, actively mixing zone that entrains an additional 20–100\% of the source flux depending on source width. Rescaled with this measured effective flux, the downstream intrusion collapses reasonably well onto the similarity solution for a wide-source intrusion.
%for all subcritical cases, failing only when the front becomes critical with respect to the ambient's fastest internal wave. 
The narrow-source intrusion is thus a wide-source intrusion fed by an entraining and expanding near-source region.%, and we quantify that region's entrainment and momentum budget.
\end{abstract}

\keywords{internal waves, gravity currents, stratified flows}%Use showkeys class option if keyword
\maketitle

\tableofcontents
\section{\label{sec:intro}Introduction}

Intrusions are the predominantly horizontal flows of well-mixed fluid into a density-stratified ambient at their level of neutral buoyancy \cite{Linden12}. Such flows arise in a wide range of environmental and geophysical settings, from oceanic salt tongues \cite[]{baringer1997mixing, stephens1999dynamics} to the subsurface intrusion layers generated during the Deepwater Horizon disaster \cite[]{GoMcontemporaneous, GoMreview} to volcanic ash clouds \cite[]{CasGiordano24}. Their prevalence has motivated extensive research (see \cite{Ungarish20} for a comprehensive overview). Nevertheless, important knowledge gaps remain, particularly regarding internal-wave generation and feedback on the intrusive flow, and the dynamics of continuously supplied planar intrusions.

Much of the existing literature has focused on intrusions generated by lock exchanges or constant-volume dam-break releases \cite{Wu69, AmenMaxworthy80, FaustPlate84, deRooij99, FlynnSutherland04, BolsterHang08, MunroeVoegli09, MaurerLinden14}. By contrast, continuously supplied planar intrusions have been examined in only a small number of studies, mostly experimental \cite{EPA72, Manins76, CeddiaKerr25}. These studies show that intrusion evolution depends strongly on source width and geometry, with distinct wide- and narrow-source regimes. For sufficiently wide sources, Manins~\cite{Manins76} and Ceddia \etal~\cite{CeddiaKerr25} reported that the intrusion front propagates at approximately constant speed. Ceddia \etal~\cite{CeddiaKerr25} further showed that the intrusion thickness collapses onto a self-similar \(x/t\) profile, where \(x\) is the horizontal coordinate and \(t\) is time. Vu \etal~\cite{Vu25} numerically revisited this regime and likewise found self-similar collapse of both the thickness and vertically averaged velocity.  They also developed a modified shallow-water model that accounts for isopycnal displacement in the ambient and agrees well with their observations.

For narrow inlets, however, the dynamics are markedly different. Zuluaga-Angel \etal~\cite{EPA72} performed experiments from a thin slot and found a turbulent hydraulic jump near the source that thickened the flow by more than two orders of magnitude, followed by a laminar intrusion that approximately fitted a self-similar parabolic profile, with a slowly growing thickness at the jump tapering towards the front. Both the cross-sectional area and the front location scaled as \(t^{3/4}\). Ceddia \etal~\cite{CeddiaKerr25} revisited this regime using a diffuser aligned with the neutral buoyancy level. Their measurements agreed with the sublinear propagation reported by Zuluaga-Angel \etal~\cite{EPA72}, but they suggested a different interpretation: the near-source region remains quasi-steady, whereas only the tip evolves self-similarly. Their data collapsed well under a non-dimensionalization based on source flux and buoyancy. This regime appears not to have been studied numerically, and at present no model describes the behavior.

%Yet the near-source structures and coherent motions in these horizontally oriented flows remain poorly resolved because of experimental limitations. Much of the existing literature on coherent structures in stratified flows concerns buoyant jets and plumes, for which the mean motion is initially perpendicular to the isopycnals \cite{wright1976spreading,baines2001twodimensional,ganesan1994laminarjet}. In those cases, buoyancy acts primarily along the mean-flow direction. By contrast, horizontally evolving intrusions propagate approximately along isopycnals, altering the relative roles of shear, buoyancy, and stratification during the initial stages in ways that remain insufficiently understood. Furthermore, to our knowledge, no numerical work has yet resolved the late-time regime of constantly supplied intrusions at neutral buoyancy. The closely related work by Vu \etal~\cite{Vu25}, focused on the wide-source regime, found that broad large-scale evolution was only weakly affected by three-dimensionality. Whether the same remains true for narrow inlets is not yet clear. Two related questions therefore motivate the present study: how the flow develops in the near-source region immediately after the injection of the fluid, and to what extent the initially two-dimensional planar pathway remains valid at later times. 

We investigate this case using high-resolution direct numerical simulations (DNS). Our simulations resolve the full evolution of the flow from inception through the development of instability to the late-time fully developed intrusion regime. Most of the study is carried out in two dimensions, which allows us to perform a broad parameter sweep over inlet width and ambient stratification. However, we also use targeted three-dimensional simulations to determine when this approximation begins to break down and to establish the true late-time evolution. Finally, we explore the dynamics in the ambient induced by the intrusion. %Finally, we compare our results with previous theoretical and experimental studies to clarify the main dynamics associated with variations in source geometry.

The initial evolution of these intrusions is momentum‑dominated and for sufficiently narrow sources we find that the leading edge of the flow consists of a prominent vortex dipole, a pair of counter-rotating vortices. These have been studied in detail for planar jets injected into an ambient of the same density.  Here separation of the shear layer at the edge of the source leads to roll‑up and the formation of a coherent dipole (Didden~\cite{didden1979formation}; Gao and Yu~\cite{gao2016formation}; Zheng \etal~\cite{zheng2025formation}). As the discharge continues, circulation is fed from the trailing jet into the leading dipole until a finite time known as the formation number is reached, beyond which the dipole saturates.  This echoes the classical finite-formation-time picture established for starting vortex rings by Gharib \etal~\cite{gharib1998universal}. Behind the vortex dipole, the trailing jet becomes unstable to Kelvin–Helmholtz disturbances in the separated shear layers, giving rise to spanwise‑coherent wave‑like structures and secondary vortices that interact with the leading dipole. The instability mechanisms governing this stage are well established in planar jet theory: linear stability analysis by Sato~\cite{sato1960stability} demonstrated that a two‑dimensional jet supports both antisymmetric (sinuous) and symmetric (varicose) unstable modes, with the sinuous mode corresponding to lateral flapping about the jet centerline. This behavior is consistent with experimental observations of Kelvin--Helmholtz roll-up and vortex pairing in free shear layers and related planar jet flows \cite{BrownRoshko74,WinantBrowand74,thomas1989experimental}.  The evolution of a horizontally propagating vortex dipole in a stratified ambient appears not to have been studied.

Internal-wave generation is another key feature of intrusions advancing through a stratified ambient. Vertical displacements of isopycnals excite internal gravity waves that may feed back on the flow by redistributing energy and momentum, modulating the near-source shear layers and front geometry, and perturbing the local density field. Examples of such wave-intrusion coupling include the large-amplitude waves generated by river plumes \cite{nash2005river} and atmospheric internal bores such as the Morning Glory \cite{rottman1989formation}. For continuously supplied narrow intrusions, Zuluaga-Angel \etal~\cite{EPA72} reported an ambient response consisting of layered, alternating-direction horizontal currents, with no reported wave feedback on the intrusion. In contrast, under weak ambient stratification, Ceddia \etal~\cite{CeddiaKerr25} found clear wave feedback on the intrusion. The waves propagate faster than the front, and the wave troughs compress the intrusion tip from behind, producing a thin, flattened tip. As successive wave crests reach the front, this process repeats, generating a pulsing modulation in the front propagation. Nevertheless, their measurements suggested that internal waves had only a limited effect on the thickness profiles and overall propagation speed. Resolving the fine-scale dynamics of this coupled intrusion-wave system is therefore another objective of the present work.

We formulate our geometry and governing equations in \S\ref{sec:methodology} and give a description of our scalings, numerical approach and key diagnostic quantities.  We then describe the evolution of strictly two-dimensional simulations in \S\ref{sec:Phasesummary}.  The intrusions develop through three distinct phases: an initial dipole-dominated phase, a developing instability phase, and a final well-mixed phase.  We systematically explore the effect of stratification on the behavior. The two-dimensional instability is unstable to perturbations in the third dimension and so for the later evolution we present and describe simulations for select parameter values in \S\ref{sec:3D}. The intrusion passes through the same three phases, although there are both quantitative and qualitative differences in the final well-mixed stage.  At sufficiently long times, the flow consists of a short laminar jet  neighboring the source, followed by an entraining and expanding turbulent jet, and a non-entraining buoyancy-dominated far-field. We show that the far-field appears to be reasonably described by the wide-source model of Vu \etal~\cite{Vu25}.  In \S\ref{sec:conclusions}, we summarize our results.

\section{\label{sec:methodology}Formulation and numerical method}
\subsection{\label{sec:governing equation}Mathematical model}

\begin{figure}    
\includegraphics[]{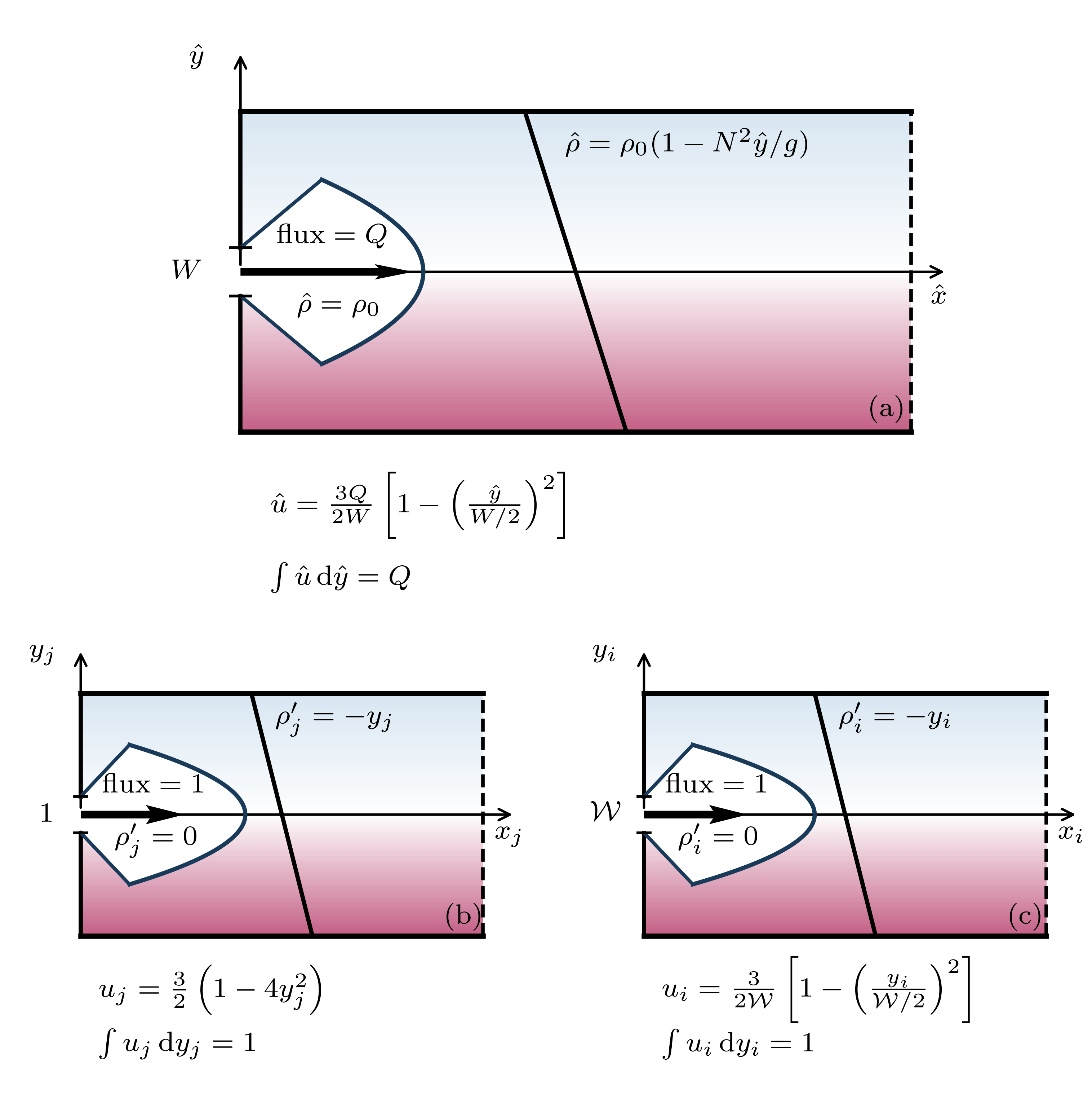}
    \caption{Configuration and key variables in (a) dimensional form, (b) dimensionless form with jet scalings, and (c) dimensionless form with intrusion scalings.  The spanwise coordinate is directed out of the page in all cases.}
    \label{fig:config}
\end{figure}

We consider the flow generated by a constant supply of well-mixed fluid into a density stratified ambient through a narrow slot. The injected fluid has density $\Drho$ and is supplied at a fixed flux $Q$ per unit span. The injection begins at time $\DIM{t}=0$ through an inlet of vertical extent $\Dwidth$, positioned symmetrically about the level where the ambient density equals $\Drho$, and extending uniformly across the span.  This configuration is shown in FIG.~\ref{fig:config}a.

The subsequent flow takes place in a channel of dimensional length $\DLa$ and height $\DHa$. Cartesian coordinates $\DIMbx = (\DIM{x}, \DIM{y}, \DIM{z})$ are used to describe the domain, with $\DIM{x}$ streamwise, $\DIM{y}$ vertical and $\DIM{z}$ spanwise (out of the page in FIG.~\ref{fig:config}a), and with the origin located on the centerline of the inlet. Dimensional quantities are indicated with hats throughout. 

The ambient is initially quiescent and linearly stratified, with density distribution profile
\[
\DIM{\rho} = \Drho\left(1 - N^2\DIM{y}/g\right),
\]
so that $\Drho$ is the density at $\DIM{y}=0$. Here $N$ denotes the buoyancy frequency associated with the background stratification, and gravity of magnitude $g$ acts in the negative $\DIM{y}$ direction.

The imposed inflow velocity field is
\[
\DIMbv = (\DIM{u},\DIM{v},\DIM{w}) = \left(\DIM{u}_\text{inlet}(\DIM{y}),0,0\right), %% 3D
\]
where $\DIM{u}_\text{inlet}(\DIM{y})$ has a parabolic profile as shown in FIG.~\ref{fig:config}a.

The flow is governed by the incompressible Navier-Stokes equations under the Boussinesq approximation given by
\begin{align} \label{eq:gov}
    \DIM{\bnabla} \cdot \DIMbv &= 0, \\
    \rho_0 \left( \frac{\partial \DIMbv}{\partial \hat{t}} + (\DIMbv \cdot \DIM{\bnabla}) \DIMbv \right) &= -\DIM{\bnabla} \DIM{p} + \rho_0 \nu \DIM{\nabla}^2 \DIMbv -  \DIM{\rho}' g \,\mathbf{e}_y, \quad \text{and} \quad\\
    \pdiff{\DIM{\rho}'}{\DIM{t}} +  \DIM{\bnabla}\cdot \left(\DIM{\rho}' \DIMbv \right) &=  D \DIM{\nabla}^2 \DIM{\rho}',
\end{align}
where $\DIM{p}$ is the pressure relative to the hydrostatic contribution $\rho_0 g \DIM{y}$, $\DIM{\rho}'=\DIM{\rho}-\rho_0$ is the density relative to the intrusion density, $\nu$ is the kinematic viscosity, $D$ is the scalar diffusivity, and $\mathbf{e}_y$ is a unit vector in the $\DIM{y}$ direction.   

We keep track of the intruded fluid through a color function $c(\hat{x},\hat{y},\hat{z},\hat{t}),$ which evolves according to %% 3D 
\begin{equation} \label{eq:color}
    \frac{\partial c}{\partial\hat{t}}+\DIM{\bnabla} \cdot \left(c\DIMbv\right)= D \DIM{\nabla}^2 c.
\end{equation}
Initially the ambient fluid has $c=0$. The injected fluid has $c=1$ at the inlet.

Boundary conditions are specified as follows.  At the inflow boundary $\DIM{x}=0$, a prescribed velocity profile is applied for $|\DIM{y}|\le W/2$, as given in FIG.~\ref{fig:config}a. At the downstream boundary $\DIM{x}=\DLa$, an outflow condition is imposed. All remaining solid boundaries satisfy no-penetration and free-slip conditions. For the scalar fields $\DIM{\rho}'$ and $c$, we impose no-flux conditions. All fields are periodic in $\DIM{z}$ with periodicity $L_z$. %% 3D

\subsection{Rescaling}\label{sec:rescaling}
To provide a clear characterization of the flow across different dynamical regimes, we employ two distinct non-dimensionalizations. During the initial stage, the motion is controlled by the source. In this regime, the appropriate reference scales are those prescribed by the source and associated with the initial inertia-dominated jet propagation. We therefore present the early-time results using what we term jet scaling. As the flow evolves, buoyancy progressively influences the dynamics more strongly and the relevant scales become those from an inertia-buoyancy balance. Accordingly, the results are rescaled using buoyancy-based reference quantities, which we refer to as intrusion scaling.

\subsubsection{Jet scaling}\label{sec:jet rescaling}
For the jet scaling, we use the length scale $W$, velocity scale $Q/W$, and time scale $W^2/Q$ to set:
\begin{equation}
\bv = (\ujet,\vjet,\wjet) =\frac{W}{Q}\DIM{\mathbf{v}}, \quad \bx = (\xjet,\yjet,\zjet) = \frac{1}{W}\DIM{\mathbf{x}} \quad \text{and} \quad \tjet = \frac{Q}{W^2} \DIM{t}. %% 3D
\end{equation}
In addition, we use the density variation across the width of the source $\rho_0N^2 W/g$ as a density scale and $\rho_0 Q^2/W^2$ as a scale for pressure variations to set
\begin{equation}
\rhojetnd = \frac{g}{\rho_0N^2 W} \, \DIM{\rho}' \quad \text{and} \quad \pjetnd = \frac{W^2}{ \rho_0 Q^2}\,\DIM{p}.
\end{equation}
The configuration in this non-dimensionalization is shown in Fig.~\ref{fig:config}b.

The governing equations in dimensionless form are now
\begin{subequations} \label{eq:gov_dimless}
\begin{align}
\bnabla \cdot \bv &= 0, \label{eq:gov_dimless_cty}\\
\frac{\partial \bv}{\partial \tjet} + (\bv \cdot \bnabla) \bv &= -\bnabla \pjetnd + \frac{1}{Re} \nabla^2 \bv - Ri\, \rhojetnd \mathbf{e}_{y}, \label{eq:gov_dimless_mom}\\
\frac{\partial \rhojetnd}{\partial \tjet} + \bnabla\cdot (\rhojetnd \bv) &= \frac{1}{Re\,\Sc} \nabla^2 \rhojetnd, \label{eq:gov_dimless_rho}\\
\frac{\partial c}{\partial \tjet} + \bnabla \cdot (c \bv) &= \frac{1}{Re\,\Sc} \nabla^2 c \label{eq:gov_dimless_c}
\end{align}
\end{subequations}
with an initial density profile in the ambient relative to the intrusion of $\rhojetnd = -y$ at $\tjet = 0$.

This rescaling leaves six dimensionless parameters: the Reynolds number $Re = Q/\nu$, the Schmidt number $Sc = \nu/D$, the Richardson number $Ri$, and the three geometric quantities
\[
\Ha = \DHa/W, \quad \La = \DLa/W \quad \text{and} \quad \Lzjet = \DLz/W. %% 3D
\]
The spanwise period $\Lzjet$ plays no role in the two-dimensional simulations. %% 3D
The Richardson number compares buoyancy and inertial effects.  The limit $Ri \to 0$ represents an unstratified ambient. As $Ri$ increases, buoyancy effects strengthen.

\subsubsection{Intrusion scaling}\label{sec:Intrusion rescaling}
Eventually, buoyancy becomes dominant, and thus we also present results rescaled for buoyancy-driven propagation (see also \cite{CeddiaKerr25}).  Specifically, we use the velocity scale $\sqrt{NQ}$, length scale $\sqrt{Q/N}$, and time scale $1/N$ to set
\begin{equation}
\mathbf{v}_\mathrm{int} = (\uint,\vint,\wint) =\DIM{\mathbf{v}}/\sqrt{NQ}, \quad \mathbf{x}_\mathrm{int}= (\xint,\yint,\zint) = \DIM{\mathbf{x}}/\sqrt{Q/N}, \quad \text{and} \quad \tint = N\DIM{t}. %% 3D
\end{equation}
We use the initial ambient density difference across the typical buoyancy length scale $\Drho \sqrt{N^3Q}/g$ as a density scale and $\rho_0 N Q$ as a scale for pressure variations to set \[
\rhointnd = \frac{g}{\rho_0\sqrt{N^3Q}}\,\DIM{\rho}'\quad \text{and} \quad \pintnd = \DIM{p}/(\rho_0 NQ).
\]
The configuration in this non-dimensionalization is shown in Fig.~\ref{fig:config}c.

The governing equations in dimensionless form are now
\begin{subequations}\label{eq:gov_dimless_int}
\begin{align} 
    \bnabla \cdot \bvint &= 0, \label{eq:gov_dimless_cty_int}\\
    \frac{\partial \bvint}{\partial \tint} + (\bvint \cdot \bnabla) \bvint \ &= -\bnabla \pintnd + \frac{1}{\Rey} \nabla^2 \bvint -  \rhointnd \,\mathbf{e}_{\yint} \quad \label{eq:gov_dimless_mom_int}\\
    \pdiff{\rhointnd}{\tint} +  \bnabla\cdot \left(\rhointnd \bvint \right) &= \frac{1}{\Rey \, \Sc} \nabla^2 \rhointnd \qquad \text{and} \label{eq:gov_dimless_rho_int}\\
    \frac{\partial c}{\partial{\tint}}+{\bnabla} \cdot \left(c{\bvint}\right) &= \frac{1}{\Rey \, \Sc}{\nabla}^2 c. \label{eq:gov_dimless_c_int}
\end{align}
\end{subequations}
The initial density difference profile is $\rhointnd = -\yint$ at $\tint=0$.  

The six dimensionless parameters are now: the Reynolds number $\Rey=Q/\nu$, the Schmidt number $\Sc = \nu/D$, and the four geometric quantities
\[
\width = \Dwidth \sqrt{N/Q}, \quad \Hint = \DHa \sqrt{N/Q}, \quad \Lint = \DLa \sqrt{N/Q} \quad \text{and} \quad \Lzint = \DLz \sqrt{N/Q}. %% 3D
\]
%When $\Wint$ is small, then the source is jet-like, much narrower and faster than the natural buoyancy scale. 
The parameter $\Hint$ represents the dimensionless ambient height, or, alternatively, $\pi$ times the ratio of the speed of mode-1 long-waves $N\DHa/\pi$ to the speed of the intrusion $\sqrt{NQ}$. If $\Hint$ is relatively small, then the intrusion propagates faster than the fastest internal waves.  Conversely, if $\Hint$ is relatively large, then internal waves can propagate ahead of the intrusion. In this study we focus on this latter case. 

\subsubsection{Relation between the jet and intrusion scalings}\label{relations}
We use two non-dimensionalizations because the early evolution is governed by jet-like dynamics, whereas the late-time propagation follows buoyancy-inertia intrusion scaling. The two are linked by %the dimensionless inlet width
% \[
% \Wint \equiv W\sqrt{N/Q},
% \]
% which implies
% \[
% Ri=\frac{N^2W^4}{Q^2}=\Wint^4,
% \qquad\text{and hence}\qquad
% \Wint = Ri^{1/4}.
% \]
% Consequently, the non-dimensional variables satisfy
\[
\xint=\Wint \xjet,\qquad
\yint=\Wint \yjet,\qquad
\zint=\Wint \zjet,\qquad
\tint=\Wint^{2}\tjet,
\qquad
\mathbf{v}_{i}=\mathbf{v}_j/\Wint. %% 3D: z added
\]
The dimensionless domain sizes follow $\Hint=\Wint \Ha$, $\Lint=\Wint \La$ and $\Lzint=\Wint \Lzjet$. Most importantly, the Richardson number of the jet scaling and the inlet width of the intrusion scaling are related by %% 3D: Lzint added
\begin{equation}
Ri = \Wint^4.
\end{equation}
Throughout, we label cases using $\Wint$ in preference to $Ri$ for consistency with the work of Vu \etal~\cite{Vu25}. We use the jet scaling when discussing the early, near-source jet-like adjustment, and the intrusion scaling when discussing the late-time buoyancy-inertia intrusion regime.

\subsection{Direct numerical simulations}

The simulations were run using the spectral element code \textsc{Semtex} \cite{blackburn2019semtex}.
Details of the specific application to the intrusion configuration are given by Vu \etal~\cite{Vu25}.

The two-dimensional simulations were run on a single core of an Intel Xeon Gold 5220R with multi-threaded execution disabled, the high-resolution cases requiring approximately seven days of computational time each; the three-dimensional simulations were run on $60$ to $80$ CPU cores and required approximately $30$ to $40$ days each.  All simulations used a polynomial order of $N_p = 9$ within each element and a time step $\Delta t \approx \mathcal{O}(10^{-4})$. Resolution requirements are documented in Appendix~\ref{app:3D}.

The two-dimensional implementation was also verified against the conservative finite-volume code of Rudman~\cite{rudman1998volume}, modified to carry a continuously varying density field in place of a discontinuous interface.  The intrusion profiles show good agreement \cite{Vu25}.

% For the two-dimensional simulations, the high-resolution DNS runs required approximately 7 days of computational time while running on a single core. The specific processor used was an Intel Xeon Gold 5220R, with multi-threaded execution disabled. Details of code convergence and verification are given in Appendix~\ref{app:3D} (FIG.~\ref{fig:convergence}).  
% \CV{The paragraph above and below this need fixing}
% The code was further verified against a version of the two-dimensional conservative finite-volume code of Rudman~\cite{rudman1998volume}. The only difference with Rudman's original code was the inclusion of a continuously varying density field instead of a discontinuous density interface. The profiles of the intrusions using the two methods show excellent agreement.  

\subsection{\label{sec:key variables} Key diagnostic variables}
To characterize the evolution of the intrusion, we define diagnostic variables based on generic dimensionless coordinates $\mathbf{x} = (x,y,z)$ and velocity field $\mathbf{v} = (u,v,w)$, which correspond to either the jet or intrusion scaling as appropriate for the analysis context, with $\mathcal{L}_z$ the corresponding dimensionless spanwise period.  Three-dimensional fields are reduced to the $(x,y)$ plane by the spanwise average %% 3D
\begin{equation} \label{eq:spav}
\spav{f}(x,y,t) = \frac{1}{\mathcal{L}_z}\int_0^{\mathcal{L}_z} f(x,y,z,t)\,\mathrm{d}z,
\end{equation}
which reduces to the field itself for the two-dimensional simulations. %The diagnostics below are defined in terms of the spanwise-averaged fields, so that they apply unchanged to both. %% 3D

We use two different definitions for the intrusion thickness.  The first is based on integrals of the color function: we define the upper and lower intrusion thicknesses, \(h_t\) and \(h_b\), as the integral of the spanwise-averaged scalar function $\spav{c}$ from the centerline to the top and bottom domain boundaries respectively: %% 3D
\begin{equation} \label{eq:h}
   h_t(x,t) = \int_{0}^{\Harbitrary/2} \spav{c}(x,y,t)\,\mathrm{d}y, \quad \text{and} \quad h_b(x,t) = \int_{-\Harbitrary/2}^{0} \spav{c}(x,y,t)\,\mathrm{d}y. 
\end{equation}
%where $\Harbitrary$ denotes the dimensionless ambient height ($\Ha$ or $\Hint$). 
The second definition is based on the outermost contour on which a threshold color function is encountered: %% 3D
\begin{equation} \label{eq:h_contour}
   h_t(x,t) = \max\{\,y : \spav{c}(x,y,t) = c_\mathrm{th}\,\}, \quad \text{and} \quad h_b(x,t) = -\min\{\,y : \spav{c}(x,y,t) = c_\mathrm{th}\,\},
\end{equation}
with threshold $c_\mathrm{th}=0.1$ throughout. In both cases, the averaged half-thickness of the intrusion is $h = (h_t+h_b)/2$ and the full thickness is $2h = h_t+h_b$, and both are applied to the spanwise-averaged field.  %Both reduce to the corresponding two-dimensional quantities for the two-dimensional simulations.  
The integral definition corresponds to the form used by Vu \etal~\cite{Vu25}; the contour definition is more akin to the experiments of Ceddia \etal~\cite{CeddiaKerr25}, where a dye intensity threshold was used.

% \CV{Please complete here - is it the outside contour, or is it an integral similar to the area integral? update the area below so it works for 2D and 3D - I think $\iint_<c>>c_th$ would work?  also update the vertically averaged horizontal velocity - I think there shouldn't be a c in there for the contour based velocity.  Make sure clear everywhere which h is which}  

% In both cases, the averaged half-thickness $h$ of the intrusion is $h = (h_t+h_b)/2$ and the full thickness is $2h = h_t+h_b$.  With the first definition, this corresponds to the form used by Vu \etal~\cite{Vu25}.  The second is more akin to that of the experiments of Ceddia \etal~\cite{CeddiaKerr25} where a dye intensity threshold was used.

Related to the second thickness definition, we quantify the apparent intrusion area by the volume per unit span of fluid with $c$ above a threshold, %% 3D
\begin{equation} \label{eq:area}
A(t) = \iint_{\spav{c} \,>\, c_\mathrm{th}} \mathrm{d}x\,\mathrm{d}y.
\end{equation}
%\textcolor{red}{which for the two-dimensional simulations is the area of the region in which $c > c_\mathrm{th}$}.
%where the threshold \(c_\mathrm{th}=0.1\); for the two-dimensional simulations this is the area of the region in which $c > c_\mathrm{th}$. 
%\AS{Can you check the reds}
The threshold $c_\mathrm{th}=0.1$ is chosen to be small enough to include the bulk of the intrusion and large enough to exclude the tracer that diffusion spreads into the ambient.  In the narrowest and most strongly entraining cases, varying $c_\mathrm{th}$ by a factor of four, from $0.05$ to $0.2$, changes the inferred thickness and area by up to approximately $15\%$, and the fitted late-time rate of area growth by up to approximately $10\%$.  These quantities therefore retain some sensitivity to the chosen threshold.  Nevertheless, the identified flow regimes, the large-scale intrusion structure and the trends across simulations remain unchanged across the tested range.
%% AS: if the 3D area in the 3D section was in fact computed from the spanwise-averaged field (area where <c> > c_th) rather than per plane, swap the definition to that one -- the two differ where the interface is corrugated in z. 
% is insensitive to small variations, while filtering out ambient background contamination.

The tip location $x_f$ at time $t$ is calculated as the largest grid coordinate $x_n$ for which 
\begin{equation}
    \max (h_t(x_n,t), h_b(x_n,t)) > \varepsilon,
\end{equation}
with $\varepsilon = 0.01$. Reducing $\varepsilon$ did not noticeably affect the identified tip location. The choice of definition for $h$ will be clear from context.

% \CV{Do you use anything aside from the vertically averaged velocity?  I don't think so.  If that's the case, let's just give the "in particular".  Again, I think there should be two calculations there - one with c and one without.  This might actually improve the velocity plots later...}
% Finally, we introduce variables averaged across the intrusion according to
% \begin{equation} \label{eq:av}
% \bar{f}(x,t) = \frac{1}{2h} \int^{\Harbitrary/2}_{-\Harbitrary/2}\spav{c f}(x,y,t)\,\text{d}y. %% 3D
% \end{equation}

The horizontal velocity averaged across the intrusion, $\bar u(x,t)$, is defined consistently with whichever thickness is in use.  With the integral thickness it is the color-weighted average
\begin{equation} \label{eq:ubar}
    \bar u(x,t)=\frac1{2h}\int^{\Harbitrary/2}_{-\Harbitrary/2}\spav{cu}\,\text{d}y , %% 3D
\end{equation}
and with the contour thickness it is the plain average of the velocity between the contours,
\begin{equation} \label{eq:ubar_contour}
    \bar u(x,t)=\frac1{2h}\int^{\;h_t}_{-h_b}\spav{u}\,\text{d}y . %% 3D
\end{equation}

In the ambient, a key variable is the vertical displacement $\eta(x,y,z,t)$ of the isopycnals passing through the point $(x,y,z)$ at time $t$.  These can be obtained via
\begin{equation}
    \eta(x,y,z,t) = y + \rho'(x,y,z,t). \label{eq:isopycnal displacement} %% 3D
\end{equation}
Its spanwise average is $\spav{\eta} = y + \spav{\rho'}$. %% 3D

% The broad scale structures that we observe for a jet-like intrusion is the symmetry-breaking phenomenon for $\Wint < 2.0$. As $\Wint$ approaches $2.0$, the flow has become symmetric and transitioned to a wide, `lazy' regime (see \cite{Vu25}). For small $\Wint \leq 1$, after the symmetry-breaking phenomenon, the flow we  a significant entrainment happening. We begin by describing a typical (baseline) planar jet for each regime, and how finer-scale features arise and vary across the phases will be discussed in section XXX.

\section{Two-dimensional simulations}\label{sec:Phasesummary}

\begin{figure}
    \centering
    \includegraphics[width=\linewidth]{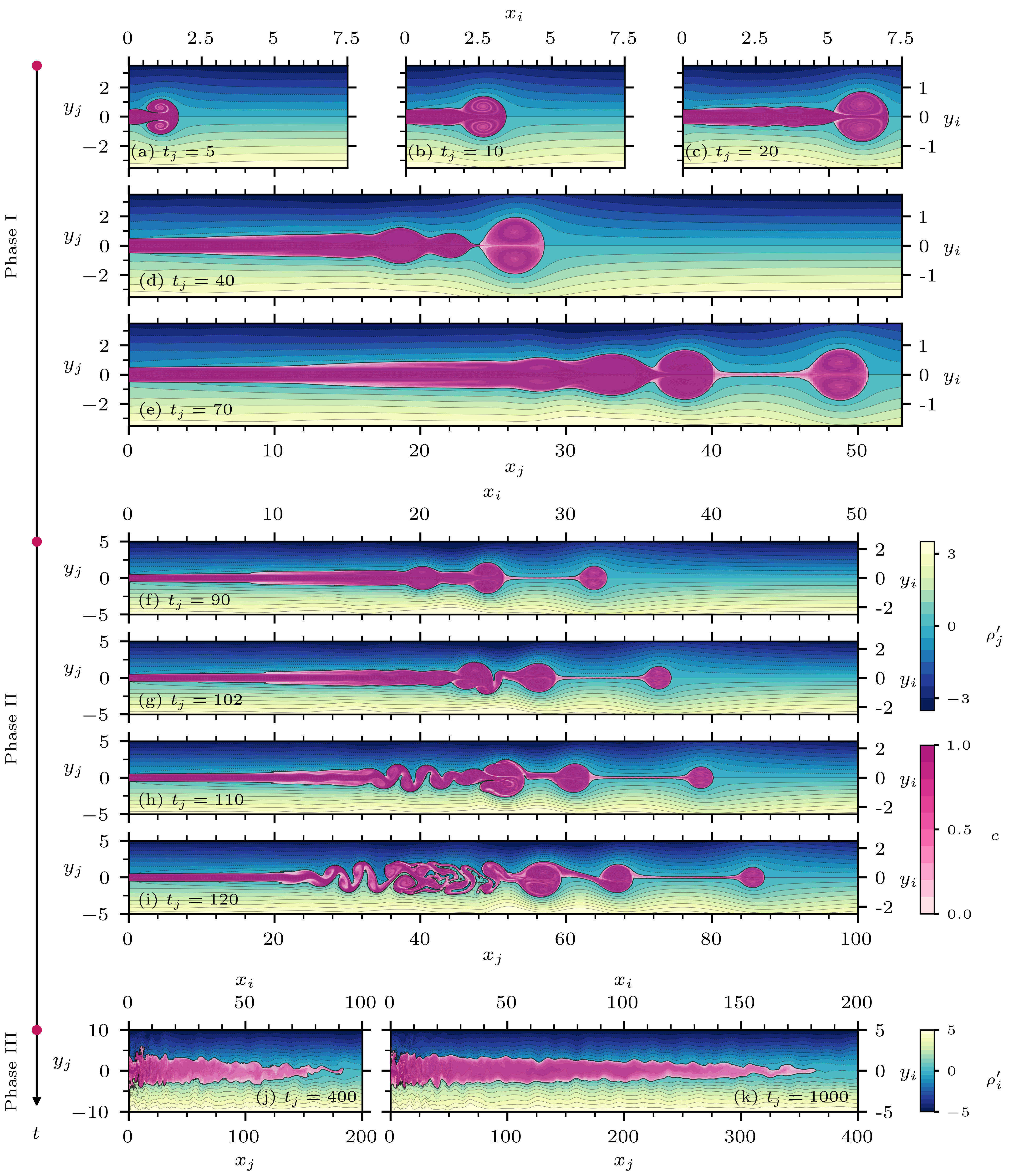}
    \caption{Summary of the three main phases of the evolution. The snapshots are for the simulation for $Re =1000$, $\Hint=10$, $\Wint = 0.5$, and $\Lint = 400$. All panels show the color function $c$ over the ambient density relative to the intrusion $\rhojetnd$ ($\rhointnd$ in Phase~III): phase I in (a--e), Phase II in (f--i) and Phase III in (j,k), the last from the intrusion-scaled run and with the vertical axis stretched by a factor of $3.5$.}
    \label{fig:2Dsnaps}
\end{figure}

In this section, we consider strictly two-dimensional intrusions, exploring the temporal phases of their evolution and how their behavior varies with parameters.  Across the parameter space that we consider, the flow transitions through three distinct phases: an initial dipole-dominated phase, followed by a developing instability phase, and a final mixed phase.  The intrusion profile for these three phases is shown in FIG.~\ref{fig:2Dsnaps} for a baseline simulation. A movie of the first two phases is provided with the Supplementary Material.  The parameter values for this simulation under both scalings are summarized in Table~\ref{tab:baseline_params}.  We describe the phases in detail and how the behavior varies across parameter space in the next three subsections.

%\CV{Add Ri for jet in the table below (empty for intrusion)?}
\begin{table}[h]
\caption{Dimensionless parameters for the baseline simulation under both scaling frameworks. Definitions of the jet and intrusion scalings and their relations are given in \S~\ref{sec:rescaling}. The Richardson number $Ri$ is a parameter of the jet scaling only; in the intrusion scaling its role is taken by the source width, $\Wint = Ri^{1/4}$, which we use to label cases throughout (see \S\ref{relations}).}
\centering
\renewcommand{\arraystretch}{1.25}   % more vertical spacing between rows
\setlength{\tabcolsep}{10pt}         % more horizontal padding between columns
\begin{tabular}{l|c|c|c|c|c|c}
Jet scaling
& $\Wint=0.5$
& $Ri=0.0625$
& $\Ha=20$
& $Re=1000$
& $Sc=10$
& $\La=800$ \\
\hline
Intrusion scaling
& $\Wint=0.5$
&
& $\Hint=10$
& $Re=1000$
& $Sc=10$
& $\Lint=400$ \\
\end{tabular}
\label{tab:baseline_params}
\end{table}

\subsection{Phase~I: Initial dipole-dominated phase}\label{sec:Phase1}
\begin{figure}
    \centering
    \includegraphics[width=\linewidth]{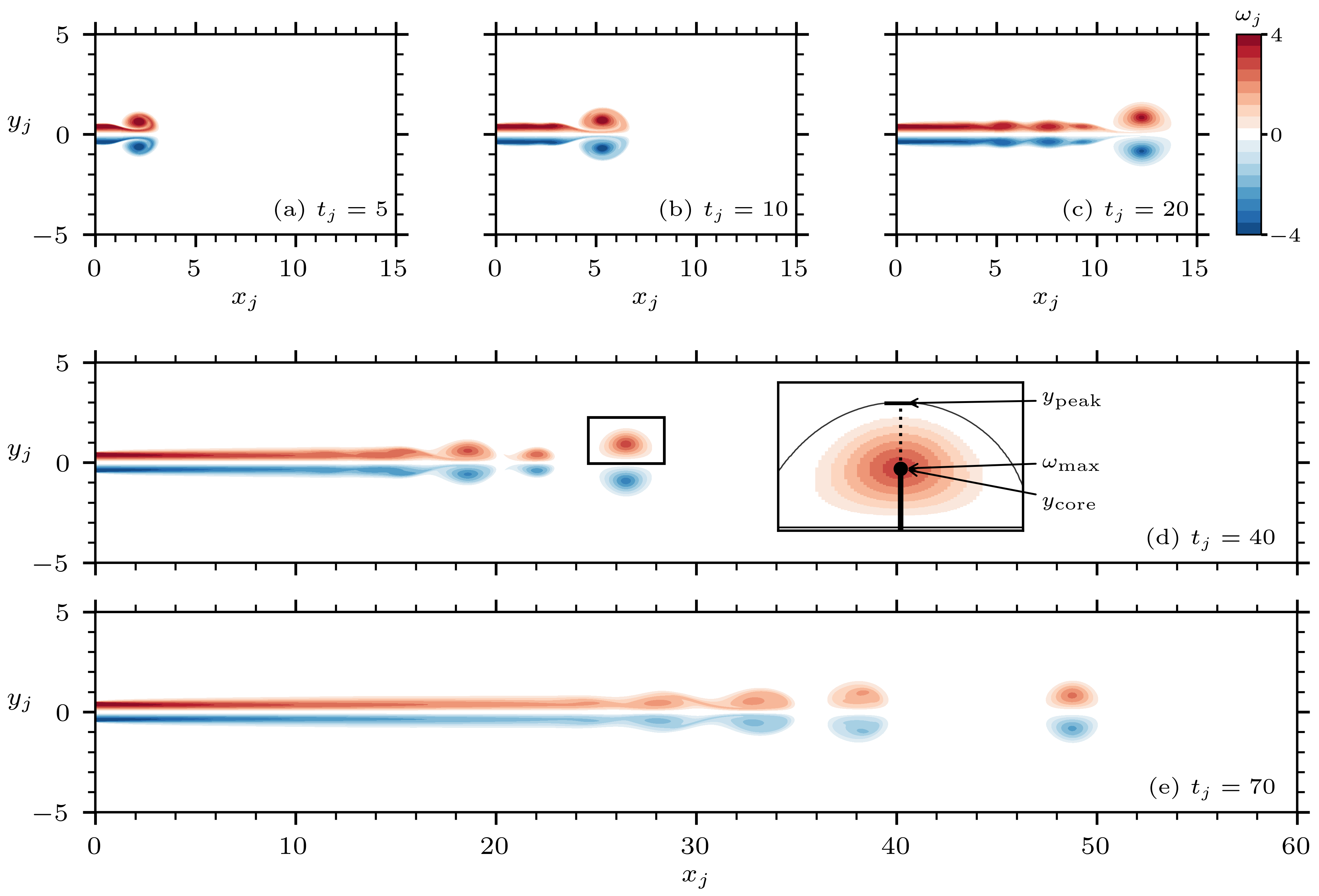}
    \caption{Evolution of the leading vortex dipole for the baseline case $\Wint = 0.5$. (a--e) Vorticity $\omega_j$ at $\tjet = 5$, $10$, $20$, $40$ and $70$; the inset in (d) defines $y_\mathrm{peak}$, $\omega_{\max}$ and $y_\mathrm{core}$. }
    %(a) Maximum vorticity magnitude within the vortex core region as a function of time. (b) Intrusion half-thickness evaluated at the vortex core location (solid, $y_\mathrm{peak}$) and the height of the vorticity maximum (dashed, $y_\mathrm{core}$). 
    \label{fig:vorticity_profile}
\end{figure}
%\CV{Please remove the top two panels of the vorticity figure and relabel the other panels - not critical}

\begin{figure}
    \centering
    \includegraphics[width=\linewidth]{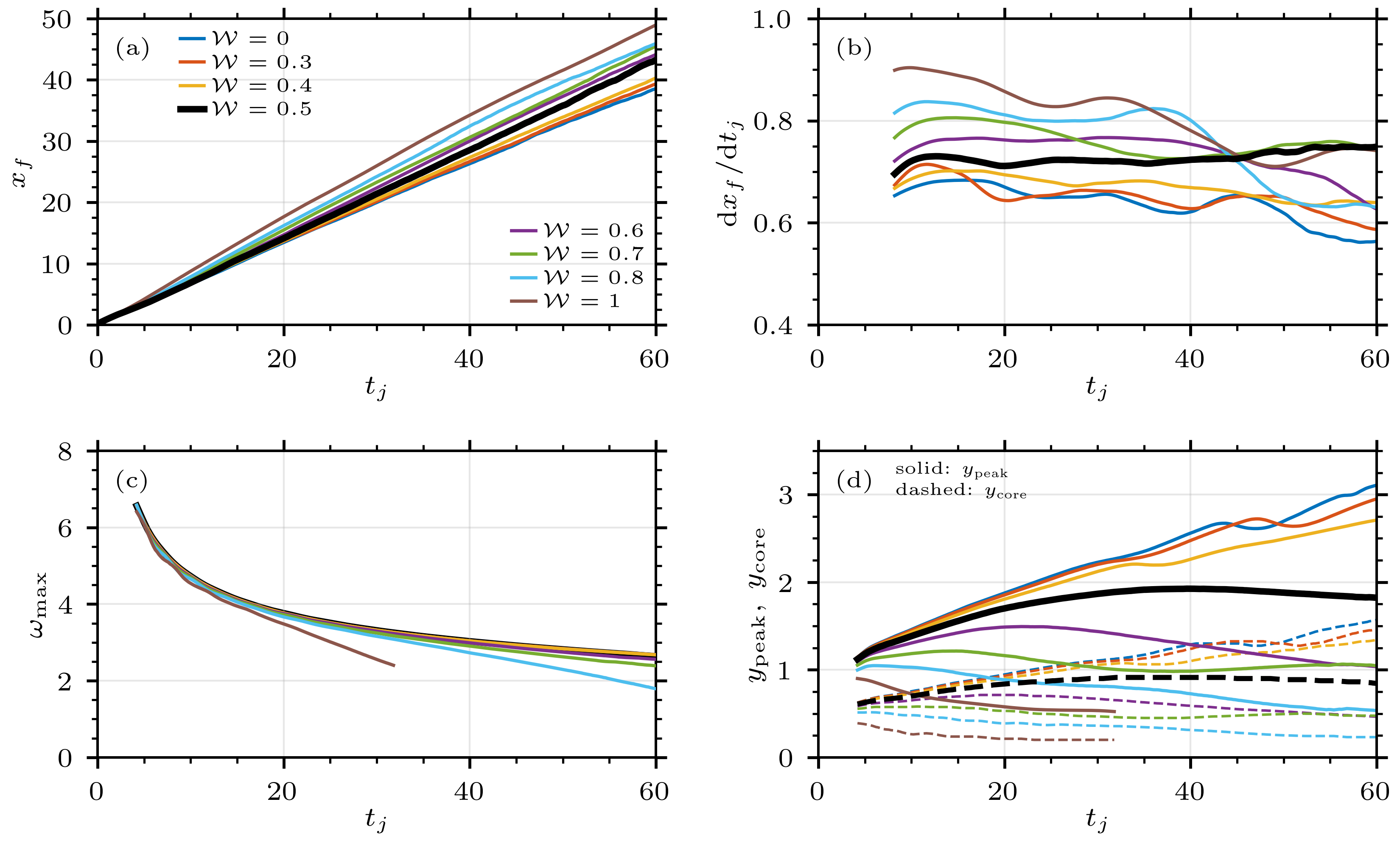}
    \caption{Front position and leading-dipole properties against time for $\Wint = 0$, $0.3$, $0.4$, $0.5$, $0.6$, $0.7$, $0.8$ and $1$, with $\Wint = 0.5$ in bold black. (a) Intrusion front position $x_f$. (b) The corresponding front speed $\mathrm{d}x_f/\mathrm{d}\tjet$. (c) Maximum vorticity $\omega_{\max}$ of the leading dipole. (d) Half-thickness of the leading dipole $y_\mathrm{peak}$ (solid) and height of its vorticity maximum $y_\mathrm{core}$ (dashed), as defined in FIG.~\ref{fig:vorticity_profile}(d). In (c,d), for $\Wint = 1$ the tracker loses the dipole at $\tjet \approx 32$.}
    \label{fig:tip_vortex_varyingW}
\end{figure} 
% \begin{figure}
%     \centering
%     \includegraphics[width=\linewidth]{paper_figures/figJ_crest_heatmaps.png}
%     \caption{Spatiotemporal evolution of the upper intrusion half-thickness $h_t(\xjet,\tjet)$ for each stratification $\Wint$, with local thickness maxima appearing as bright ridges. The solid white line marks the intrusion tip. Panel borders are colored by $\Wint$. 
%     {We plot $h$ as the integral of the color function.}}
%     \label{fig:crests_varyingW}
% \end{figure}

% \begin{figure}
%     \centering
%     \includegraphics[width=\linewidth]{paper_figures/figN_crest_heatmap.png}
%     \caption{Spatiotemporal evolution of the local thickness maxima (crests) on the upper surface of the intrusion; the color indicates the local intrusion thickness $h_t$, as shown by the color bar. The full field $h_t(\xjet,\tjet)$ is shown, in which the crests appear as bright ridges. The solid white line marks the intrusion tip, identified by the leading vortex dipole. The first peak of the trailing jet propagates at approximately the same speed as the intrusion tip for $\tjet \lesssim 45$, indicating phase-locked motion between the vortex head and the trailing surface undulations. We plot $h$ as the integral of the color function.}
%     \label{fig:crests}
% \end{figure}

\begin{figure}
    \centering
    \includegraphics[width=\linewidth]{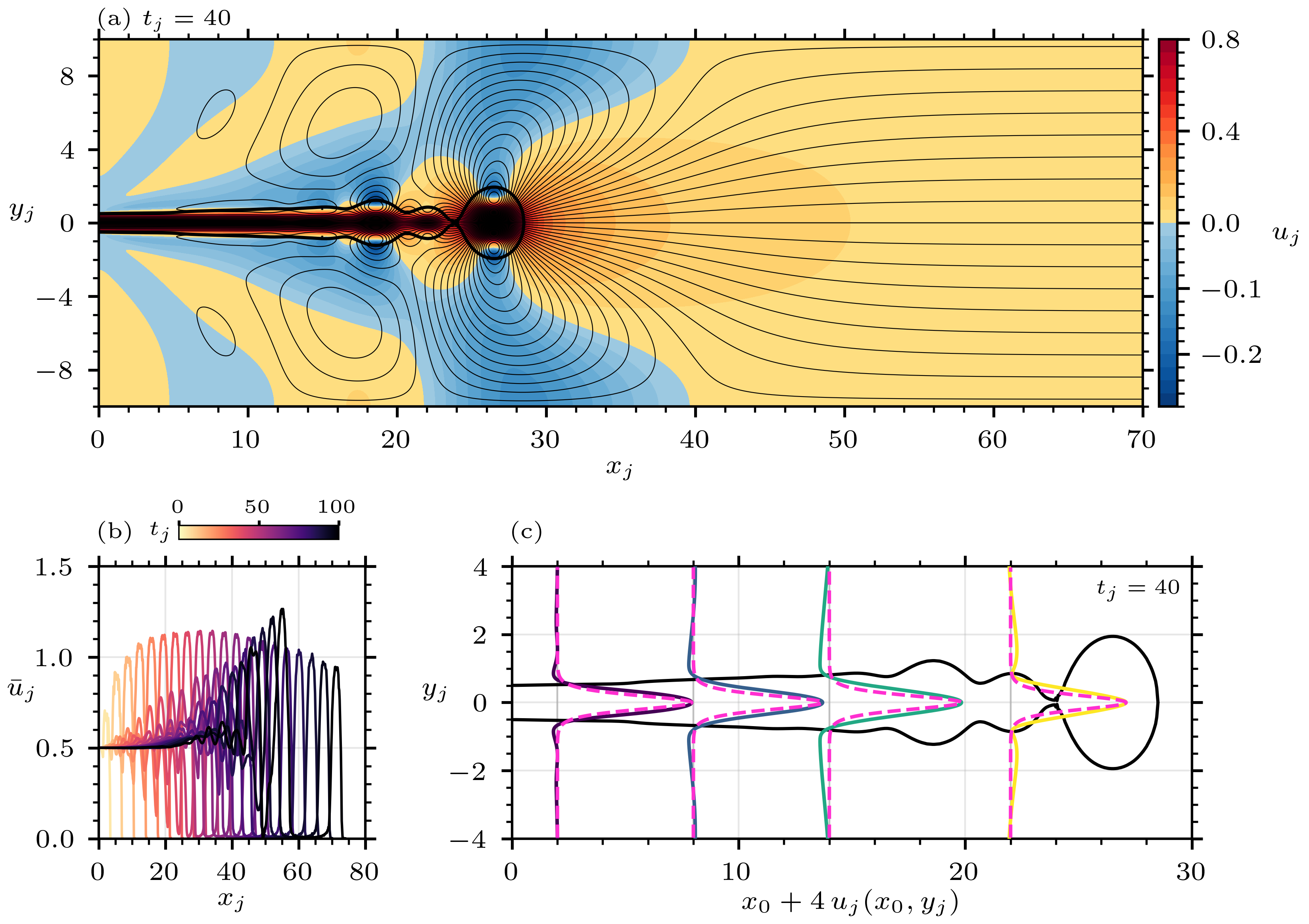}
    \caption{(a) Streamlines superimposed on the horizontal velocity field (color map) at $\tjet=40$ for the baseline case; the intrusion envelope is marked by the bold contour that extends to $\xjet \approx 28$. The leading vortex dipole and the attached trailing jet are clearly visible, with predominantly horizontal flow within the intrusion body. (b) Depth-averaged horizontal velocity $\bar u(\xjet)$ at times $\tjet = 5$ to $100$. (c) Velocity profiles $u(\yjet)$ at stations $x_0 = 4, 10, 16, 22$ at $\tjet = 40$, offset by $x_0$ and scaled by a factor of $4$, with the intrusion envelope in black. Dashed curves are the Bickley profile $u_m\,\mathrm{sech}^2(\yjet/\ell)$, with $u_m$ taken from the data and $\ell = 5/18$ set by the inlet fluxes, so the width is predicted rather than fitted. {We plot {$\bar{u}$} defined from $h$ as the integral of the color function.}}
    \label{fig:velocity_profiles}
\end{figure}

%\AS{Check the definition of the vortex extent below - that it matches with the figure}
Figure~\ref{fig:2Dsnaps}a-e shows the initial evolution of an intrusion with $\width = 0.5$.  The vorticity at the corresponding times is shown in FIG.~\ref{fig:vorticity_profile}.  Throughout this phase, the flow remains symmetric about the centerline.  The profile is dominated by a prominent leading vortex dipole. This forms at early times (FIG.~\ref{fig:2Dsnaps}a) from roll up of the two shear layers extending from the upper and lower edges of the source and it persists throughout this phase.  

The front position and properties of the leading vortex over time are shown in the bold black curves of FIG.~\ref{fig:tip_vortex_varyingW}.
We identify the location of the dipole core by determining the position \(x_{\text{core}}\) at which the magnitude of the vorticity, $\omega = \partial v/\partial x - \partial u/\partial y$, is locally maximal in the region near the intrusion front. This location corresponds to the core of the vortex. The half-thickness of the dipole is then defined as $h(x_{\mathrm{core}})$. The maximum vorticity in the dipole gradually decays over time (FIG.~\ref{fig:tip_vortex_varyingW}a), while its thickness initially grows during formation and subsequently decreases slightly (FIG.~\ref{fig:tip_vortex_varyingW}d). Its speed remains effectively constant with $x_f \approx 0.72t$.

%The vortex strength decays monotonically, while the dipole thickness initially grows during formation and subsequently decreases as stratification suppresses the dipole.

Behind the vortex dipole, a small amplitude top/bottom symmetric undular wave train develops (FIG.~\ref{fig:2Dsnaps}a--e). These waves originate as Kelvin--Helmholtz billows on the upper and lower shear layers (see the vorticity profile in FIG.~\ref{fig:vorticity_profile}c). Their wavelength is approximately $2.2$ and they are first apparent at $\tjet=20$ at a distance $\xjet \approx 4$ from the source (FIG.~\ref{fig:2Dsnaps}c). The distance between their appearance and the dipole increases over time so that more waves progressively become visible.  The spatiotemporal evolution of these waves is highlighted in FIG.~\ref{fig:crests_varyingW}d.  They propagate forwards at nearly the same speed as the dipole for $\tjet \lesssim 45$ and subsequently mergers occur between pairs that slows them slightly (FIG.~\ref{fig:crests_varyingW}d).

Between the source and these undulations, the profile is flat with unit thickness. This matches the source width in the jet non-dimensionalization, albeit with a diffuse appearance (FIG.~\ref{fig:2Dsnaps}d,e). The velocity profile in this region and in the undular region ahead of it is qualitatively similar to a Bickley jet \cite{Bickley1937} (FIG.~\ref{fig:velocity_profiles}c). %\textcolor{red}{To be more specific, taking $t_j=40$, the profiles at $\xjet \lesssim 12$, normalized by the local centerline speed are close to $\mathrm{sech}^2(\yjet/\ell)$, though wider than the width $\ell = 5/18$ implied by matching the volume and momentum flux of the source.  
Further downstream, where the interfacial waves are present ($x_j \gtrsim 16$), the exterior of the jet has a wavy structure and a noticeable return flow over some crests (FIG.~\ref{fig:velocity_profiles}c). %\CV{Which t. I don't understand the end of this sentence - it feels like something is missing}

%\subsection*{Ambient feedback}
\begin{figure}
    \centering
    \includegraphics[width=\linewidth]{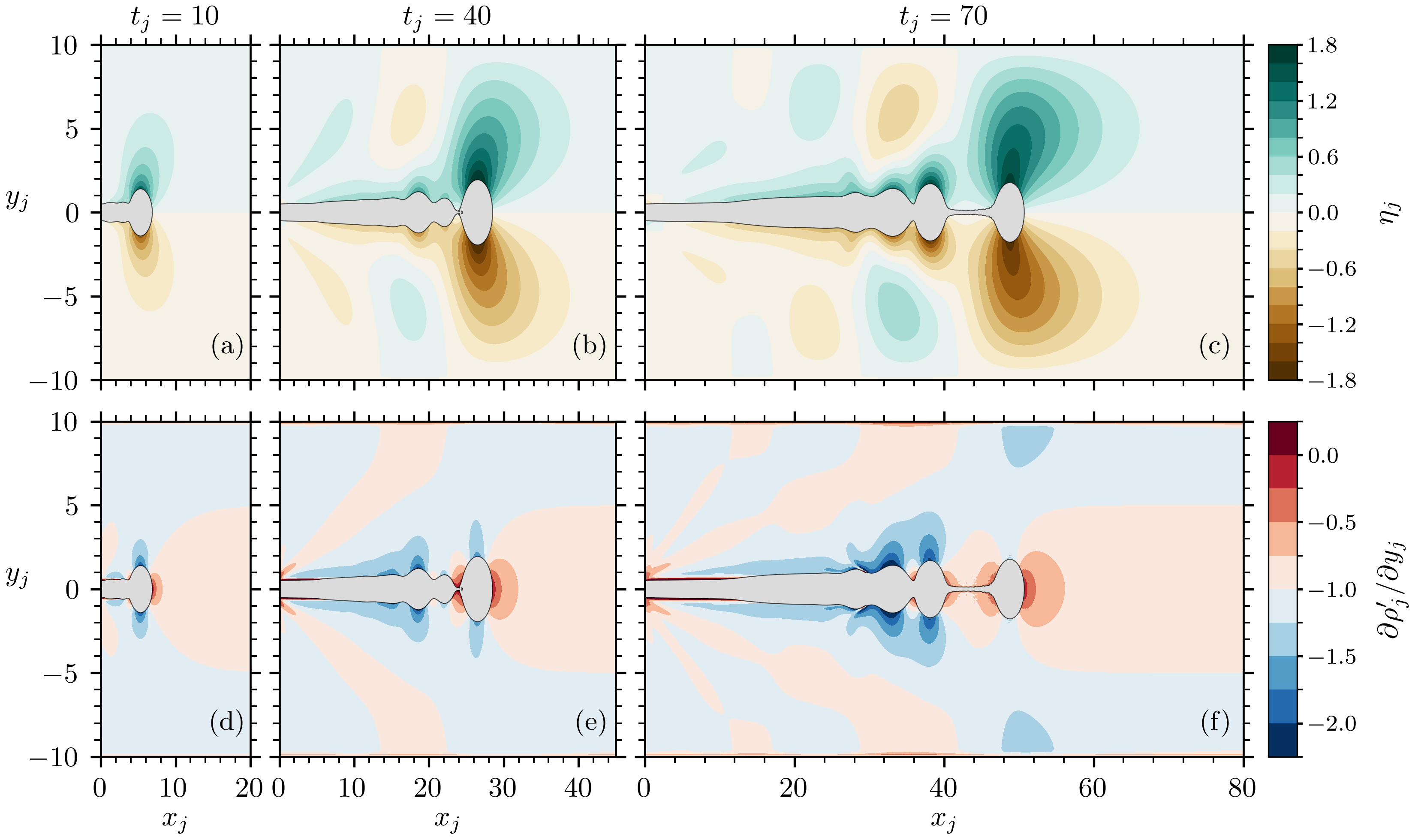}
    \caption{{Ambient response for the baseline simulation at $\tjet = 10$, $40$, and $70$, from left to right. Panels (a-c) show the isopycnal displacement $\eta_j = y_j + \rho'_j$, and panels (d-f) show the vertical density gradient $\partial\rho'_j/\partial y_j$. In (d-f), the color scale is centered on the undisturbed ambient value $-1$: blue indicates locally strengthened stratification and red indicates weakened stratification. The intrusion is masked in gray with a black solid curve envelope. Panel widths are proportional to their streamwise extents ($\xjet \leq 20$, $45$ and $80$).}
    % Vertical density gradient $\partial \rhojetnd / \partial \yjet$ at (a) $\tjet=10$, (b) $\tjet=40$ and (c) $\tjet=70$, the same times as FIG.~\ref{fig:vorticity_profile}(b,d,e). The color scale diverges about the undisturbed ambient value $\partial \rhojetnd / \partial \yjet = -1$: white is the unperturbed ambient, blue indicates locally strengthened stratification and red indicates weakened stratification. The intrusion is masked in gray with a black solid curve envelope. Panel widths are proportional to their streamwise extents ($\xjet \leq 20$, $45$ and $80$).%, so a given $\Delta\xjet$ is the same length in each panel.
    }
    \label{fig:perturbation_density_gradient}
\end{figure}
In the ambient, the dominant feature is uplift of the isopycnals. Figure~\ref{fig:2Dsnaps}a-e shows the isopycnals while FIG.~\ref{fig:perturbation_density_gradient} shows the isopycnal uplift and evolution of the vertical density gradient. The isopycnals are lifted immediately above the intrusion and lowered beneath it. Although internal waves are generated, they do not propagate ahead of the vortex dipole but appear coupled to it. 
%FIG.~\ref{fig:perturbation_density_gradient} show that there is no clear evidence of internal-wave propagation ahead of the intrusion CHECK. 
Immediately ahead of the intrusion tip, the local stratification is weakened: the vertical gradient of the relative density changes from its background value $\partial \rhojetnd/\partial \yjet = -1$ to $\partial \rhojetnd/\partial \yjet \approx -0.5$ (a $50\%$ reduction in magnitude; FIG.~\ref{fig:perturbation_density_gradient}c). A similar reduction is seen on the left side of the dipole, whereas directly above it the gradient intensifies, indicating a locally strengthened stratification. Behind the dipole and its surroundings, near the body of the intrusion, there is some weak evidence of higher-order internal wave modes (FIG. \ref{fig:perturbation_density_gradient}c), although the dominant feature is the significant uplift of the isopycnals immediately above the intrusion.

This wave field is qualitatively reminiscent of phase-locked lee waves observed by Baines and Hoinka~\cite{BainesHoinka85}. The reduction in gradient upstream (FIG.~\ref{fig:perturbation_density_gradient}b,c) is consistent with the upstream blocking observed in stratified flow over topography. The structure of the dipole is also consistent with laboratory observations in which vortex rings radiate phase-locked waves during their initial translation through a stratification \cite{grayson2024transient}. In this sense, the dipole acts like a moving topographic obstacle, forcing the surrounding density field to deform as if the ambient flow were passing over a fixed ridge. The crests remain attached to the intrusion front, and their orientation is nearly stationary in the tip's reference frame. 

\subsubsection{Varying stratification}\label{sec: varyingW}

\begin{figure}
    \centering
    \includegraphics[width=\linewidth]{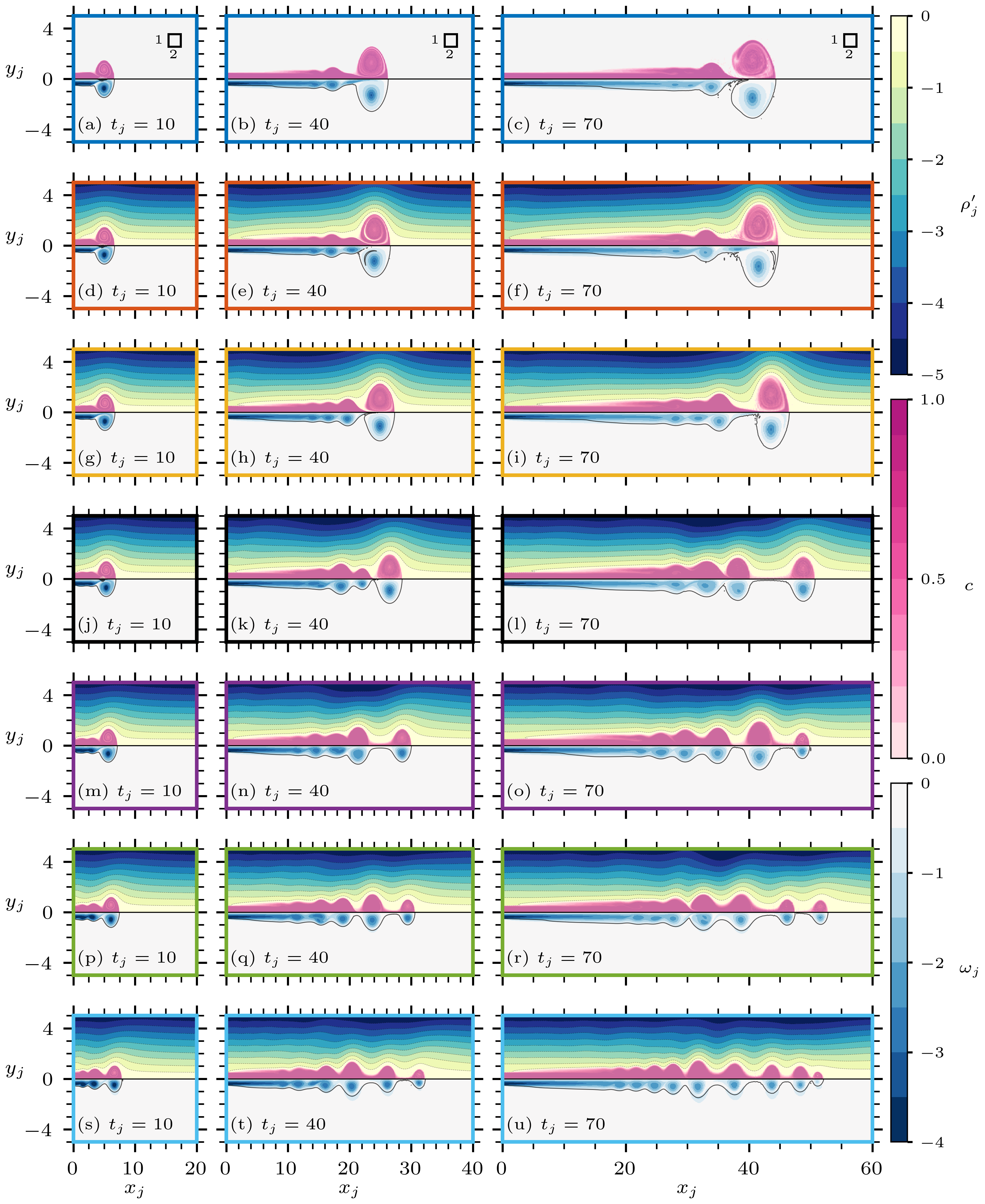}
    \caption{Effect of stratification on the early-time evolution. (a--u) Intrusion profiles at $\tjet=10$, $40$ and $70$ (left to right) for $\Wint = 0$, $0.3$, $0.4$, $0.5$, $0.6$, $0.7$ and $0.8$ (top to bottom), with the color function {superimposed on the ambient density relative to the intrusion $\rho_j'$,} above $\yjet=0$ and the vorticity $\omega_j$ below.  Panel borders are colored by $\Wint$ as in FIG.~\ref{fig:crests_varyingW}.  %Increasing stratification suppresses the vertical growth of the leading dipole and shortens the flat region near the source.  
    The box in the top row is square in data units and indicates the aspect-ratio distortion of each column.}
    \label{fig:varying_strat}
\end{figure}

\begin{figure}
    \centering
    \includegraphics[width=\linewidth]{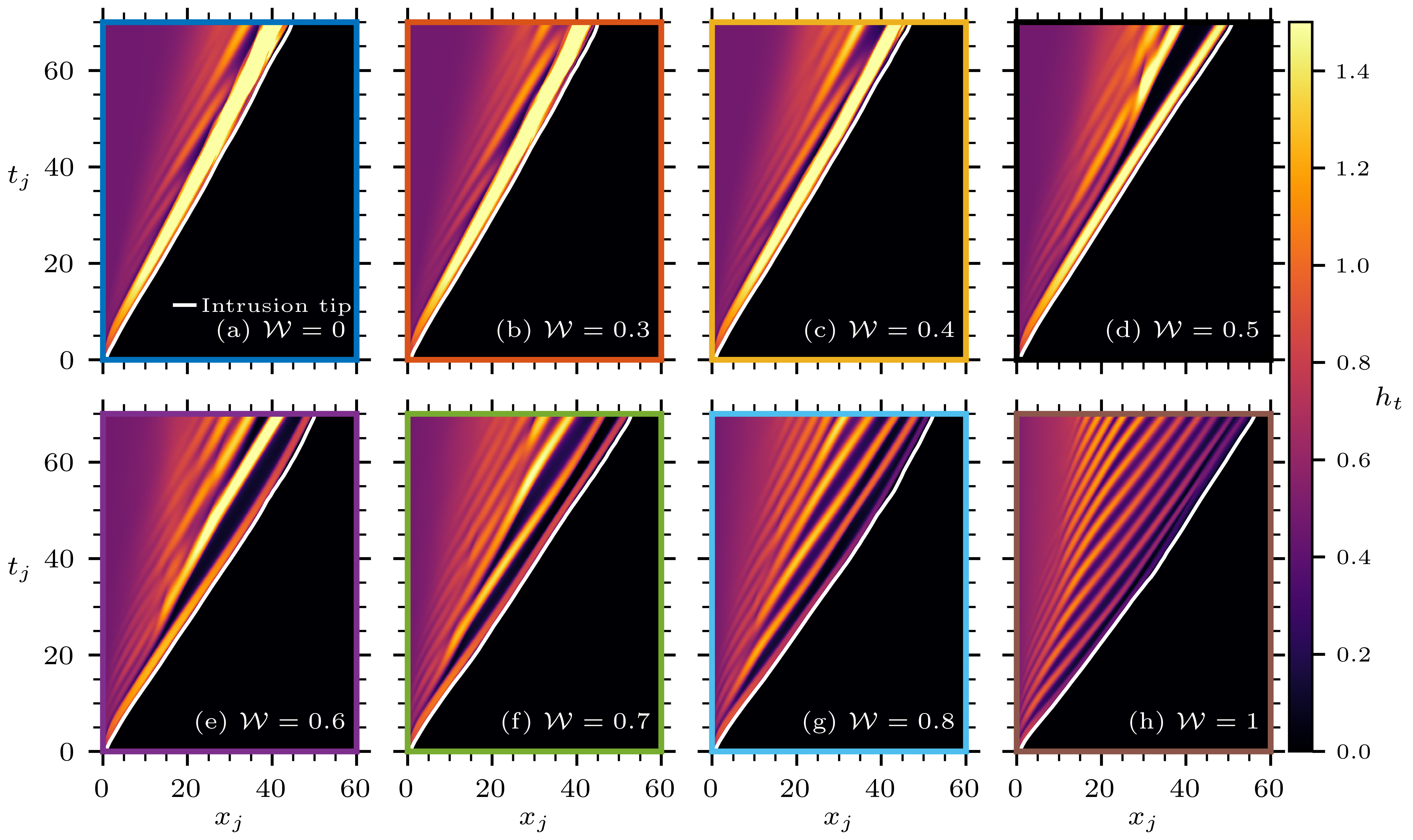}
    \caption{Spatiotemporal evolution of the upper intrusion half-thickness $h_t(\xjet,\tjet)$ for each stratification $\Wint$, with local thickness maxima appearing as bright ridges. The solid white line marks the intrusion tip, identified by the leading vortex dipole. Panel borders are colored by $\Wint$. 
    For $\Wint \lesssim 0.5$ the first peak of the trailing jet propagates at approximately the same speed as the intrusion tip, indicating phase-locked motion between the vortex head and the trailing surface undulations. {We plot $h$ as the integral of the color function.}}
    \label{fig:crests_varyingW}
\end{figure}

% \begin{figure}
%     \centering
%     \includegraphics[width=\linewidth]{paper_figures/figI_phase1_tip_and_vortex.png}
%     \caption{Front position and leading-dipole properties against time for $\Wint = 0$, $0.15$, $0.3$, $0.5$, $0.65$, $0.8$ and $1$, with $\Wint = 0.5$ in bold black. (a) Intrusion front position $x_f$. (b) The corresponding front speed $\mathrm{d}x_f/\mathrm{d}\tjet$. (c) Maximum vorticity $\omega_{\max}$ of the leading dipole. (d) Half-thickness of the leading dipole $y_\mathrm{peak}$ (solid) and height of its vorticity maximum $y_\mathrm{core}$ (dashed), as defined in FIG.~\ref{fig:vorticity_profile}(f). In (c,d), for $\Wint = 1$ the tracker loses the dipole at $\tjet \approx 32$.}
%     \label{fig:tip_vortex_varyingW}
% \end{figure} 

Despite the source dominating the flow during this phase, several details are strongly dependent on the strength of the stratification, or equivalently the value of $\width$, although others are minimally affected.  For all $\width$, the flow remains symmetric in this phase for our configuration. The changing profiles over time with varying $\width$ are shown in FIG.~\ref{fig:varying_strat} for $\Rey=1000$ and $\Hjet = 20$; the top half shows the color function and intrusion shape while the lower half shows the vorticity of the flow. The spatiotemporal evolution of the half-thickness is shown in FIG.~\ref{fig:crests_varyingW} and the properties of the leading vortex dipole are shown in FIG.~\ref{fig:tip_vortex_varyingW}.

The leading dipole is largest for the unstratified case $\width=0$ and becomes progressively smaller as $\width$ increases and stratification suppresses vertical expansion (FIG.~\ref{fig:tip_vortex_varyingW}d).  For $\width \lesssim 0.4$, the dipole grows in extent throughout the time shown (for $t\le 60$; FIG.~\ref{fig:tip_vortex_varyingW}d), reaching a half-thickness of $3.1$, $3.0$ and $2.7$ for $\width = 0$, $0.3$ and $0.4$ respectively.  For $0.5 \lesssim \width \lesssim 0.7$, the dipole initially grows before decaying, with the maximum reached progressively earlier as $\width$ increases: at $\tjet \approx 40$, $22$ and $15$ for $\width = 0.5$, $0.6$ and $0.7$, attaining half-thicknesses of $1.9$, $1.5$ and $1.2$ respectively. For $\width \gtrsim 0.8$, it decays from inception.  Notably the dipole half-thickness at formation is almost independent of stratification ($1.00$--$1.14$ across all $\width$ at $\tjet=5$); the differences develop entirely through the subsequent growth.

In contrast to this strong dependence on the intensity of the stratification, the maximum vorticity in the dipole is minimally affected by $\width$ (FIG.~\ref{fig:tip_vortex_varyingW}c): the spread across all $\width$ is within $2\%$ at $\tjet=10$ and $11\%$ at $\tjet=40$.  This is presumably because, in the jet non-dimensionalization, the velocity profile of the source and hence vorticity in the shear layers is independent of $\width$ and so the rolled up vorticity at the front remains comparable.  It also implies stratification does not impact vorticity decay in the dipole and instead diffusive effects are solely responsible.

\begin{figure}
    \centering
    \includegraphics[width=\linewidth]{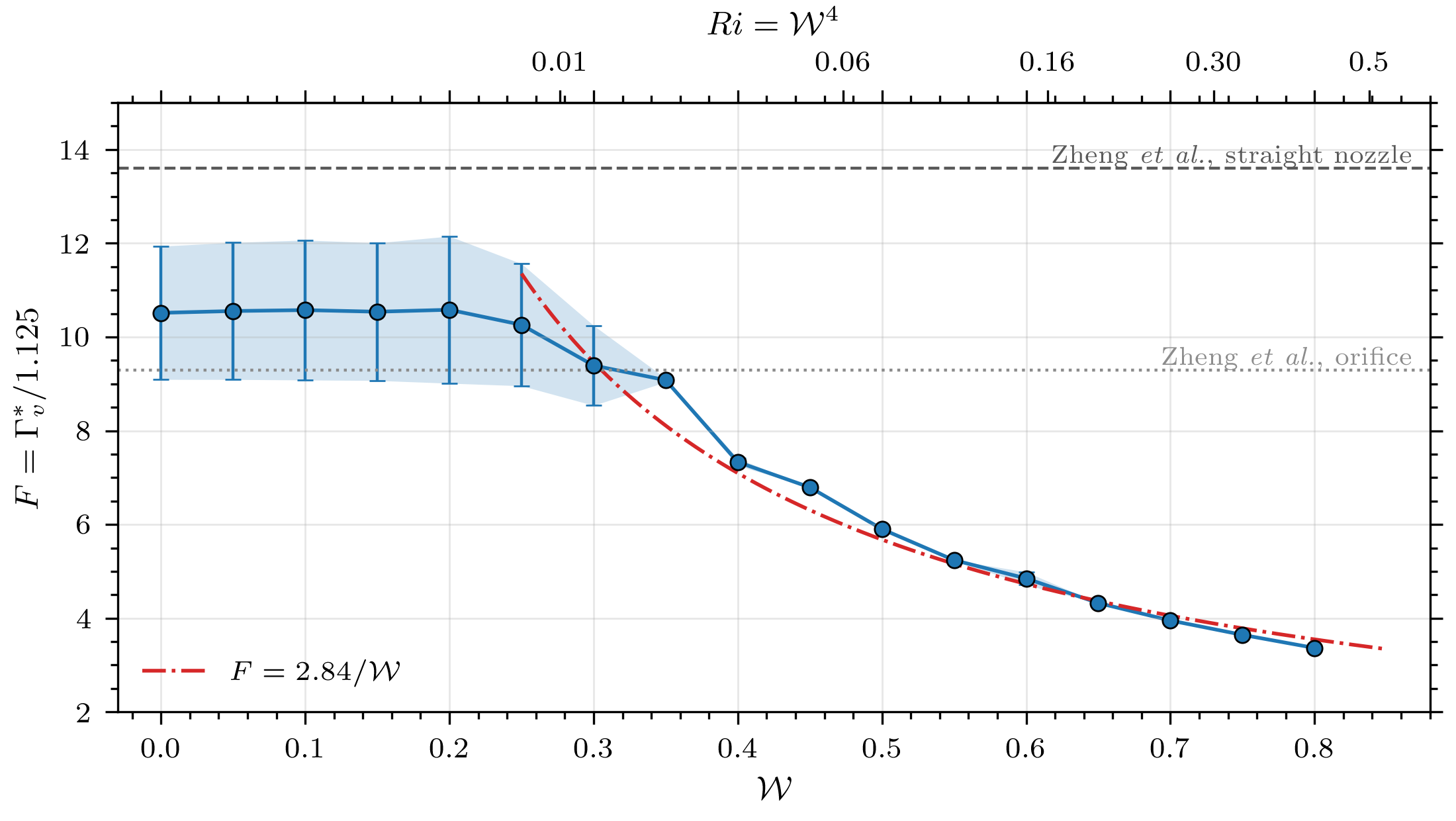}
    \caption{Formation number $F$ versus source width $\Wint$ (lower axis) and $Ri=\mathcal{W}^4$ (upper axis) at $Re=1000$, $\Hjet=20$, $\Sc=10$. The formation number $F$ is defined by $F= \Gamma^*_v/1.125$, where $\Gamma^*_v$ is the circulation of the leading dipole evaluated using the $5\%$-vorticity core method of \cite{zheng2025formation}, and 1.125 the rate at which circulation is discharged by the parabolic inlet. Error bars span the range over which the plateau in $\Gamma_v$ may be read, from its value before the rear stagnation point is first lost to its late-time level. %The red dash-dotted curve is $F=2.57\width^{-1.2}$. 
    The gray horizontal lines indicate the reference formation numbers $9.3$ and $13.6$ for the orifice and straight-nozzle configurations of Zheng~\etal~\cite{zheng2025formation}, respectively.
    %identified by: find local maximum pressure as boundaries, then compute the integral of w over that region. Then the bands are for taking LOCAL points where the circulation is plateauing (there's some jumps due to fluid feeding into the vortex that might make the algorithm detection
    }
    \label{fig:formation_time}
\end{figure}

A key observation for classical vortex dipoles and vortex rings is the formation number, the time by which the source has discharged as much circulation as the dipole or ring ultimately holds. Beyond this point, further discharge feeds the trailing jet rather than the dipole or ring \cite{gharib1998universal}. This time is shown in FIG.~\ref{fig:formation_time} across a range of stratifications.  For $\Wint \lesssim 0.3$, the formation
number is $F \approx 10.5$, within the range $9.3$--$13.6$ reported by Zheng \emph{et al.}~\cite{zheng2025formation} for two dimensional starting jets from an orifice and a straight nozzle respectively. The error bars span the range over which the saturated dipole circulation may be read. For $\width \gtrsim 0.3$ the formation number decays rapidly with $\width$, as $F \approx 2.84/\width$, consistent with the reducing prominence of the dipole itself: at fixed time the stratified core carries essentially the same peak vorticity as the unstratified one but occupies $40\%$ less area, and so saturates at a smaller circulation.

The front propagates at an approximately constant speed throughout this phase, but that speed increases systematically with $\width$ (FIG.~\ref{fig:tip_vortex_varyingW}a,b): fitting $x_f = ct$ over $\tjet<20$ gives $c = 0.67$, $0.68$, $0.69$, $0.72$, $0.75$, $0.80$ and $0.84$ for $\width = 0$, $0.3$, $0.4$, $0.5$, $0.6$, $0.7$ and $0.8$, so that the most strongly stratified case advances approximately $25\%$ faster than the unstratified one.  This is presumably because vertical confinement from stratification squeezes the intrusion further forwards. 

%\AS{When compare to below to figure consider DO FOR NUMBER OF CRESTS INSTEAD} 
In contrast to the trend for the leading dipole, the waves on the trailing jet behind it grow in amplitude and extent as the stratification increases, although the wavelength is minimally affected.  In the unstratified case, $\width=0$, only a few crests are apparent and have small amplitude.  In this case, the waves propagate significantly faster than the dipole and they are subsumed into it.  For $\width \lesssim 0.5$, the waves grow in amplitude towards the front, the largest crest sitting within one wavelength of the dipole, while for $\width \gtrsim 0.6$ the waves grow to a maximum amplitude progressively closer to the source before decreasing further forward: at $\tjet=70$ the largest crest lies at $\xjet \approx 41.5$, $38.8$ and $31.7$ for $\width = 0.6$, $0.7$ and $0.8$, against front-most crests at $\xjet \approx 49.0$, $46.4$ and $48.2$ respectively. For $\width < 0.5$, the waves propagate forward faster than the dipole and are subsumed into it.  For $\width \gtrsim 0.5$, the trailing waves pair up (apparent as a braiding pattern in FIG.~\ref{fig:crests_varyingW}) and they no longer catch the leading dipole.  For $\width=0.5$ this manifests as the thinning and lengthening neck that develops between the dipole and the following wave described for our baseline simulation.  For $\width \gtrsim 0.8$, the waves no longer pair up over the time interval we consider: the number of crests in the trailing train grows monotonically (from $4$ to $7$ for $\width=0.7$ and from $5$ to $9$ for $\width=0.8$ between $\tjet=30$ and $70$), whereas for $\width = 0.5$ and $0.6$ it is non-monotonic, the drops marking pairing events.

The flat region near the source is unaffected by the stratification except that its extent reduces as $\width$ increases: at $\tjet=70$ the first crest sits at $\xjet \approx 29$, $29$, $25$, $25$, $22$, $19$ and $17$ for $\width = 0$, $0.3$, $0.4$, $0.5$, $0.6$, $0.7$ and $0.8$ respectively (FIG.~\ref{fig:crests_varyingW}).

Overall, this initial penetration of the jet appears as a vortex dipole for small $\width$ and an undular bore for moderate $\width$.

\subsubsection{Varying Reynolds number}
\begin{figure}
    \centering
    \includegraphics[width=\linewidth]{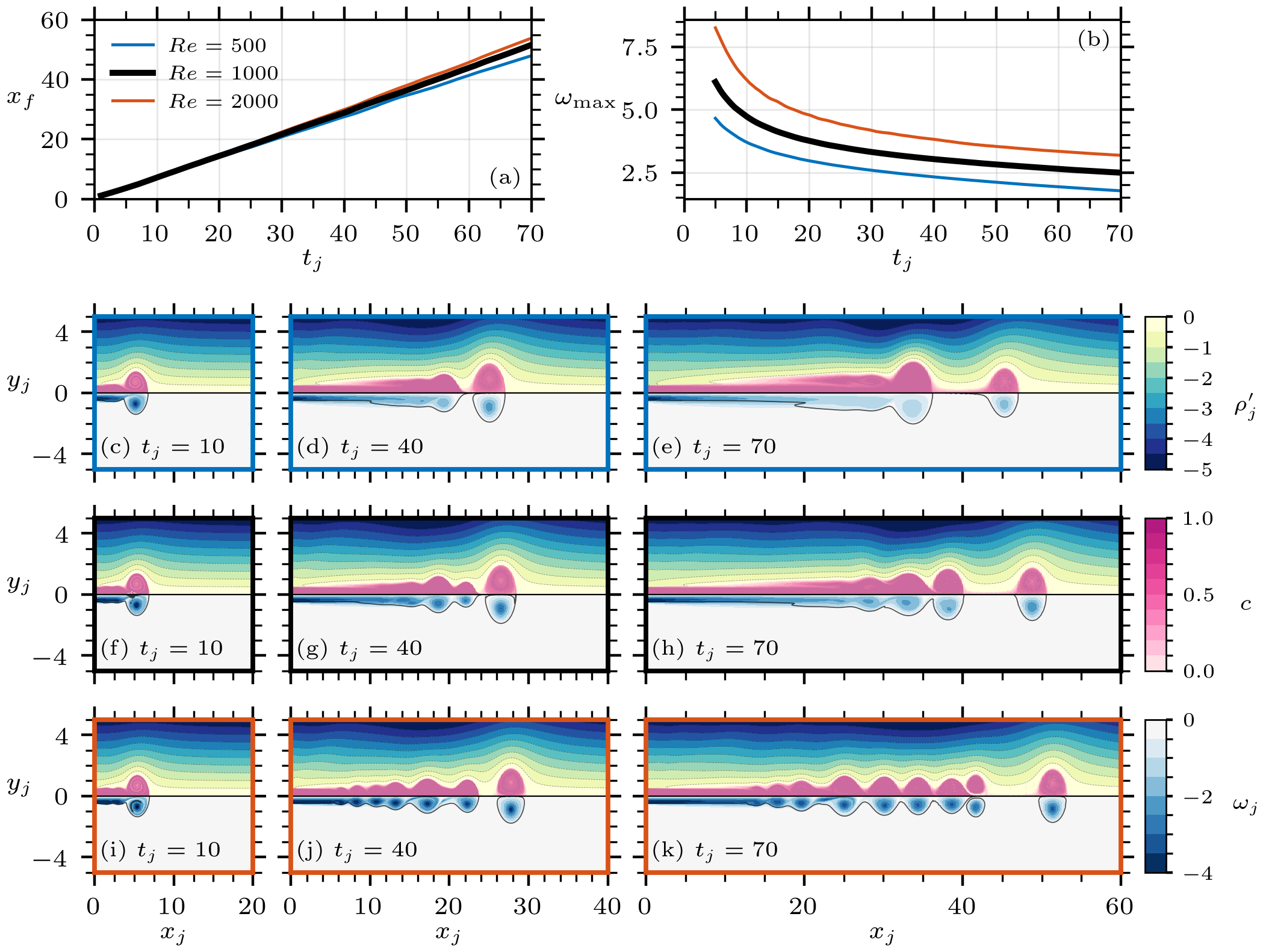}
    \caption{Effect of $\Rey$ on the early-time evolution of the intrusion. (a) Intrusion front position $x_f$ as a function of time for different values of $\Rey$. (b) Corresponding decay of the vortex strength. (c--k) Intrusion profiles at times $\tjet=10$ (c,f,i), $\tjet=40$ (d,g,j), and $\tjet=70$ (e,h,k) for $\Rey = 500$ (top row), $1000$ (middle) and $2000$ (bottom).}
    \label{fig:varying_Rey_phase1}
\end{figure}

Figure~\ref{fig:varying_Rey_phase1} shows a comparison of the tip locations and intrusion profiles for various Reynolds numbers, with all other parameters identical to those for the baseline simulation. The broad-scale qualitative structure remains the same, with a leading vortex dipole and a varicose mode on the trailing jet.  The front speed increases only from $0.68$ to $0.76$ as $\Rey$ rises from $500$ to $2000$ (FIG.~\ref{fig:varying_Rey_phase1}a), and the dipole thickness is essentially unchanged.  
The small scales, by contrast, depend strongly on $\Rey$. The peak vorticity of the dipole increases from $3.7$ to $6.2$ at $\tjet = 10$ (FIG.~\ref{fig:varying_Rey_phase1}b), consistent with the thinning of the shear layers at the source. The Kelvin--Helmholtz roll-up sets in progressively closer to the source, the first trailing crest at $\tjet = 40$ lying at $\xjet \approx 19$, $15$ and $6$ for $\Rey = 500$, $1000$ and $2000$, and the wave train becomes correspondingly more numerous, periodic and coherent as $\Rey$ increases (FIG.~\ref{fig:varying_Rey_phase1}c--k): at $\tjet = 70$ there are three trailing crests at $\Rey = 500$ and $1000$ but nine at $\Rey = 2000$. At higher Reynolds numbers, the shear layers roll up into clearer, more periodic undulations, whereas at $\Rey = 500$ viscous effects smooth the interface and leave only a few long bulges. These observations indicate that varicose interfacial waves are a robust feature of the jet-dominated stage, with the large-scale motion set by the jet and the number, spacing and sharpness of the waves controlled primarily by viscous effects.

%################################ Phase~II #######################

\subsection{Phase~II: Growing instability and transition to turbulence}\label{sec:Phase2}
\begin{figure}
    \centering
    \includegraphics[width=\linewidth]{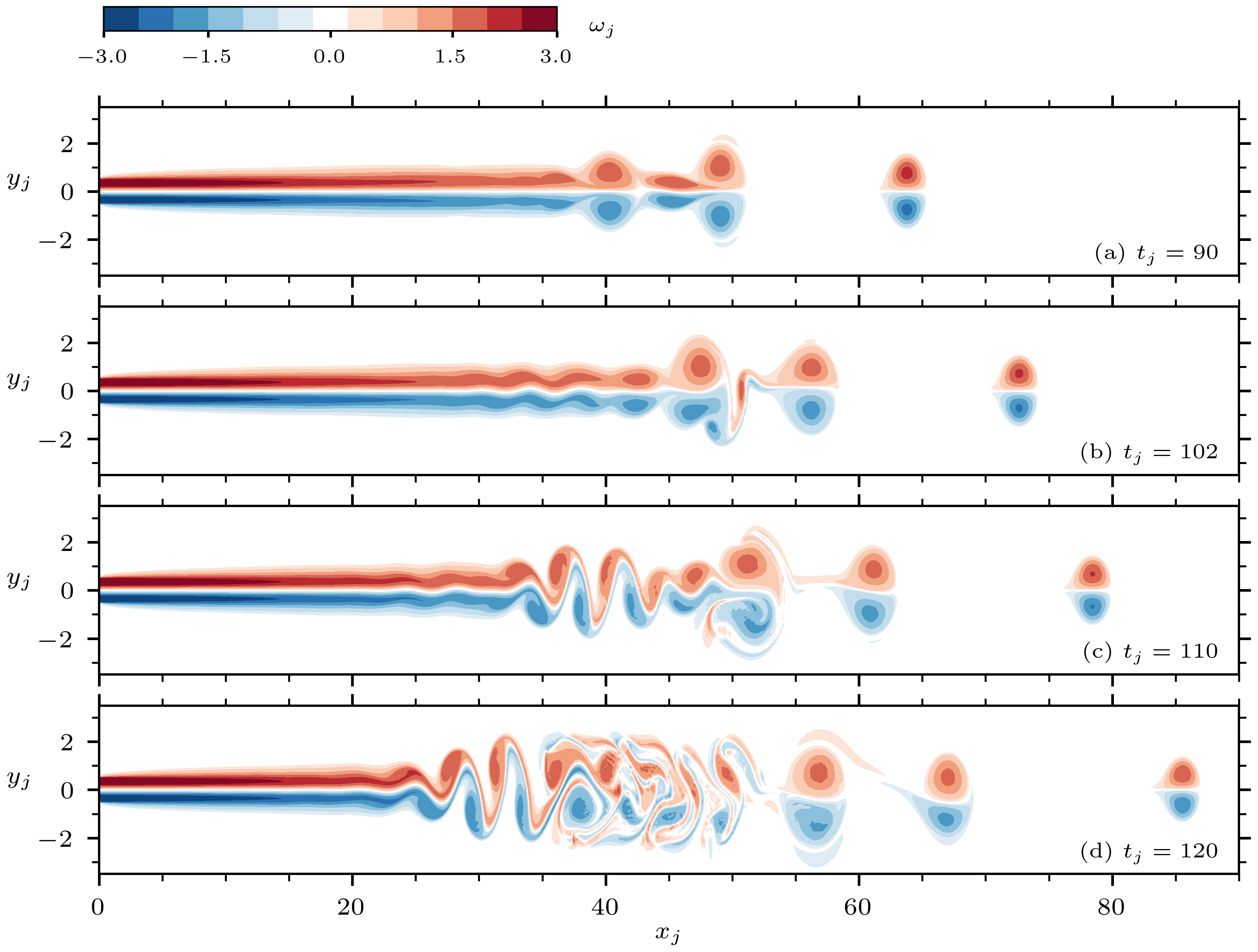}
    \caption{Vorticity $\omega_j$ during Phase~II for the baseline simulation at (a) $\tjet=90$, (b) $102$, (c) $110$, and (d) $120$. All panels share one streamwise scale, so the advance of the leading dipole can be read directly down the figure; the vertical axis is stretched. The initially symmetric trailing jet undergoes a sinuous instability and breaks down into a train of vortices, while the leading vortex dipole remains coherent and continues to propagate downstream.}
    \label{fig:phase_2_evolution}
\end{figure}

\begin{figure}
    \centering
    \includegraphics[width=\linewidth]{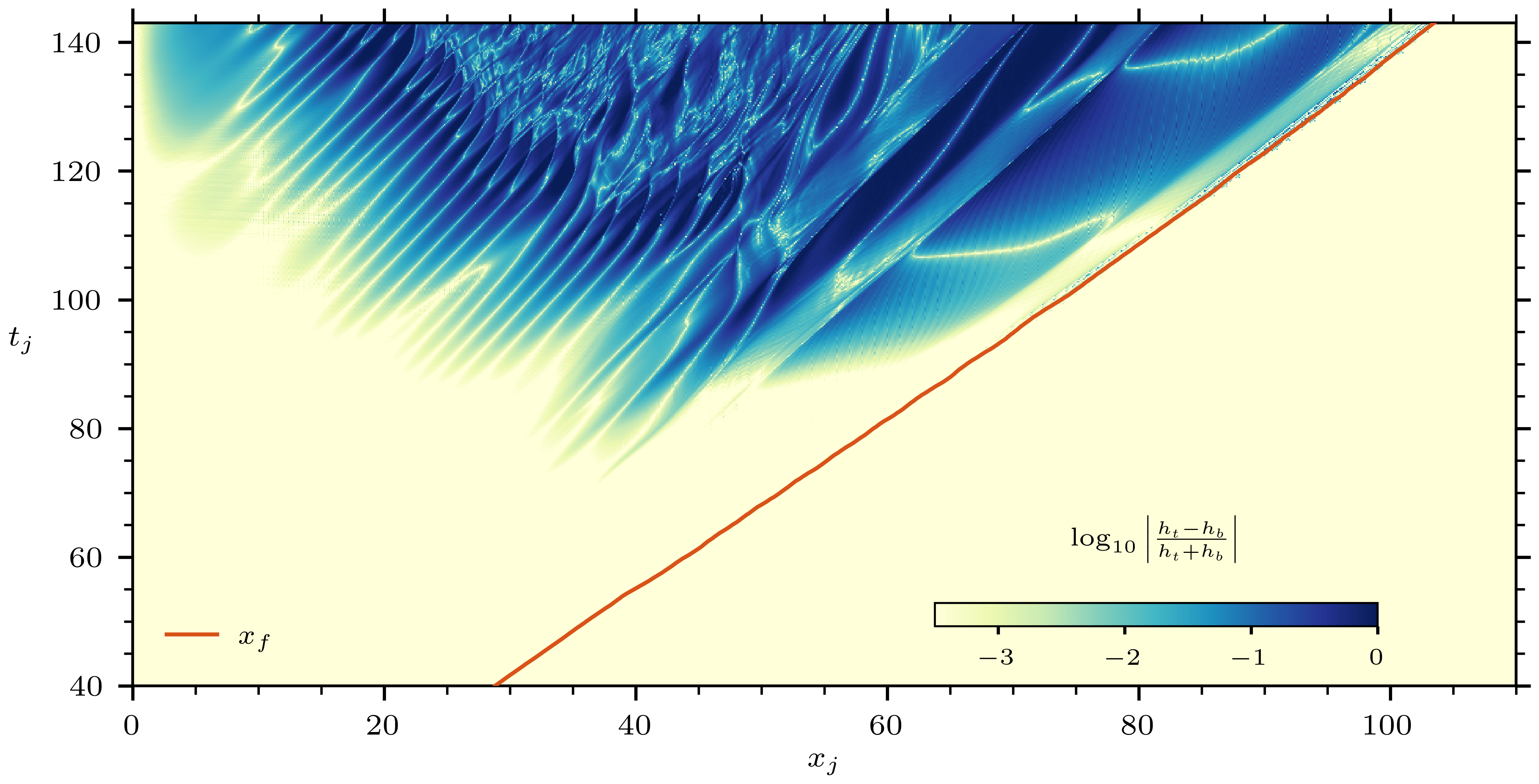}
    \caption{Spatiotemporal evolution of the normalized thickness difference $\log_{10}\Delta h(\xjet,\tjet)/2h(\xjet,\tjet)$, with $\Delta h = |h_t - h_b|$, for the baseline simulation. The red line is the intrusion tip $x_f(\tjet)$. {We plot $h$ as the integral of the color function.}}
    \label{fig:asymmetry_map}
\end{figure}

\begin{figure}
    \centering
    \includegraphics[width=\linewidth]{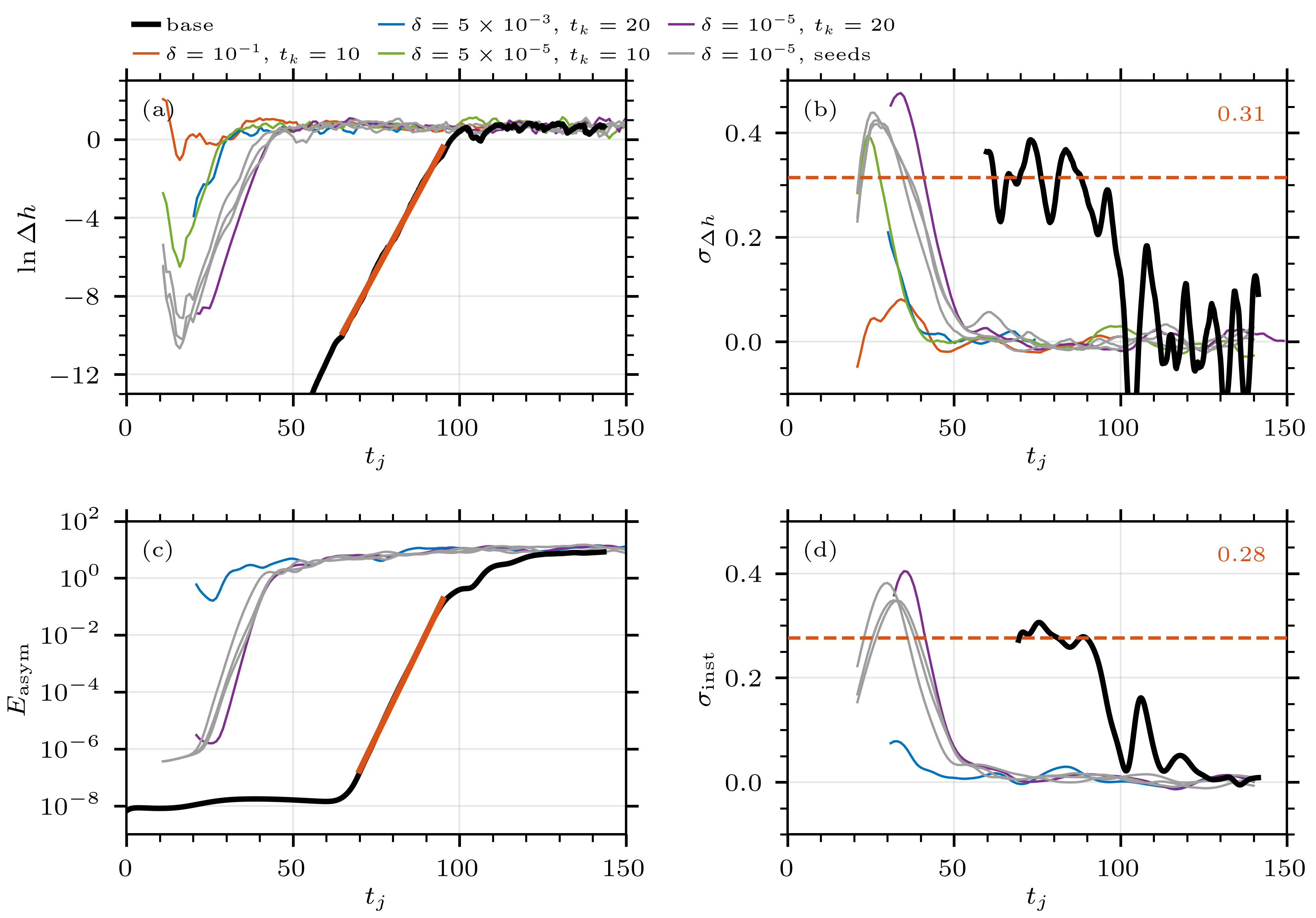}
    \caption{Growth of the asymmetry for the baseline simulation (black) and for simulations seeded with white noise of amplitude $\delta$ (colored; the three gray curves are further realizations at $\delta = 10^{-5}$). (a) $\ln \max_x \Delta h$ and (b) its growth rate $\sigma_{\Delta h} = \mathrm{d}(\ln \max_x \Delta h)/\mathrm{d}t$; (c) the asymmetric energy $E_\mathrm{asym}$ and (d) its growth rate $\sigma_\mathrm{inst} = \tfrac12\,\mathrm{d}(\ln E_\mathrm{asym})/\mathrm{d}t$. The orange segments in (a,c) are least-squares fits over $65 \leq \tjet \leq 95$, and the dashed lines in (b,d) mark the fitted rates. {We plot $h$ as the integral of the color function.}}
    \label{fig:symbreakking}
\end{figure}

For $\width \lesssim 1.6$, Phase~I eventually becomes unstable to a sinusoidal, flapping instability.  The onset and growth of this instability characterises Phase~II.  By the end of the phase, the flow appears turbulent in the sense established for two-dimensional flows.  %WHAT IS THE CUT-OFF W?  MIGHT BE BETTER TO SAY THIS.  IN PREVIOUS SECTION PLOTTED TO A GIVEN TIME - EXPLAIN THAT.  W=1.6+/-0.1

The profile evolution during this stage for our baseline simulation is shown in FIG.~\ref{fig:2Dsnaps}.
Figure~\ref{fig:phase_2_evolution} shows corresponding vorticity fields.  The instability is visually apparent by $\tjet=90$, most evidently at $\xjet \approx 45$.  This mode progressively grows and spreads outwards from its inception points, with a clearly sinuous structure further back along the trailing jet.  %By $\tjet = 104$ the shear layers have rolled up into a staggered train of vortices, and by $\tjet = 120$ this train has broken down over $25 \lesssim \xjet \lesssim 50$ while the leading dipole, now well ahead, remains symmetric.

The development of asymmetry is more clearly visible in FIG.~\ref{fig:asymmetry_map}, which shows $\Delta h(x,t) = |h_t(x,t) - h_b(x,t)|$, the difference in the thickness profiles above and below the centerline, normalized by the overall thickness $2h$.  Inception occurs at a few locations in the body of the intrusion, at $\xjet \approx 30$--$45$ from $\tjet \approx 70$.  The wedge-shaped envelope emanating from these locations indicates upstream- and downstream-propagating fronts of asymmetry that eventually extend across the entire intrusion.  The downstream edge of the envelope advances at close to the front speed of $0.72$, so that the asymmetry reaches the leading dipole by $\tjet \approx 100$, while the upstream edge moves back towards the source at approximately $0.5$ and reaches it by $\tjet \approx 130$.

We quantify the growth rate of the instability by considering both a global geometric measure $\max_x \Delta h(x,t)$ and a global energy measure 
\begin{equation}
    E_{\mathrm{asym}} = \iint \tfrac{1}{2}\bigl(u'^2 + v'^2\bigr) \, \mathrm{d}A, %\\
    %u'(x,y) &= \tfrac{1}{2}, \\
    %v'(x,y) &= \tfrac{1}{2}\left[v(x,y) + v(x,-y)\right],
\end{equation}
where $u'=\left[u(x,y,t) - {u}(x,-y,t)\right]/2$ and $v'=\left[{v}(x,y,t) + v(x,-y,t)\right]/2$ are the asymmetric components of the streamwise and vertical velocity fields.  These measures and their growth rates are shown in FIG.~\ref{fig:symbreakking}.  While the flow is initially symmetric, both measures are small, but they increase sharply once the instability begins and then grow exponentially until saturation.  For the geometric measure, the growth rate is approximately $\sigma_{\Delta h} = 0.31$ (FIG.~\ref{fig:symbreakking}b).  As the energy is a quadratic quantity, we define the growth rate of the instability from the energy as $\sigma_{\mathrm{inst}}=\tfrac12\,d(\ln E_{\mathrm{asym}})/dt$.  This has an approximate value of $0.28$ (FIG.~\ref{fig:symbreakking}d).  The two measures grow over more than five decades before saturating at $\tjet \approx 100$, when $\max_x |h_t - h_b|$ reaches about $2$, comparable to the thickness of the leading dipole.

The instability shown in FIGs.~\ref{fig:phase_2_evolution}--\ref{fig:symbreakking} develops from numerical error alone.  To investigate the robustness of the instability to the particular realisation of the perturbation and its amplitude, in FIG.~\ref{fig:symbreakking} we include the growth rate analysis for simulations seeded with white noise of different amplitudes $\delta$, added to the velocity field on restarting from the baseline solution at $\tjet = 10$ or $20$ (as implemented in \textsc{Semtex} \cite{blackburn2019semtex}). For three different realizations with the same amplitude (gray curves with $\delta = 10^{-5}$), the growth rates vary minimally while the earliest and latest onset differ by less than $10$ units of time.  Perturbing the system later does not change the growth rate (purple curve).  As the amplitude increases, onset is sooner but the growth rates remain effectively the same for $\delta \leq 5 \times 10^{-3}$; at $\delta = 10^{-1}$ the disturbance is already nonlinear when it is introduced and no exponential stage is observed.  The growth rates of the seeded runs are comparable to those of the baseline, although reaching higher values, perhaps because the instability then develops on a younger and thinner trailing jet.  All runs saturate at comparable amplitude.

The observed instability is consistent with the classical planar-jet instability studies of Tatsumi and Kakutani~\cite{tatsumi1958stability} and Sato and Sakao~\cite{SatoSakao1964} for unstratified flows.  The classical theory describes the temporal growth of a disturbance on a parallel, steady jet, whereas the trailing jet in our simulations develops both in space and in time.  A comparison is possible by relating our observed growth rates to the theory based on local properties of the jet.  For a Bickley jet, Sato's growth rate for the most-amplified sinuous mode is $0.142 u_0/b$ and the wavelength is $2\pi b/0.79$, where $u_0$ is the centerline velocity and $b$ is the distance from the centerline where $u=u_0/2$.
We estimate $u_0$ and $b$ from the profile averaged over three wavelengths centered on the unstable wave packet, following it as it moves. Over $70 \leq \tjet \leq 84$, while the disturbance is still small enough ($\max_x |h_t - h_b| \lesssim 10^{-2}$, FIG.~\ref{fig:symbreakking}a) for this averaged profile to represent the base flow rather than the instability itself, $u_0 \approx 1.2$ and $b \approx 0.55$--$0.58$. The spread is not measurement scatter: the jet continues to develop along its length and in time, so $u_0$ and $b$ depend on where and when they are evaluated, and the parallel-flow assumption underlying the theory is only approximately satisfied. 
%At later times the averaged profile widens rapidly, but this reflects the flapping of the jet visible in FIG.~\ref{fig:phase_2_evolution}(b,c) rather than its spreading, and those times are excluded. 
Thus the theoretical rate is $\sigma \approx 0.29$--$0.31$, and the corresponding wavelength $\lambda \approx 4.4$--$4.6$. These are to be set against the measured $0.31$ from $\max_x \Delta h$ and $0.28$ from the energy, and against the measured wavelength of approximately $4.7$, the spacing of the sinuous undulations. % and of the vortices they roll into in FIG.~\ref{fig:phase_2_evolution}(c,d), which is roughly twice the spacing of the Kelvin--Helmholtz billows at their inception in Phase~I and consistent with the mode developing on the billows after they have paired. 
This agreement suggests that the instability is the sinuous mode of the local velocity profile, and that neither buoyancy nor viscosity controls its growth at this parameter point.%, consistent with the weak dependence of the growth rate on $\Wint$ reported in \S\ref{sec:phase2_varyingW}.  
A proper comparison would require a linear stability analysis of the developing base flow including the stratification, which we leave for future work.

%DON'T UNDERSTAND WAVEPACKET.  SPEED STILL LESS THAN FRONT.  WHY DIFFERENT FROM DELTA H?

%Physically, the overall picture suggests a localized onset region that seeds disturbances spreading along the shear layer in both directions. This can be confirmed by isolating the sinuous mode directly: $v$ on the centerline vanishes identically for any varicose disturbance, so $v(\xjet, 0)$ contains only the sinuous component. Tracked in this way (not shown), the disturbance is a wavepacket rather than a global mode, born near $\xjet \approx 34$ at $\tjet \approx 70$ and advected downstream at approximately $0.54$ while it amplifies, consistent with the wedge in FIG.~\ref{fig:asymmetry_map}; its amplitude grows over two and a half decades at $\sigma \approx 0.28$, consistent with the $0.31$ obtained from $\max_x \Delta h$ in FIG.~\ref{fig:symbreakking}(b). 

Throughout Phase~II the leading dipole is unaffected by the instability behind it: it remains symmetric to within $\Delta h < 10^{-2}$, continues to advance at $0.72$, and its vorticity and thickness continue the slow decay of Phase~I.  In the ambient, the isopycnals remain lifted above and depressed below the dipole as in Phase~I, while the breaking region behind it sheds smaller-scale disturbances into the surrounding fluid (FIG.~\ref{fig:vary_strat_growth_rate}).

%\AS{AMBIENT RESPOSNE}
%% AS: ambient-response paragraph is a placeholder -- check against the displacement version of figE before keeping it.

\subsubsection{Varying stratification}

\begin{figure}
    \centering
    \includegraphics[width=\linewidth]{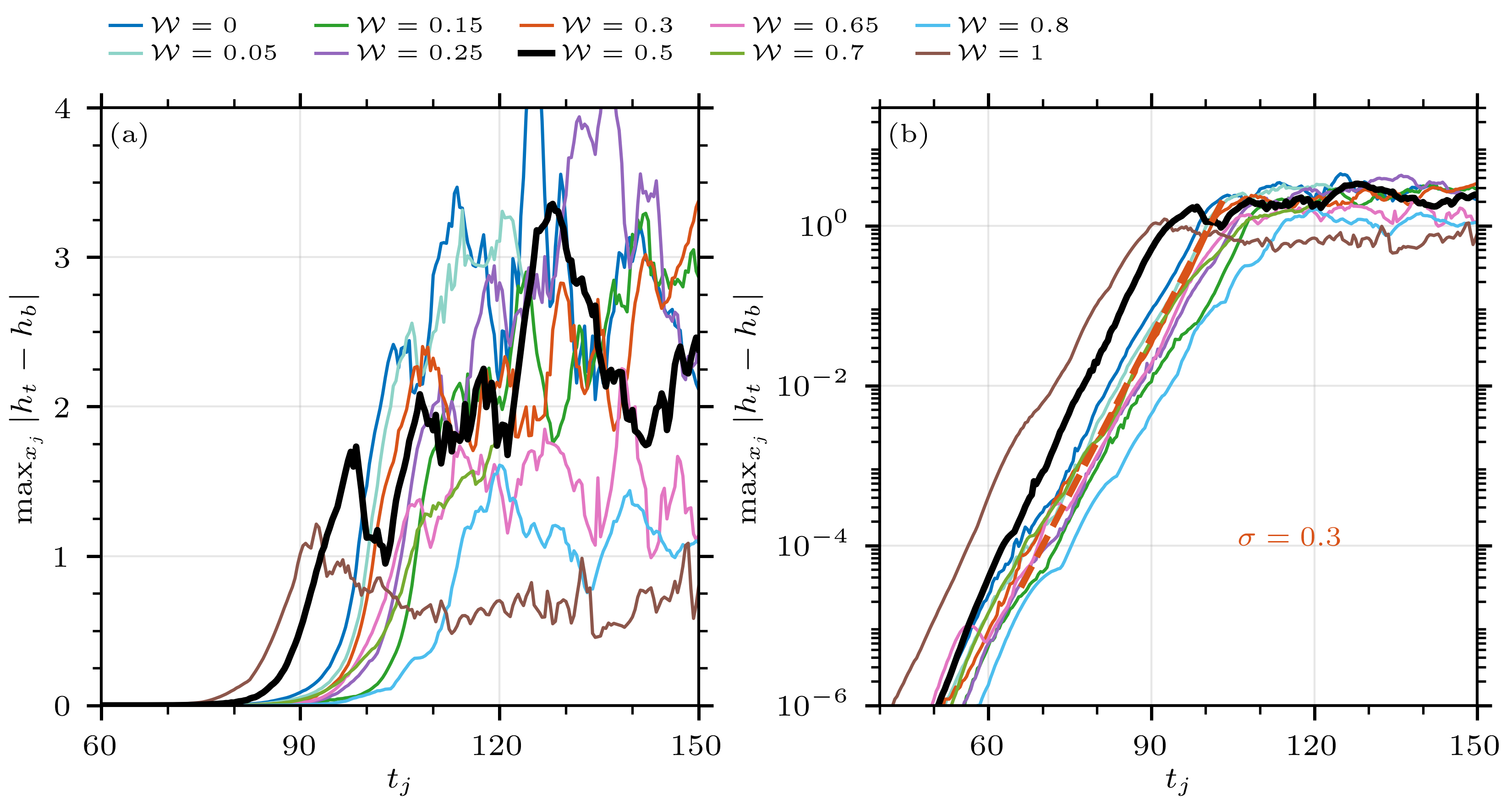}
    \caption{Growth of the asymmetry $\max_x |h_t - h_b|$ against time for each stratification $\Wint$, on (a) linear and (b) semilogarithmic axes. The dashed line in (b) has slope $\sigma = 0.3$, the growth rate of the baseline case, and is a reference for the gradient only. {We plot $h$ as the integral of the color function.}}
    \label{fig:asym_vs_W}
\end{figure}

\begin{figure}
    \centering
    \includegraphics[width=\linewidth]{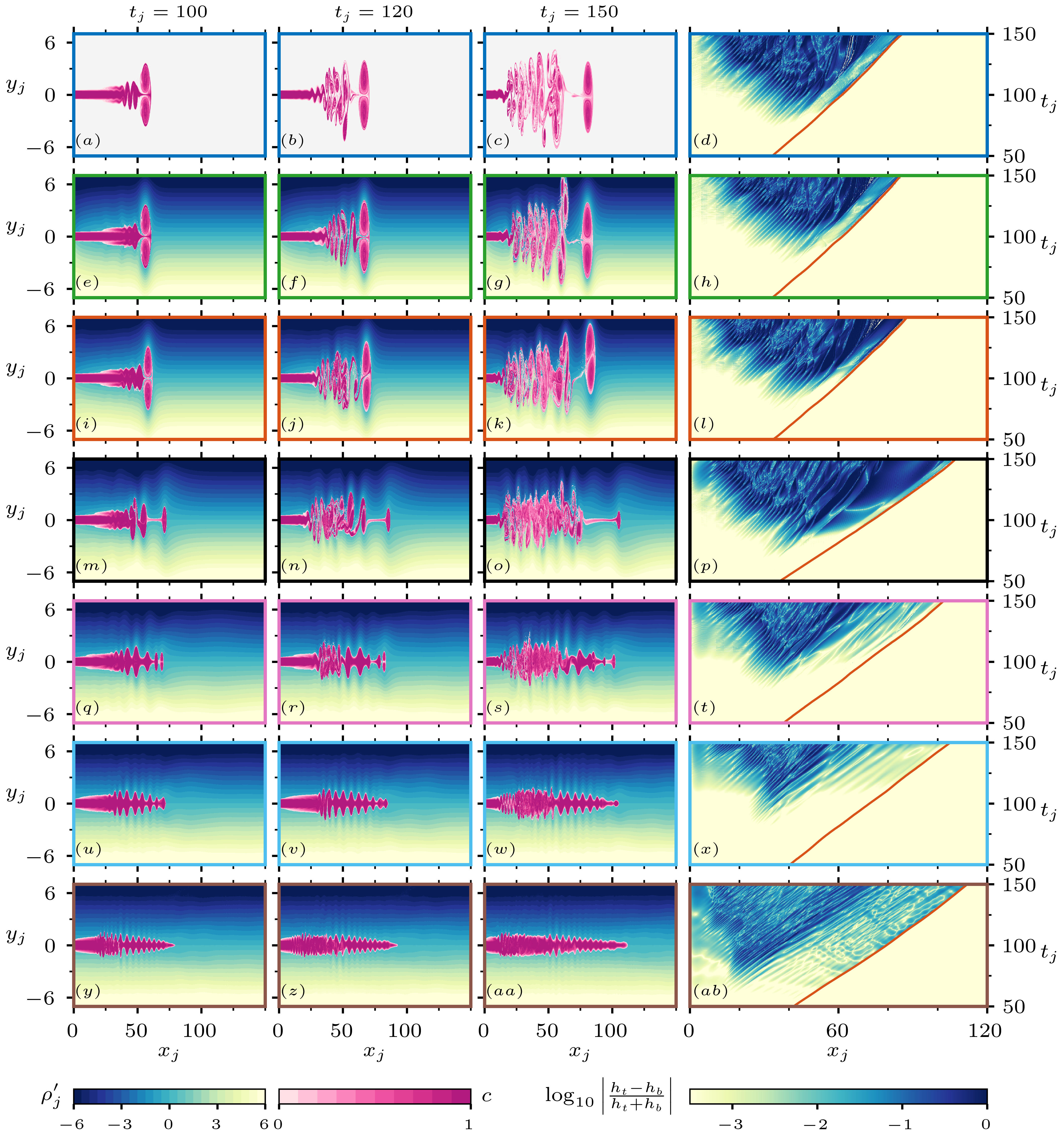}
    \caption{Phase~II against stratification. Each row is one value of $\Wint$ ($0$, $0.15$, $0.3$, $0.5$, $0.65$, $0.8$ and $1$ from top to bottom), identified by its border color as in FIG.~\ref{fig:asym_vs_W}. The first three columns show the intrusion at $\tjet = 100$, $120$ and $150$; the last column shows $\log_{10}\Delta h(\xjet,\tjet)/2h(\xjet,\tjet)$ with the front $x_f$ in red, as in FIG.~\ref{fig:asymmetry_map}. The vertical axis of the snapshots is stretched. {We plot $h$ as the integral of the color function.}}
    \label{fig:vary_strat_growth_rate}
\end{figure}

\begin{figure}
    \centering
    \includegraphics[width=\linewidth]{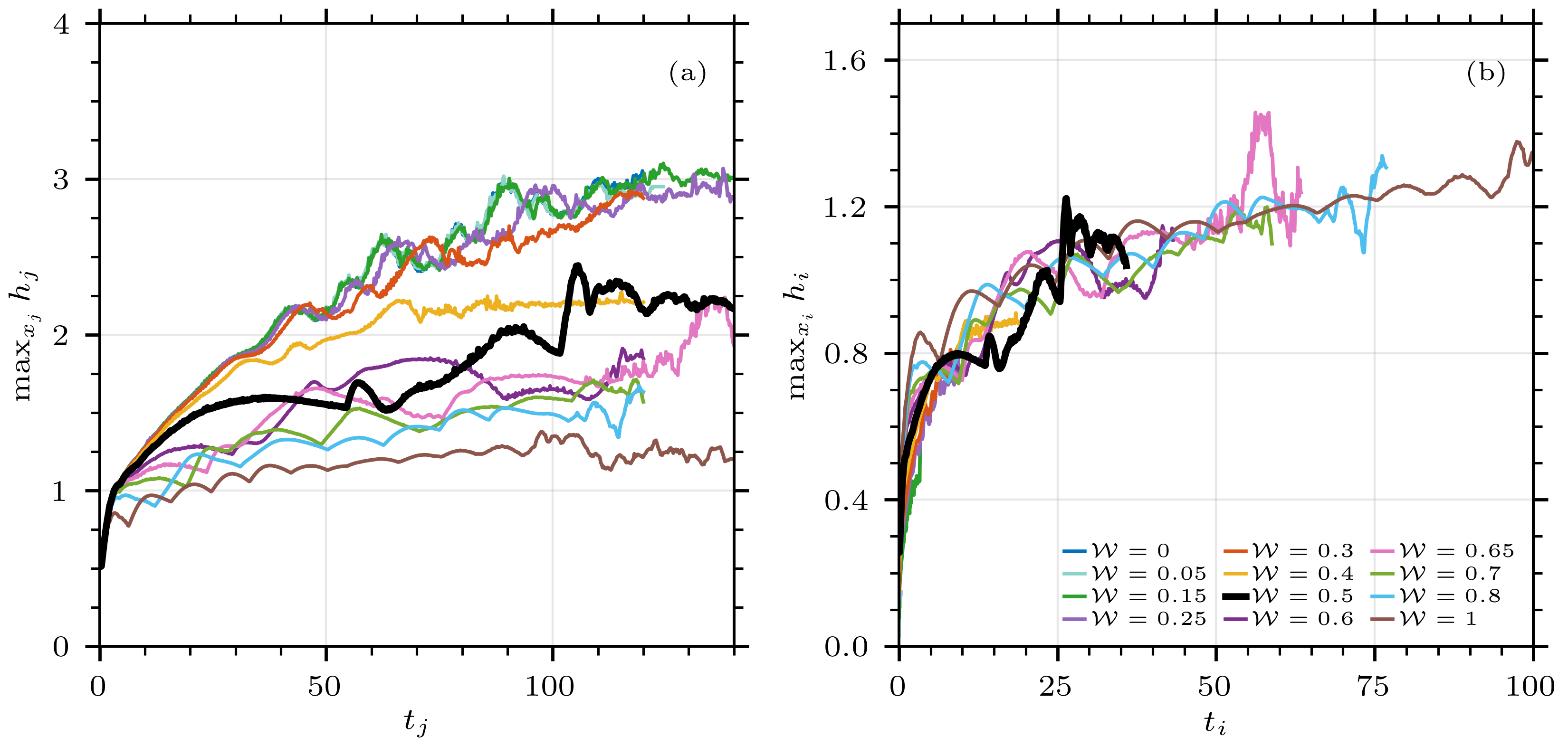}
    \caption{Maximum half-thickness of the intrusion, $\max_x h$ with $h = (h_t + h_b)/2$, against time for each $\Wint$, in (a) the jet scaling and (b) the intrusion scaling ($h_i = \Wint h_j$, $t_i = \Wint^2 t_j$). $\Wint = 0$ has no intrusion scale and appears in (a) only. {We plot $h$ as the integral of the color function.}}
    \label{fig:thickness_vs_W}
\end{figure}

Stratification initially appears to play a limited role in Phase~II.  Figure~\ref{fig:asym_vs_W} shows the geometric measure $\max_x \Delta h(x,t)$ against time for various $\width$.  There is no clear trend in the onset time across the different simulations, and the difference between the earliest and latest onsets is only about $20$ units of time, roughly twice the range for the repeated simulations described above, suggesting $\width$ is not a significant factor in onset time.  More significantly, the growth rates of the instability are essentially identical across all values of $\width$, at $0.30 \pm 0.03$.  This suggests that the instability mechanism is primarily shear-driven, rather than controlled by buoyancy, consistent with the good comparison of the growth rate for the baseline case against the unstratified analysis.

%MOVE SATO DISCUSSION HERE?

%CHECK MAX DELTA H IS MAX DELTA H, NOT MAX 2h BELOW
Figure~\ref{fig:vary_strat_growth_rate} shows profiles at individual instants of time and the evolution of $\Delta h(\xjet,\tjet)/2h(\xjet,\tjet)$ over time for a selection of $\width$.  The qualitative evolution remains the same in all cases, with symmetry breaking originating within the intrusion body and propagating both upstream and downstream.  However, the saturation amplitude is significantly suppressed at larger $\width$. %, decreasing from $\max_x \Delta h \approx 3$ for $\width = 0$ to about $2$ for $0.15 \leq \width \leq 0.5$, $1.5$ for $\width = 0.65$, $1.2$ for $\width = 0.8$ and $0.6$ for $\width = 1$.  
The final state also changes qualitatively: for $\width \lesssim 0.3$ the leading dipole persists and is followed by a broad turbulent wake, whereas for $\width \gtrsim 0.8$ the dipole has disappeared and the intrusion remains a thin, irregular train of waves.
%% AS: "DOUBLE CHECK ONCE FIGURE FINAL" -- numbers above are from jet_scaling integrated_data over 108 <= t_j <= 120; re-read from the final figE/figF.

Finally, in FIG.~\ref{fig:thickness_vs_W}, we plot $\max_x h$ against time.  This data collapses well in the intrusion scaling for late times, suggesting that the vertical extent of the final turbulent state is controlled by stratification.

\subsubsection{Varying Reynolds number}

\begin{figure}
    \centering
    \includegraphics[width=\linewidth]{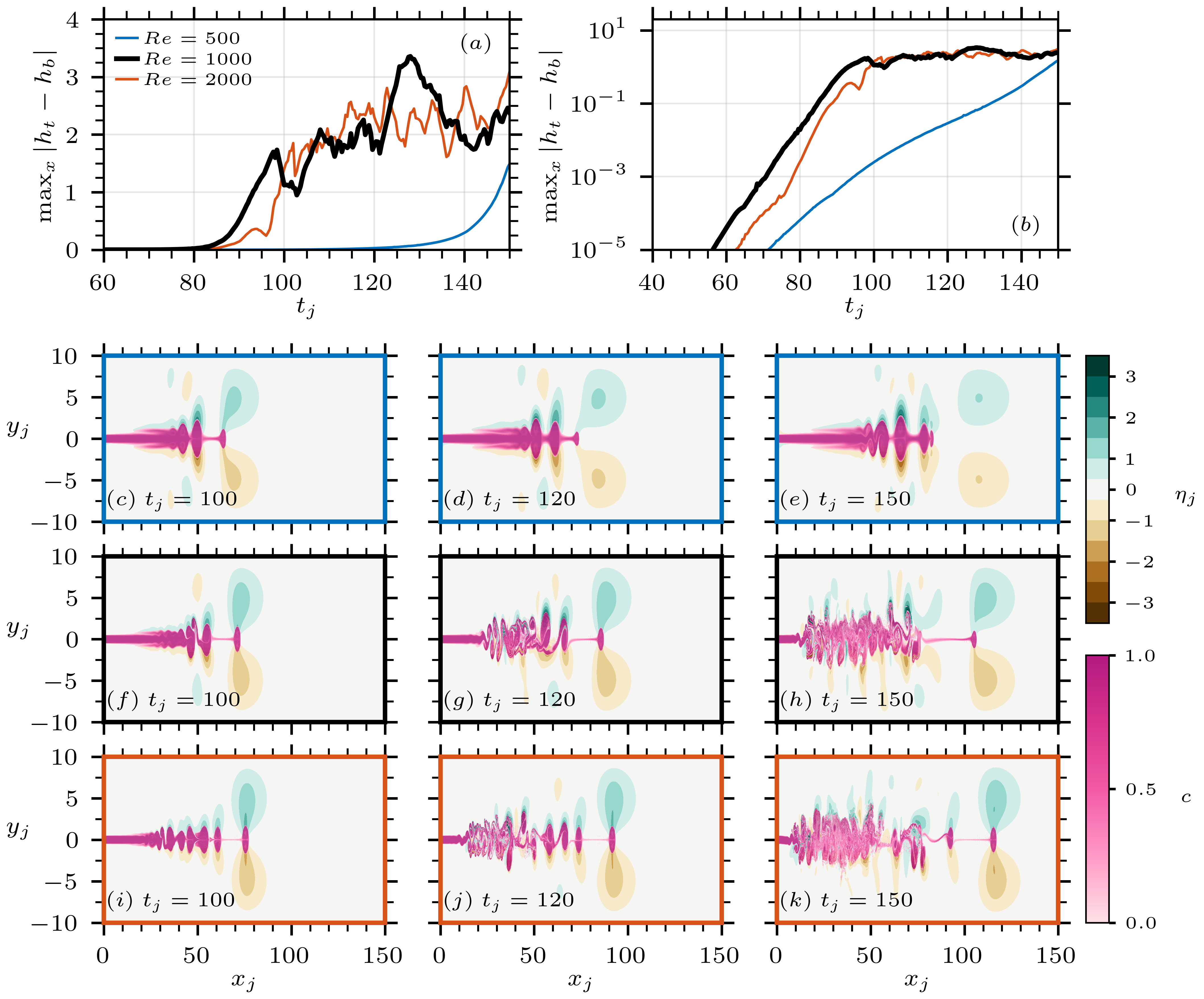}
    \caption{Growth of asymmetry during the transition to turbulence for different values of $\Rey$. Maximum thickness difference as a function of time for (a) different $\Rey$. (b) Corresponding logarithmic representation. (c--k) Snapshots of the intrusion superimposed on the isopycnal displacement $\eta_j$ at $\tjet = 100$, $120$ and $150$ for $\Rey = 500$ (c--e), $1000$ (f--h) and $2000$ (i--k); rows are identified by border color as in (a). {We plot $h$ as the integral of the color function.}}
    \label{fig:vary_Re_growth_rate}
\end{figure}

%% AS: paragraph below left as is, per instruction.  Note that it is contradicted by the data:
%% the fitted growth rates are sigma = 0.105 (Re = 500), 0.32 (1000) and 0.28 (2000), so Re = 500
%% grows three times more slowly and has not saturated by t_j = 150; and the Re = 2000 row of
%% figH is (i--k), not (f,g).
The effect of the Reynolds number during Phase~II is shown in FIG.~\ref{fig:vary_Re_growth_rate}. In all cases, the asymmetry exhibits a clear exponential growth phase.  The initial growth rate is broadly comparable across the three Reynolds numbers considered, although the later growth appears to increase with $\Rey$. This suggests that viscous effects do not significantly alter the initial instability mechanism although do play a role at later times. At lower Reynolds number $\Rey = 500$, viscous damping leads to smoother interfaces and a delay in the onset of instability (FIG. \ref{fig:vary_Re_growth_rate}b). At $\Rey = 2000$, the flow exhibits sharper interfacial structures and more pronounced vortex interactions during the nonlinear stage (FIG. \ref{fig:vary_Re_growth_rate}f,g), though the onset of instability is comparable with that in $\Rey = 1000$. Despite these quantitative differences, the overall evolution remains qualitatively similar in all cases, with symmetry breaking originating within the intrusion body and propagating both upstream and downstream. %These results suggest that the sinuous instability is robust, with Reynolds number primarily influencing the small-scale structure of the transition rather than its underlying nature.

\subsection{Phase~III: Well-mixed intrusion}\label{sec:Phase3}

\begin{figure}
    \centering
    \includegraphics[width=\linewidth]{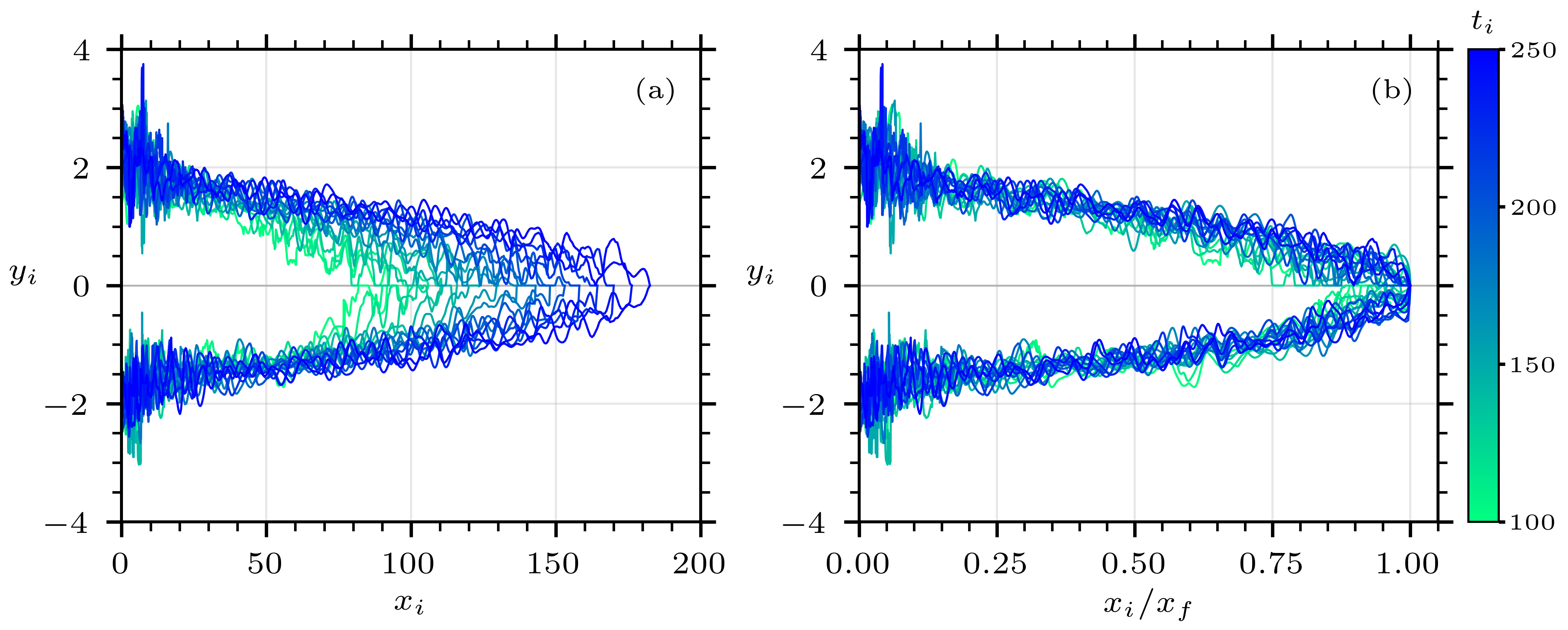}
    \caption{Late-time thickness profiles for the baseline simulation. (a) $h_t(\xint)$ and $-h_b(\xint)$ every $10$ time units over $100 \leq \tint \leq 250$, colored by time. (b) The same profiles plotted against $\xint/x_f$, with $x_f$ the front position; the downstream part of the intrusion collapses approximately. {We plot $h$ as the maximum vertical extent of the contour $c = 0.1$.}}
    \label{fig:phase3_thickness}
\end{figure}

\begin{figure}
    \centering
    \includegraphics[width=\linewidth]{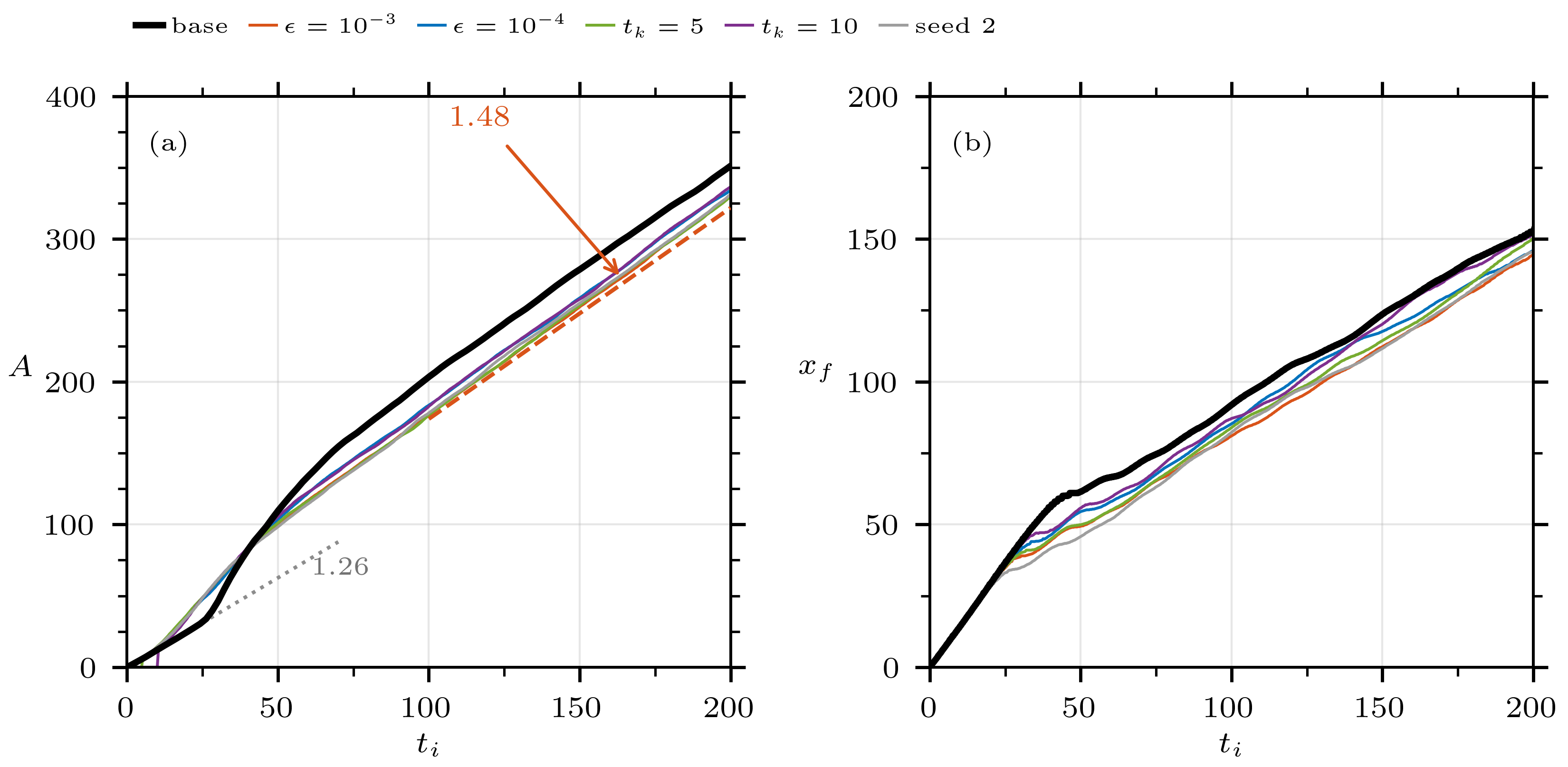}
    \caption{(a) Cross-sectional area $A(\tint)$ and (b) front position $x_f(\tint)$ for the baseline simulation (bold black) and for simulations seeded with white noise (colored): amplitude $\epsilon$, or $\epsilon = 10^{-2}$ added at $\tint = t_k$ or with a second random seed. The dotted line in (a) is the initial growth rate $1.26$, fitted over $0 \leq \tint \leq 25$, and the dashed line the late-time rate $1.48$, fitted over $100 \leq \tint \leq 200$; both are drawn offset below the curve for clarity. %\textcolor{red}{The front is the furthest $\xint$ where $c=0.1$ reaches right most location, and is not monotonic: the leading tip pulses, and when it thins below that value the detected front falls back to the last thickened detectable region. (see phase III 2D snapshots plot)}
    }
    \label{fig:phase3_area_front}
\end{figure}
%% AS: the seeded runs here are the intrusion-scaled set (W=0.5_noise001, noise0001, noise01_t5, noise01_t10, noise01_seed2), not the jet-scaled set of Phase~II. Please confirm the amplitude/restart time of each for the caption -- the session files do not record them.

Once the instability of Phase~II has spread across the whole intrusion, the flow relaxes into a late-time regime in which the fluctuations decay under diffusion and the stabilizing influence of stratification.  Now the large-scale evolution is controlled by buoyancy rather than by the initial jet momentum. This is Phase~III.  We note that the two-dimensional near-source structure of this phase does not survive the addition of the spanwise direction.  We nevertheless include a discussion of the two-dimensional simulations because the general behavior does carry over and because the contrast with the three-dimensional near-source structure is itself informative.

As foreshadowed in FIG.~\ref{fig:thickness_vs_W}, the intrusion scaling of \S\ref{sec:Intrusion rescaling} now provides a better collapse of the data and so we switch to this scaling (see also \cite{CeddiaKerr25}).  The parameters in this scaling are listed in Table~\ref{tab:baseline_params}. For the baseline simulation $\tint = \Wint^2 \tjet = \tjet/4$, so that the saturation of the instability at $\tjet \approx 100$--$120$ corresponds to $\tint \approx 25$--$30$.

% \AS{VORTICITY PLOT?}

% \begin{figure}
%     \centering
%     \includegraphics[width=\linewidth]{paper_figures/phase3_vorticity_W.png}
%     \caption{Phase 3 2D vorticity plot varying W}
%     \label{fig:phase3_2D_vorticity_varyingW}
% \end{figure}

% \begin{figure}
%     \centering
%     \includegraphics[width=\linewidth]{paper_figures/phase3_vorticity_Re.png}
%     \caption{Phase 3 2D vorticity plot varying $Re$}
%     \label{fig:phase3_2D_vorticity_varying_Re}
% \end{figure}

The profile during this stage is shown in FIG.~\ref{fig:2Dsnaps}(j,k) for our baseline simulation, while FIG.~\ref{fig:phase3_thickness}(a) shows the top and bottom thickness profiles $h_t$ and $-h_b$ over time calculated using the contour definition. By $\tint = 250$, the intrusion has entrained a substantial volume of ambient fluid. Its total thickness $h_t + h_b$ is largest near the source, where it fluctuates about $2.4$ with instantaneous peaks of up to $3.5$.
%; in this near-source region substantial mixing and vorticity roll-up persist, and the flow continues to interact with the left-hand boundary of the domain. 
This near-source region is the one shown in \S\ref{sec:3D} to be an artifact of the two-dimensionality of the simulations.  Farther downstream, by contrast, the strongest turbulent activity has decayed and the intrusion is comparatively well mixed.  It forms a confined wedge that thins gradually away from the source %($h_t + h_b \approx 2.1$ at $\xint = 30$ and $1.3$ at $\xint = 100$)
and more rapidly towards the front. When the profiles are replotted against the normalized coordinate $\xint/x_f$ (FIG.~\ref{fig:phase3_thickness}b) they collapse reasonably well.
%% AS: the old draft quoted h_t+h_b ~ 4.4 near the source. That figure came from the contour (c > 0.1 envelope) thickness, which is 3.3 at x=30 and ~4 near the source at t=250; with the integral definition of h_t, h_b used in the paper and in this figure the near-source total is 2.4 (smoothed) with spikes to 3.5. The numbers above are the integral ones.

Entrainment is significant throughout Phase~III. FIG.~\ref{fig:phase3_area_front}(a) shows the cross-sectional area $A(\tint)$. Since the supplied area increases as $\tint$, the excess of $A$ over $\tint$ measures the dilution of the intrusion by entrained ambient fluid. The area grows in three stages. It increases linearly over $\tint \in [0, 25]$ at the rate $\mathrm{d}A/\mathrm{d}\tint \approx 1.26$ (dotted line), already above unity because of diffusion and ambient fluid mixing into the vortices during Phase~I. It then grows more rapidly from $\tint \approx 25$, with $\mathrm{d}A/\mathrm{d}\tint$ peaking at $\approx 4$ around $\tint \approx 30$--$35$ as the sinuous instability breaks down the trailing jet. Finally it returns to linear growth for $\tint \gtrsim 80$, with $\alpha = \mathrm{d}A/\mathrm{d}\tint = 1.48$ over $100 \leq \tint \leq 200$ (dashed line), the window used for all late-time fits in this section. The late-time rate is thus about $50\%$ above the source flux, and the mean concentration of the fluid counted in $A$ is $\tint/A \approx 0.6$. %Changing the threshold to $c_\mathrm{th} = 0.05$ or $0.2$ changes $\alpha$ by about $\pm 8\%$ ($1.58$ and $1.36$) without altering this picture.
%% AS: threshold sensitivity was checked for the baseline only, from the contour thickness files (average_h*.txt, thresholds 0.05/0.1/0.2), same 100-200 window: alpha = 1.58/1.47/1.36; A(250) = 456/430/394. The other W were not re-run at other thresholds.

The front position (FIG.~\ref{fig:phase3_area_front}b) shows a contrasting behavior. The front advances at a constant speed of $\approx 1.5$ (the Phase~I speed of $0.72$ rescaled to intrusion units) until $\tint \approx 30$, then decelerates abruptly as the instability reaches it. % the speed falls to $\approx 0.35$ by $\tint \approx 50$, when the front almost stalls at $x_f \approx 62$, before recovering to $0.6$--$0.8$. 
At late times the front advances at a mean speed of $\approx 0.6$ over $100 \leq \tint \leq 200$, less than half the Phase~I speed, with a slight pulsing modulation superposed.  %Taken together, these results suggest that Phase~III is best interpreted as an emerging self-similar far-field intrusion fed by a persistent, entraining, non-self-similar adjustment region near the source, rather than as a single global similarity solution for the full flow.

The colored curves in FIG.~\ref{fig:phase3_area_front} are simulations in which the instability was triggered earlier by white noise (as described for Phase~II). The late-time behavior does not depend on how the instability was triggered: the late-time slopes are $\alpha = 1.51$--$1.53$ for all five seeded runs, the fronts advance at the same late-time speed, and the curves differ from the baseline only by an offset of about $5\%$ in $A$ and of a few units in $x_f$ resulting from the earlier transition.

\subsubsection{Varying stratification}\label{sec:phase3_varyingW}

\begin{figure}
    \centering
    \includegraphics[width=\linewidth]{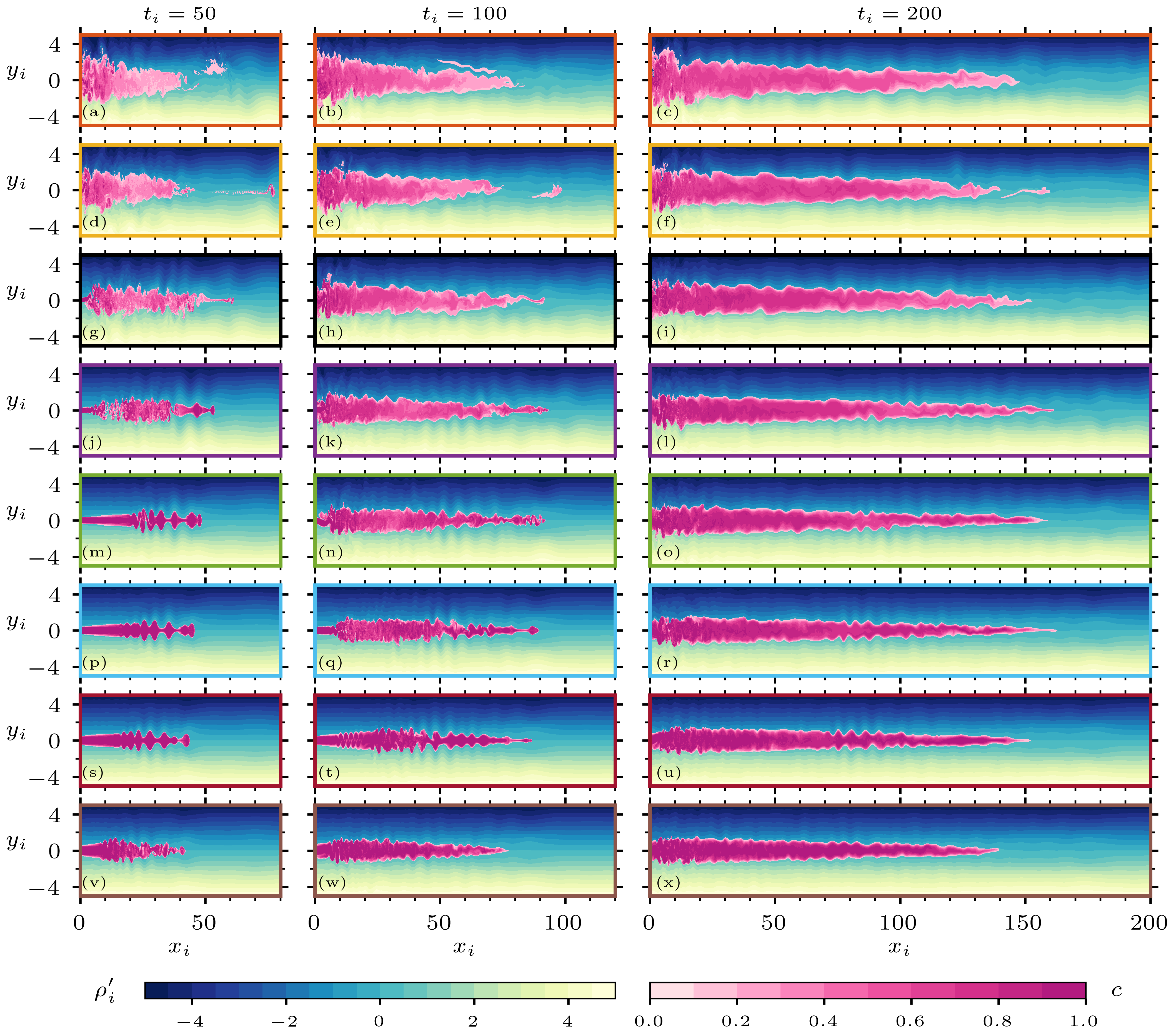}
    \caption{Color function $c$ over the ambient density perturbation $\rho'_i$ at $\tint = 50$ (left), $100$ (middle) and $200$ (right) for $\Wint = 0.3$ (a--c), $0.4$ (d--f), $0.5$ (g--i), $0.6$ (j--l), $0.7$ (m--o), $0.8$ (p--r), $0.9$ (s--u) and $1$ (v--x), with $\Hint = 10$ and $\Rey = 1000$. All panels share one streamwise scale, the columns being cropped to the extent of the intrusion, and the vertical axis is stretched; fluid with $c < 0.1$ is drawn as ambient.}
    \label{fig:phase3_profiles_W}
\end{figure}

\begin{figure}
    \centering
    \includegraphics[width=\linewidth]{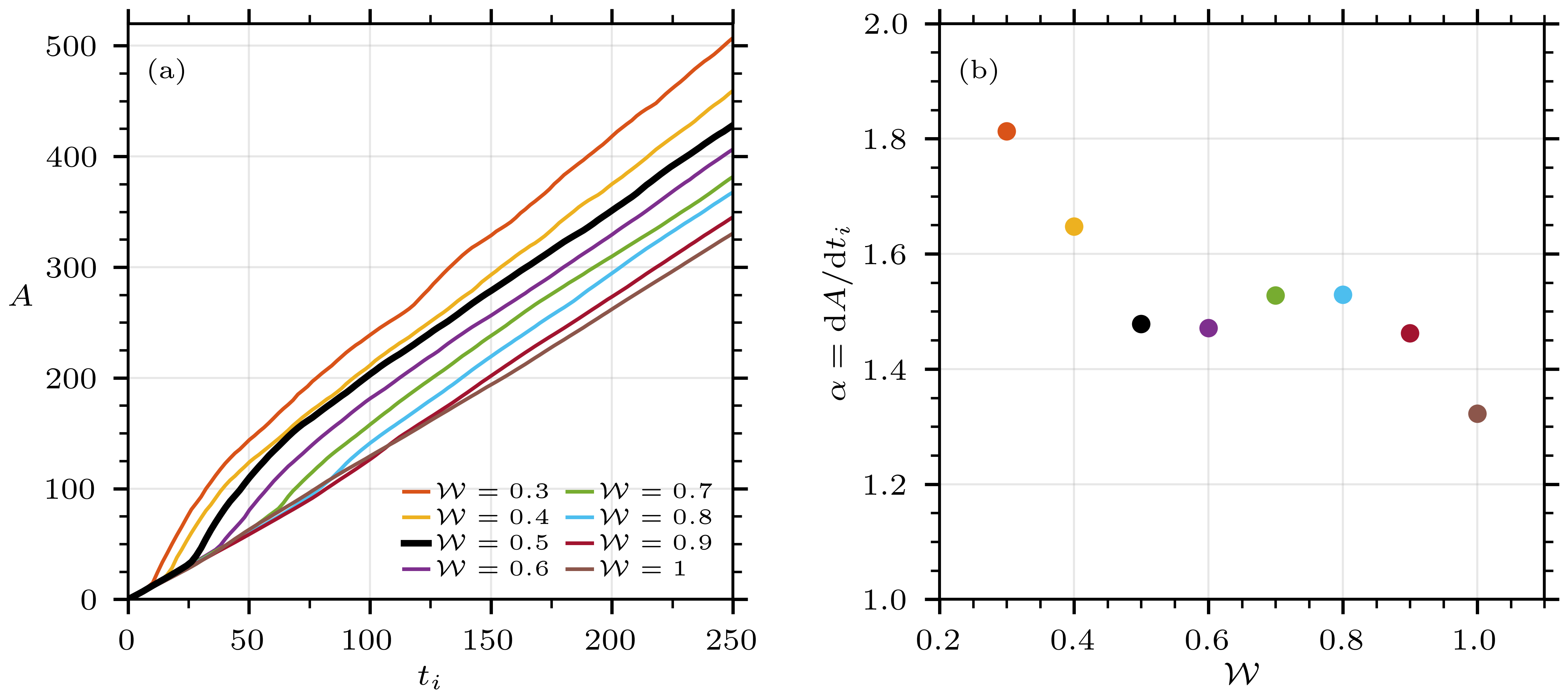}
    \caption{(a) Intrusion area $A(\tint)$ for $\Wint = 0.3$--$1$ ($\Hint = 10$, $\Rey = 1000$). (b) The late-time growth rate $\alpha = \mathrm{d}A/\mathrm{d}\tint$, fitted over $100 \leq \tint \leq 200$.}
    \label{fig:phase3_area_vs_W}
\end{figure}

FIG.~\ref{fig:phase3_profiles_W} shows the color function and the ambient density perturbation at a selection of times for $\Wint = 0.3$--$1$. In intrusion units, the transition occurs later as $\Wint$ increases, but by $\tint=200$, every case has relaxed to a wedge-shaped intrusion.
%At $\tint = 50$ the intrusions for $\Wint \leq 0.5$ (a,d,g) have already broken down along their whole length, $\Wint = 0.6$ (j) is in the midst of the sinuous breakdown, whereas for $\Wint \geq 0.7$ (m,p,s,v) the trailing jet still carries the laminar varicose wave train of Phase~I behind a coherent head. By $\tint = 100$ the breakdown has spread along the body for all $\Wint \leq 0.9$ (b,e,h,k,n,q,t) and is weakest for $\Wint = 1$ (w). By $\tint = 200$ every case has relaxed to a wedge-shaped intrusion (c,f,i,l,o,r,u,x). 
In this regime, the front position is nearly independent of $\Wint$: at $\tint = 200$ it lies at $x_f \approx 140$--$170$ for all cases, and at $\tint = 250$ at $170$--$200$, so the intrusion scaling collapses the late-time front propagation to within about $\pm 8\%$ even though the transition occurs at very different $\tint$. The more jet-like intrusions are visibly more dilute and more diffuse, with detached filaments and patches of tracer ahead of and above the main body (panels b,c,e,f), whereas at large $\Wint$ the body remains sharp-edged and beaded by the remnants of the varicose waves (panels r,u,x).

The intrusion area exhibits the same broad three-stage evolution for every $\Wint$ (FIG.~\ref{fig:phase3_area_vs_W}a): an initial approximately linear growth, a rapid increase once the instability sets in, and a return towards approximately linear growth at late times. This sequence is clearest for the smaller $\Wint$. %, which show the earliest and strongest departure from the initial linear trend and the largest area at a given time, indicating stronger instability-driven entrainment in the more jet-like cases: the peak growth rate occurs at $\tint \approx 12$, $20$, $32$, $50$, $66$, $88$ and $108$ for $\Wint = 0.3$--$0.9$ and decreases from $\mathrm{d}A/\mathrm{d}\tint \approx 4.7$ to $1.7$, while for $\Wint = 1$ there is no distinct burst. %
As $\Wint$ increases, the same progression remains apparent, although the intermediate stage becomes weaker and less sharply distinguished, so the return to late-time linear growth is less dramatic. 
The late-time slope $\alpha$ (FIG.~\ref{fig:phase3_area_vs_W}b) decreases from $1.81$ at $\Wint = 0.3$ and $1.65$ at $0.4$ to $1.46$--$1.53$ over $0.5 \leq \Wint \leq 0.9$, and to $1.32$ at $\Wint = 1$.  Nevertheless, the near-source thickness of the wedge is essentially the same for all $\Wint$; the extra area for small $\Wint$ manifests as a marginally broader region of greater thickness.
%Thus smaller-$\width$ intrusions are more dilute than : the integral thickness $h_t + h_b$ in the wedge is essentially the same for all $\Wint$ ($1.65$--$1.76$ at $\xint = 50$ and $\tint = 200$), and the extra area at small $\Wint$ is the diffuse, low-concentration fluid visible in FIG.~\ref{fig:phase3_profiles_W}(c,f).
%% AS: on the Q_eff question. Rescaling lengths with the late-time rate alpha in place of the source flux (x_f / sqrt(alpha)) does not improve the collapse of the front: x_f(250)/sqrt(alpha) = 143, 150, 151, 158, 160, 154, 150, 147 for W = 0.3 ... 1, the same +-6% spread as the raw x_f(250) = 192, 192, 184, 192, 198, 191, 181, 169. The raw intrusion scaling already collapses x_f, and alpha measures dilution rather than an extra volume flux (see the previous paragraph), so I have not introduced Q_eff.
%% AS: every late-time fit in this section (both area figures, the text, the Re numbers) uses 100 <= t <= 200, the only window all runs cover (the seeded runs end at t = 200). The old draft's 1.52 for the baseline was the 60-200 window, which still overlaps the concave part of A(t); on 100-200 it is 1.48.

\subsubsection{Varying Reynolds number}

\begin{figure}
    \centering
    \includegraphics[width=\linewidth]{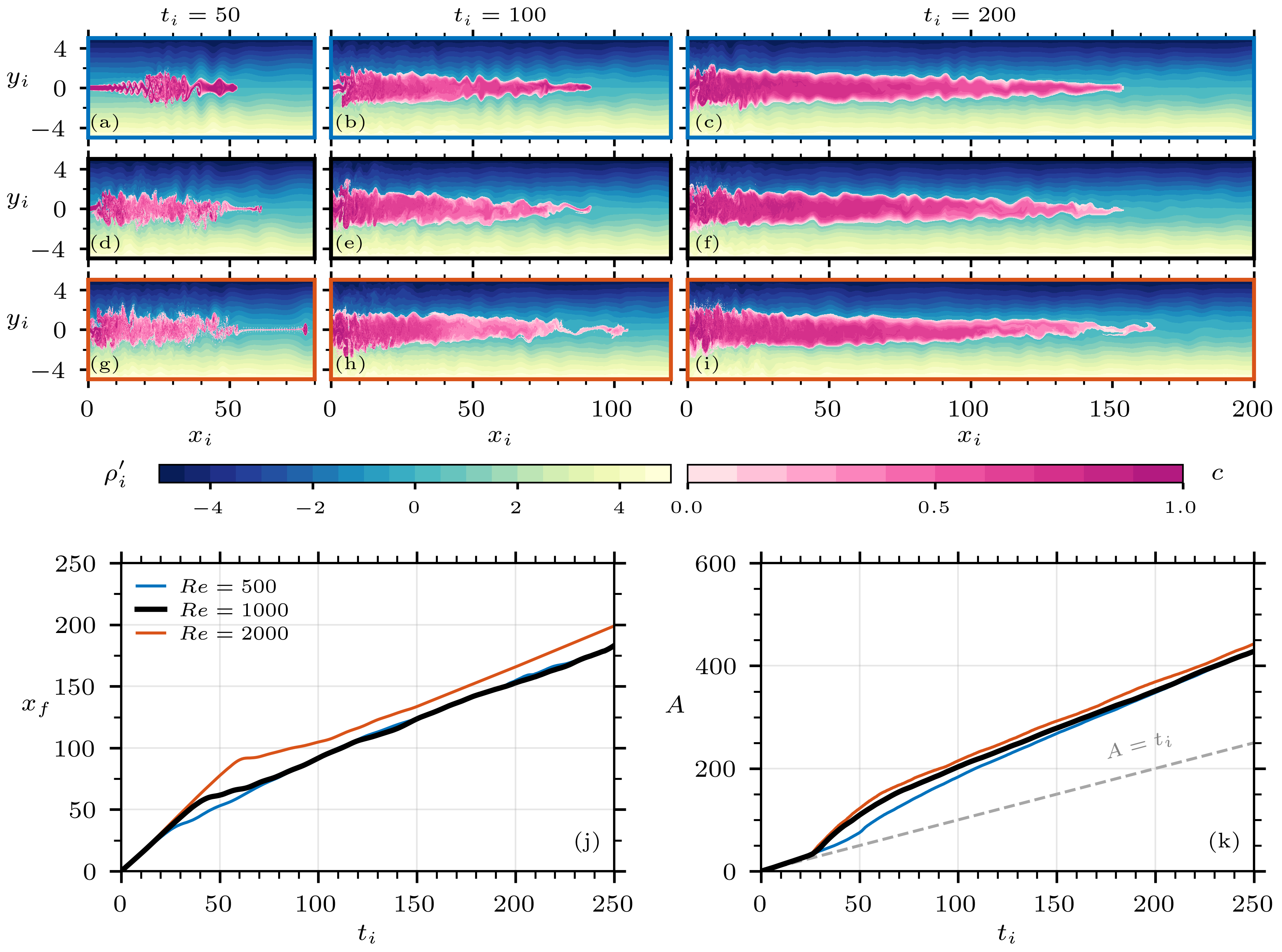}
    \caption{Effect of $\Rey$ on Phase~III, for $\Rey = 500$ (a--c), $1000$ (d--f) and $2000$ (g--i) at $\tint = 50$, $100$ and $200$, with all other parameters as in the baseline case; color function over the density perturbation as in FIG.~\ref{fig:phase3_profiles_W}. (j) Front position $x_f(\tint)$ and (k) intrusion area $A(\tint)$; the dashed line in (k) is $A = \tint$, the area of the source fluid alone.}
    \label{fig:phase3_varying_Re}
\end{figure}

FIG.~\ref{fig:phase3_varying_Re} shows the late-time evolution for $\Rey = 500$, $1000$ and $2000$, with all other parameters identical to those of the baseline case. The broad-scale structure remains the same.  In particular, at $\tint = 200$ the three intrusions have the same wedge shape and total thickness (c,f,i; $h_t + h_b \approx 2.9$--$3.4$ at $\xint = 30$ and $1.8$--$2.2$ at $\xint = 100$), and the late-time front speed of $\approx 0.6$ (j) and the late-time area growth (k) are similar.  However, some fine-scale and early-dynamics structures differ. %Given the importance of the spanwise coordinate, we do not discuss these further.

\subsubsection{Comparison with experimental studies} \label{sec:2d_comparison}

Comparing our planar results with the experiments of Zuluaga-Angel \etal~\cite{EPA72} and Ceddia \etal~\cite{CeddiaKerr25}, the clearest point of agreement is the overall wedge-like structure of the downstream intrusion. The near-source region, however, disagrees. Both experimental studies reported a rapidly expanding near-source region, interpreted as an entraining hydraulic-jump/adjustment zone. Our two-dimensional simulations instead develop a near-source region that is thickened all the way to the source, without this expansion. This motivates the three-dimensional simulations of the next section.

%% ---- Three-dimensional simulations (merged from 3D_section.tex, 8 Sep) ----
\providecommand{\ASq}[1]{\textcolor{red}{#1}}   % Anja's open notes, in red
\providecommand{\Qe}{Q_e}
\providecommand{\xs}{x_s}

\section{Three-dimensional simulations} \label{sec:3D}

The two-dimensional simulations of \S\ref{sec:Phasesummary} become turbulent at late times, and two-dimensional
turbulence is not generally representative of real, three-dimensional turbulent flows. It is
therefore imperative to validate these observations against three-dimensional simulations. We 
find that some qualitative structures are robust: the far-field wedge persists essentially
unchanged. However, the near-source region is significantly different.%% AS: suggested content for the intro (from old_ver.tex l.758): why 3D now, that the
%%     aim is not a full parameter survey, and that spanwise-averaged fields are shown.

We perform six simulations at the parameter values shown in table~\ref{tab:3D_runs}. The simulations do not become unstable in the spanwise direction through numerical error alone, so in each case we introduce a white-noise perturbation of amplitude $0.1$ at time $\tint=3$.  As our focus is the behavior at later times, we only present results using the intrusion scaling.
The spanwise domain has length $\Lzint=5$ with $N_z=160$ spanwise planes, and all other parameters are those of the baseline case in table~\ref{tab:baseline_params}.  Unless stated otherwise, the fields shown are averaged over the span.

Here we focus our discussion on the simulation with $\width=0.3$ and $\Hint=10$.  A movie of the evolution is provided with the Supplementary Material.
%% AS: L_z and N_z from old_ver.tex l.760 (H=10, W=0.3 run); please confirm they are the
%%     same for the H=6 runs.

\begin{table}[h]
\caption{Three-dimensional simulations. All runs have $\Rey=1000$, $\Sc=10$, $\Lint=400$, $\Lzint=5$ and $N_z=160$.  $t_{\max}$ is the final time reached; $U_f$ is the front speed over the second half of the run; $\Qe$ and $\xs$ are the areal flux and virtual origin used in \S\ref{sec:3D_wedge}, and $\Hint/2\pi$ is the mode-2 long-wave speed of the ambient.}
\centering
\renewcommand{\arraystretch}{1.2}
\setlength{\tabcolsep}{8pt}
\begin{tabular}{cc|cccc|c}
$\Hint$ & $\Wint$ & $t_{\max}$ & $U_f$ & $\Qe$ & $\xs$ & $\Hint/2\pi$ \\
\hline
10 & 0.2 & 63  & 0.85 & 1.96 & 6.4  & 1.59 \\
10 & 0.3 & 75  & 0.77 & 1.64 & 4.4  & 1.59 \\
10 & 0.5 & 84  & 0.73 & 1.42 & 6.7  & 1.59 \\
6  & 0.3 & 100 & 0.95 & 1.54 & 5.1  & 0.95 \\
6  & 0.5 & 122 & 0.82 & 1.42 & 6.2  & 0.95 \\
6  & 0.8 & 80  & 0.69 & 1.28 & 11.4 & 0.95 \\
\end{tabular}
\label{tab:3D_runs}
\end{table}
%% AS: U_f = linear fit of the c=0.1 front position over t >= t_max/2 (fig3D_collapse.py
%%     data); Q_e = slope of A(t) over the same window; x_s = median position of the
%%     thickness maximum (x<30) over the same window.  H=6/W=0.8 has area data only for
%%     61 <= t <= 80.

\subsection{Phase~II}

\begin{figure}
    \centering
    \includegraphics[width=\linewidth]{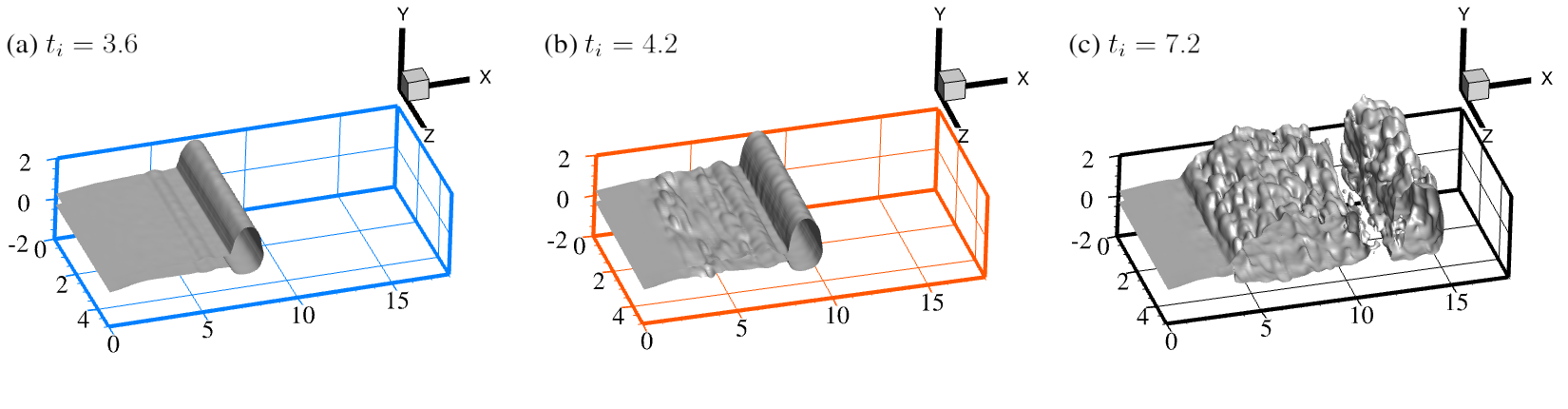}
    \includegraphics[width=\linewidth]{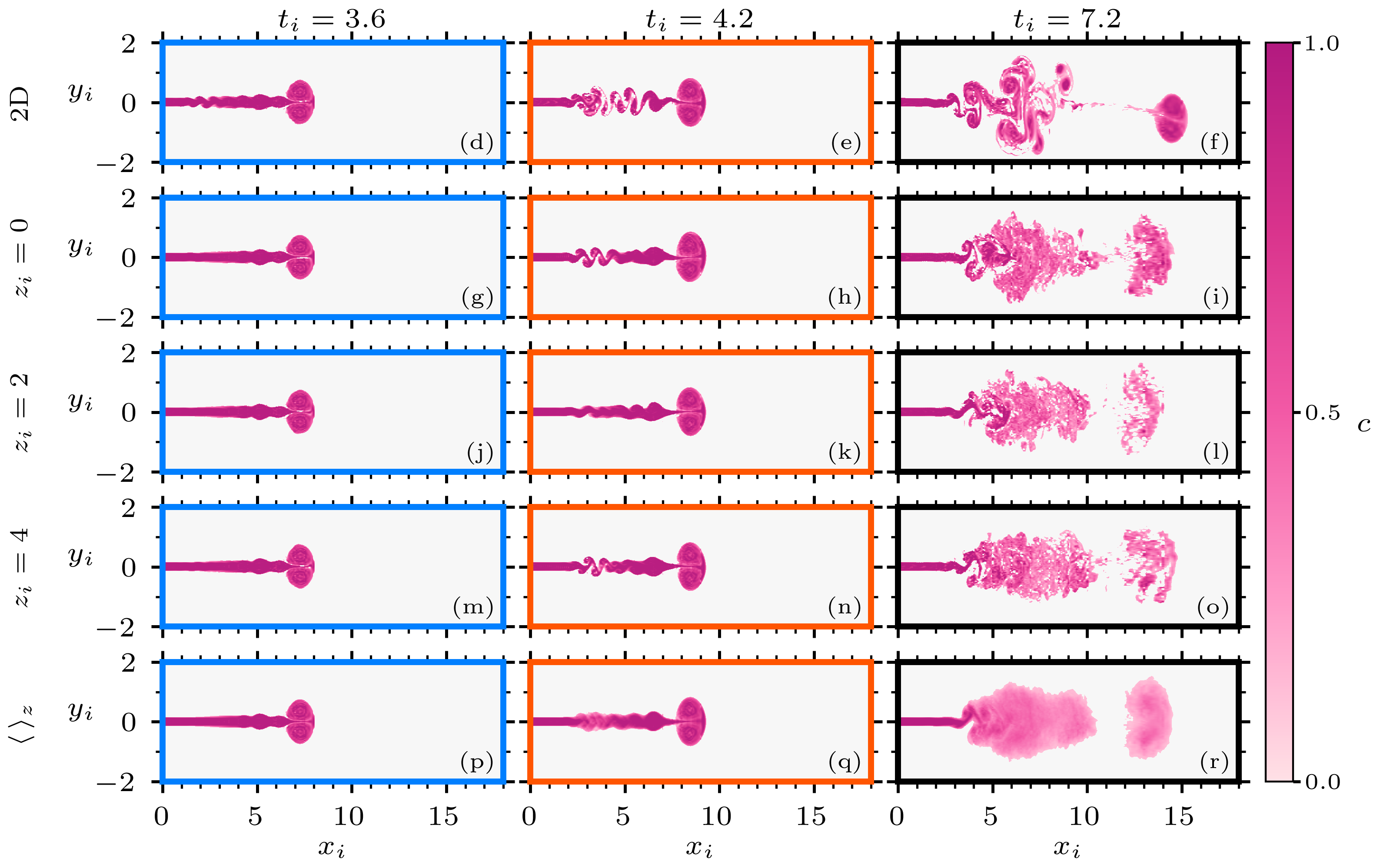}
    \caption{Development of the instability in Phase~II for $\Hint=10$, $\Wint=0.3$ after the white-noise perturbation of amplitude $0.1$ is applied at $\tint=3$.
    (a--c) Iso-surfaces $c=0.1$ of the three-dimensional intrusion at $\tint = 3.6$, $4.2$ and $7.2$, showing the emergence of spanwise structure.
    (d--f) the corresponding two-dimensional simulation, perturbed at the same time with the same noise amplitude.
    (g--o) The intrusion on the spanwise planes $\zint = 0$, $2$ and $4$ at the same times.
    (p--r) The spanwise-averaged three-dimensional fields and
    }
    \label{fig:3D_phase2}
\end{figure}

Our main focus is the flow in Phase~III, but we first briefly discuss the development of instability from $\tint=3$ until the end of Phase~II.  Profiles are shown in FIG.~\ref{fig:3D_phase2} together with the corresponding 2D simulation perturbed in the same way at the same time.  The early behavior in 3D corresponds closely with that in 2D: the flow has a prominent coherent leading vortex dipole and an instability develops on the trailing jet that is primarily sinuous with a weak spanwise modulation.  However, towards the end of Phase~II, the 2D and 3D simulations begin to differ.  The leading vortex loses its structure in the 3D simulations while that in 2D continues to propagate coherently.  The instability on the trailing jet is also more diffuse in the 3D simulations and does not extend as far vertically.
%% AS: skeleton had "$\Ha=10$"; in paper.tex \Ha is H_jet, so \Hint is used here.

%% AS: old caption gave the slices as z_jet = 0, 6, 12 (jet scaling); in intrusion units
%%     these are z_int = 0, 1.5, 3 for W=0.3 (z_int = W z_jet ... please check the factor).

\subsection{Phase~III}

\begin{figure}
    \centering
    \includegraphics[width=\linewidth]{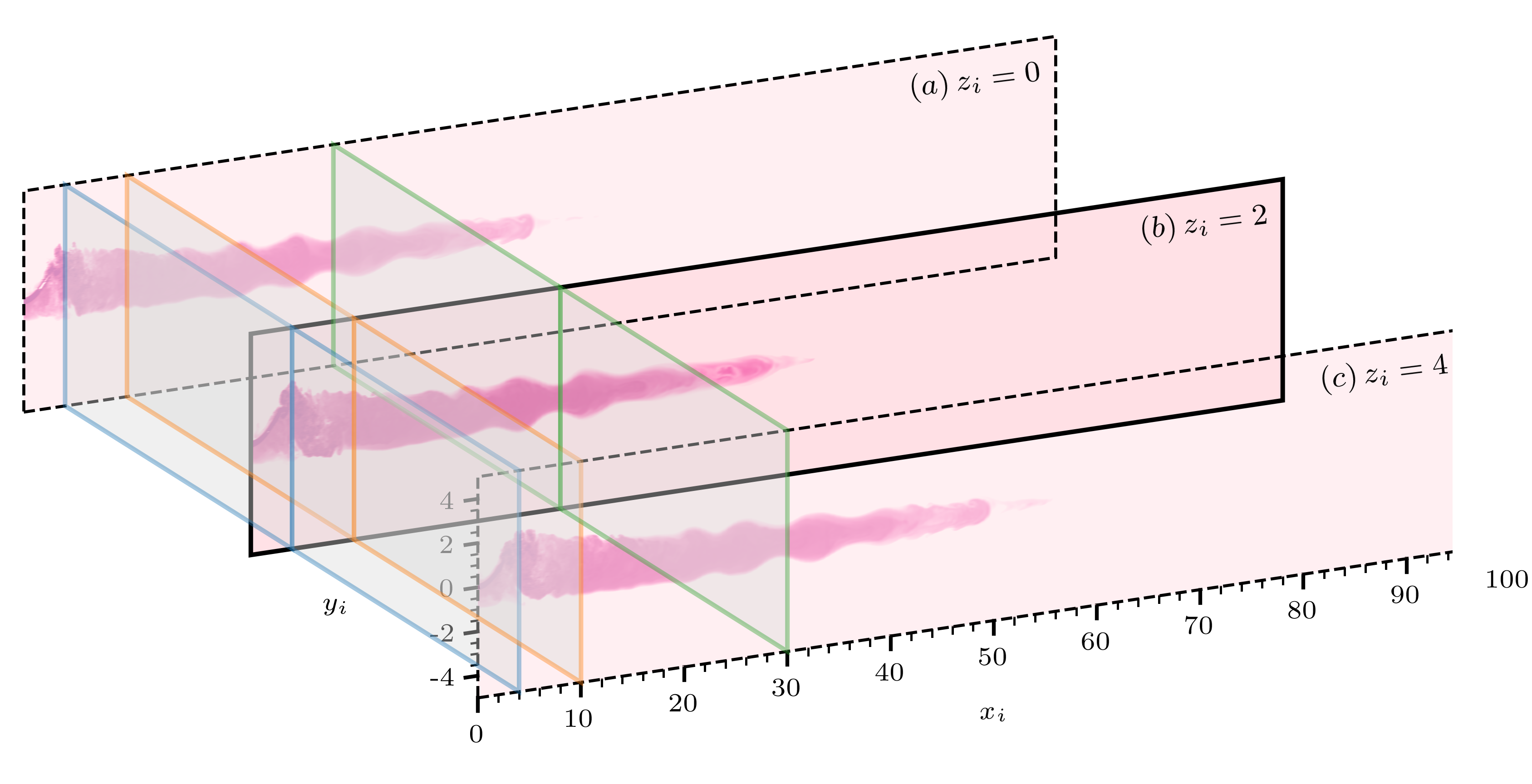}
    \includegraphics[width=\linewidth]{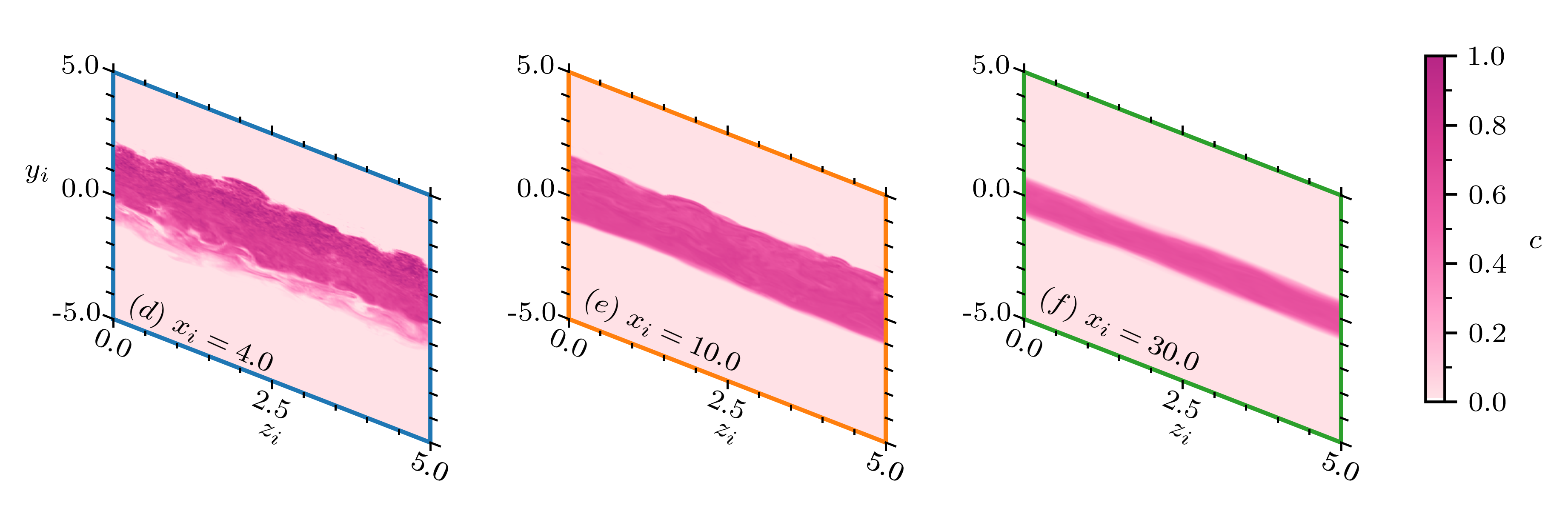}
    \includegraphics[width=\linewidth]{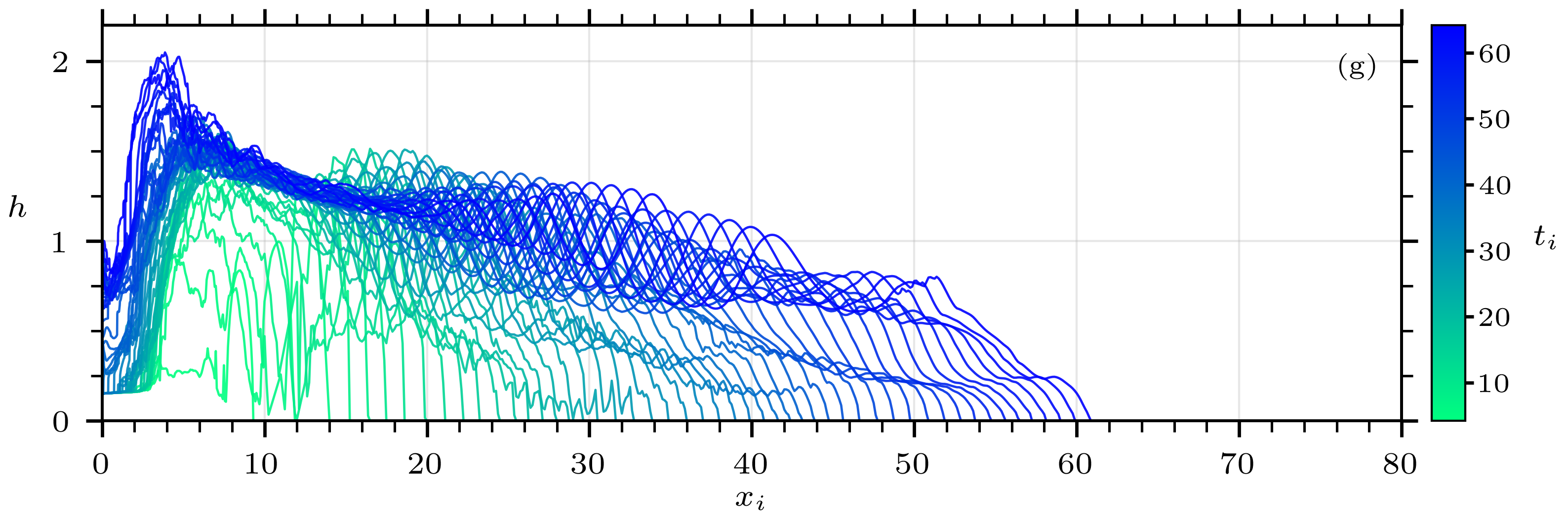}
    \caption{Three-dimensional intrusion structure for $\Hint=10$, $\Wint=0.3$. At $\tint = 56.4$, panels (a--c) show the color function on planes $\zint =0,2$, and $4$, and panels (d--f) show cross-sections at $\xint=4,10$, and $30$, respectively. Panel (g) shows the evolution of the intrusion half-thickness. {We plot $h$ as the maximum vertical extent of the contour $c = 0.1$.}}
    \label{fig:3D_late}
\end{figure}

\begin{figure}
    \centering
    \includegraphics[width=\linewidth]{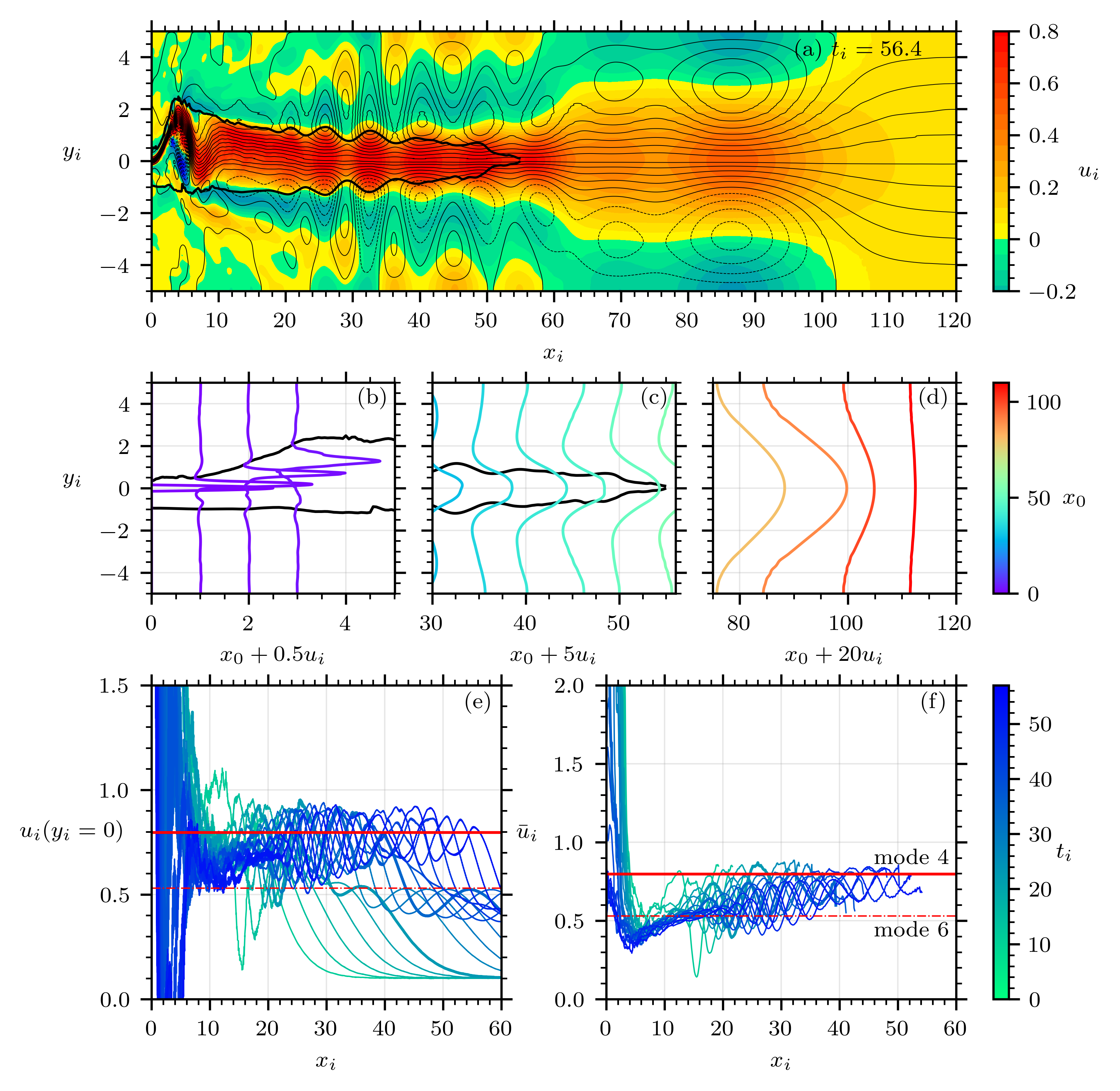}
    \caption{Spanwise-averaged velocity structure for the simulation in FIG.~\ref{fig:3D_late}. (a) Streamlines (solid black curves) superimposed on the horizontal velocity at $\tint=56.4$; the bold black curve is the $c=0.1$ contour. (b--d) Horizontal velocity profiles (b) near the inlet, (c) within the intrusion body and (d) ahead of the intrusion. (e) Horizontal velocity along the centerline, $u_i(\yint=0)$, and (f) vertically averaged horizontal velocity $\bar{u_i}$, at times from $\tint=9$ to $56.4$ in intervals of $0.6$. The dotted and dashed red lines indicate the speeds of the mode-4 and mode-6 internal waves. {We plot $\bar{u_i}$ from $h$ being extracted as the maximum vertical extent of the contour $c = 0.1$.}}
    \label{fig:3D_velocity}
\end{figure}

\begin{figure}
    \centering
    \includegraphics[width=\linewidth]{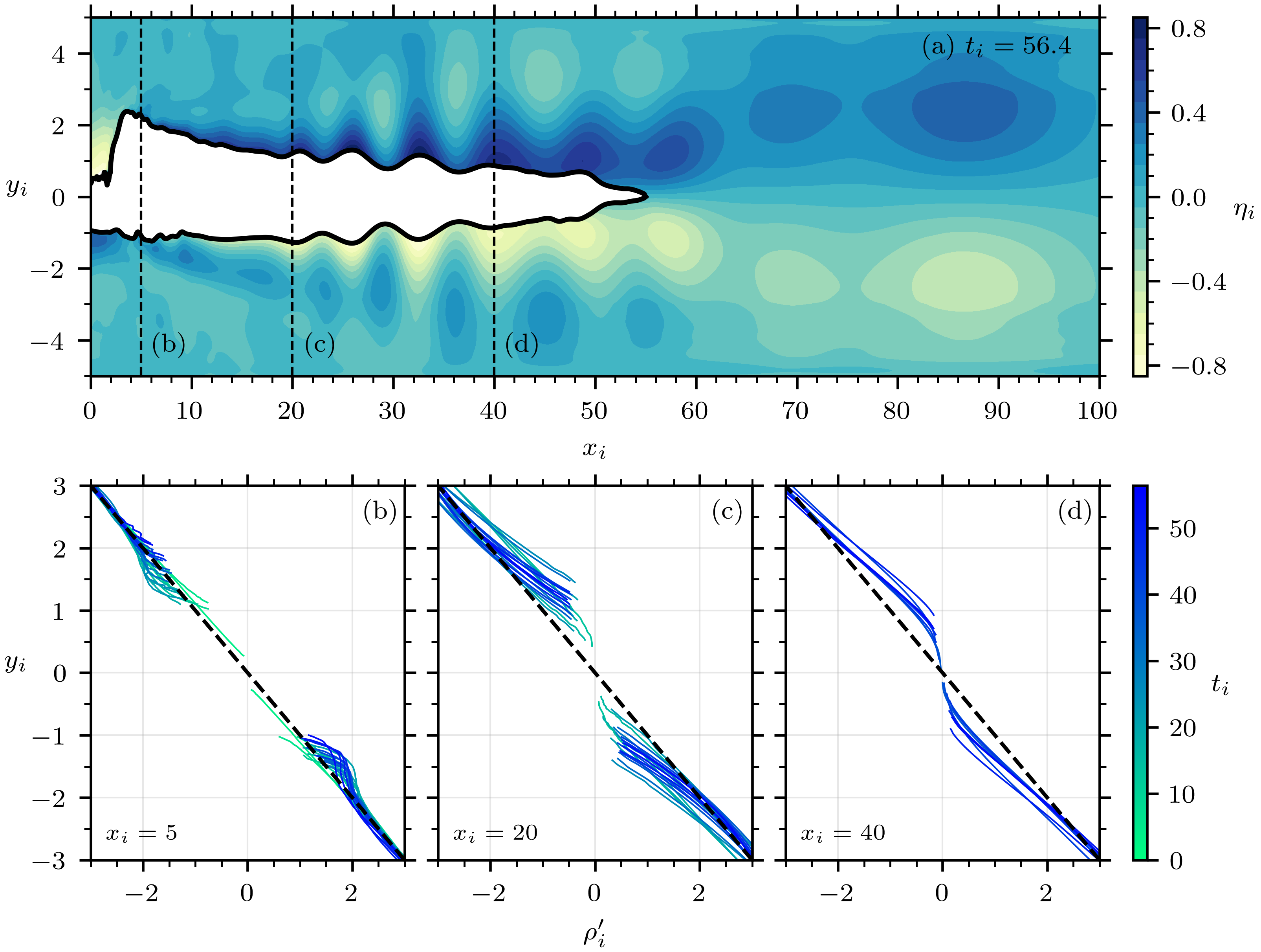}
    \caption{Ambient displacement and density structure for $\Hint=10$, and $\Wint=0.3$. (a) Spanwise-averaged isopycnal displacement $\spav{\eta_i}$ at $\tint=56.4$, with the intrusion masked in white and the intrusion envelope in solid black line. Dashed vertical lines mark the locations used in (b-d). Panels (b-d) show vertical profiles of $\spav{\rho'_i}$ at $\xint = 5,20$, and $40$, respectively, colored by time. The black dashed lines show the undisturbed profile $\rho_i'=-\yint$.
    }
    \label{fig:3D_density}
\end{figure}

\begin{figure}
    \centering
    \includegraphics[width=\linewidth]{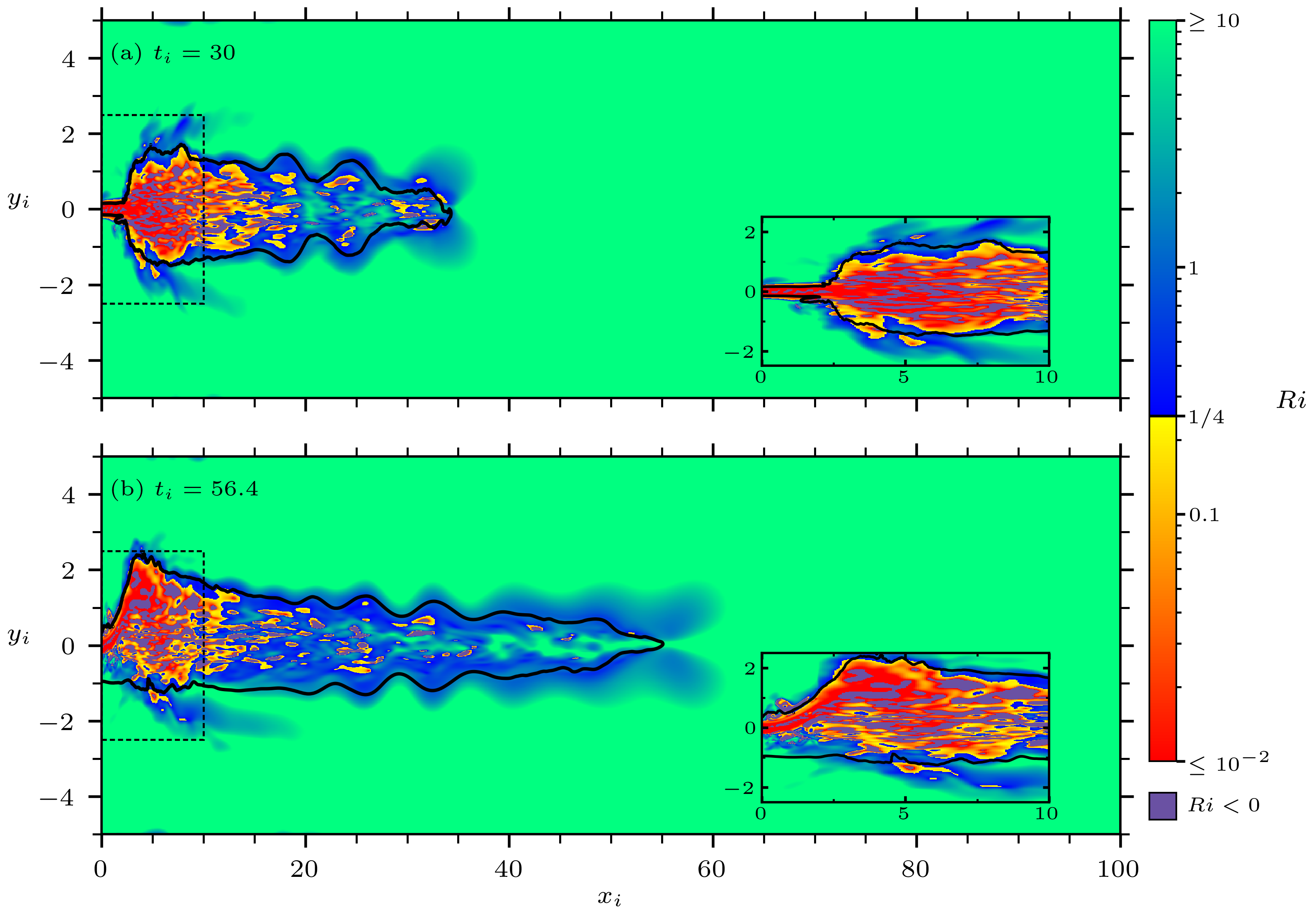}
    \caption{Gradient Richardson number for $\Hint=10$ and $\Wint=0.3$ at (a) $\tint=30$ and (b) $\tint=56.4$. The local quantity $Ri_g = -(\partial_{\yint}\rhointnd)/ [(\partial_{\yint}\uint)^2 + (\partial_{\yint}\wint)^2]$ is calculated from the full three-dimensional fields before spanwise averaging; colors show $\langle Ri_g\rangle$. Insets enlarge the near-source regions. Black contours mark $\langle c\rangle=0.1$, and purple indicates negative values of $\langle Ri_g\rangle$.}
\label{fig:3D_Ri}
\end{figure}

\begin{figure}
    \centering
    \includegraphics[width=\linewidth]{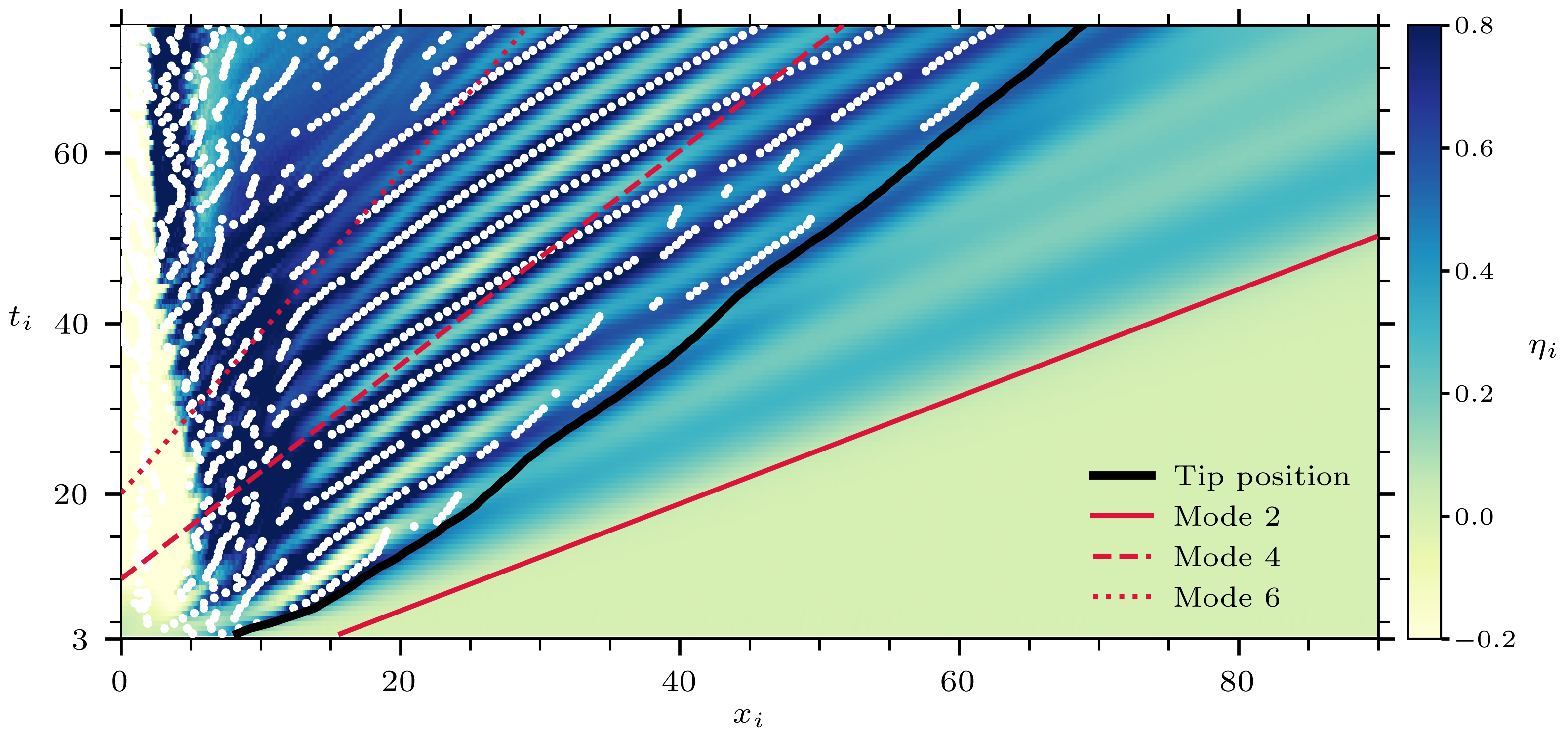}
    \caption{Temporal evolution of the isopycnal displacement along a horizontal slice $\yint=1.2$, for $\Hint=10$, $\Wint=0.3$. The solid black line is the intrusion tip. The white dots mark the positions of the wave crests on the upper surface of the intrusion. The red solid, dashed, and dotted lines have the slopes of the mode-2, mode-4 and mode-6 long-wave speeds $\Hint/2\pi$, $\Hint/4\pi$, and $\Hint/6\pi$, respectively. {We plot $h$ as the maximum vertical extent of the contour $c = 0.1$.}}
    \label{fig:3D_waves}
\end{figure}

Figures \ref{fig:3D_late}--\ref{fig:3D_waves} show details of the same simulation at later times, in Phase~III.  Figure \ref{fig:3D_late} shows a snapshot of the full intrusion structure and the evolution of spanwise-averaged thicknesses, while FIGs.~\ref{fig:3D_velocity} and~\ref{fig:3D_density} show the evolution of the spanwise-averaged velocity and density profiles respectively.  Figure~\ref{fig:3D_Ri} shows a snapshot of the gradient Richardson number.  Finally, FIG.~\ref{fig:3D_waves} shows the temporal evolution of the density profile along a horizontal slice immediately above the intrusion as well as the position of the intrusion tip and the positions of wave crests on the upper surface of the intrusion.

%% AS: the denominator uses (du/dy)^2 + (dw/dy)^2, i.e. not the usual (du/dy)^2 alone;
%%     the definition in the caption should match the script (plot_richardson_number_3d.py).

Overall the intrusion appears as an expanding, turbulent near-source region that connects to a well-mixed, gradually thinning wedge-like region ahead of it.  This behavior is qualitatively similar to the experiments of Zuluaga-Angel \etal~\cite{EPA72} and Ceddia \etal~\cite{CeddiaKerr25}, but differs fundamentally from the 2D results in the behavior at the source.

\begin{figure}
    \centering
    \includegraphics[width=\linewidth]{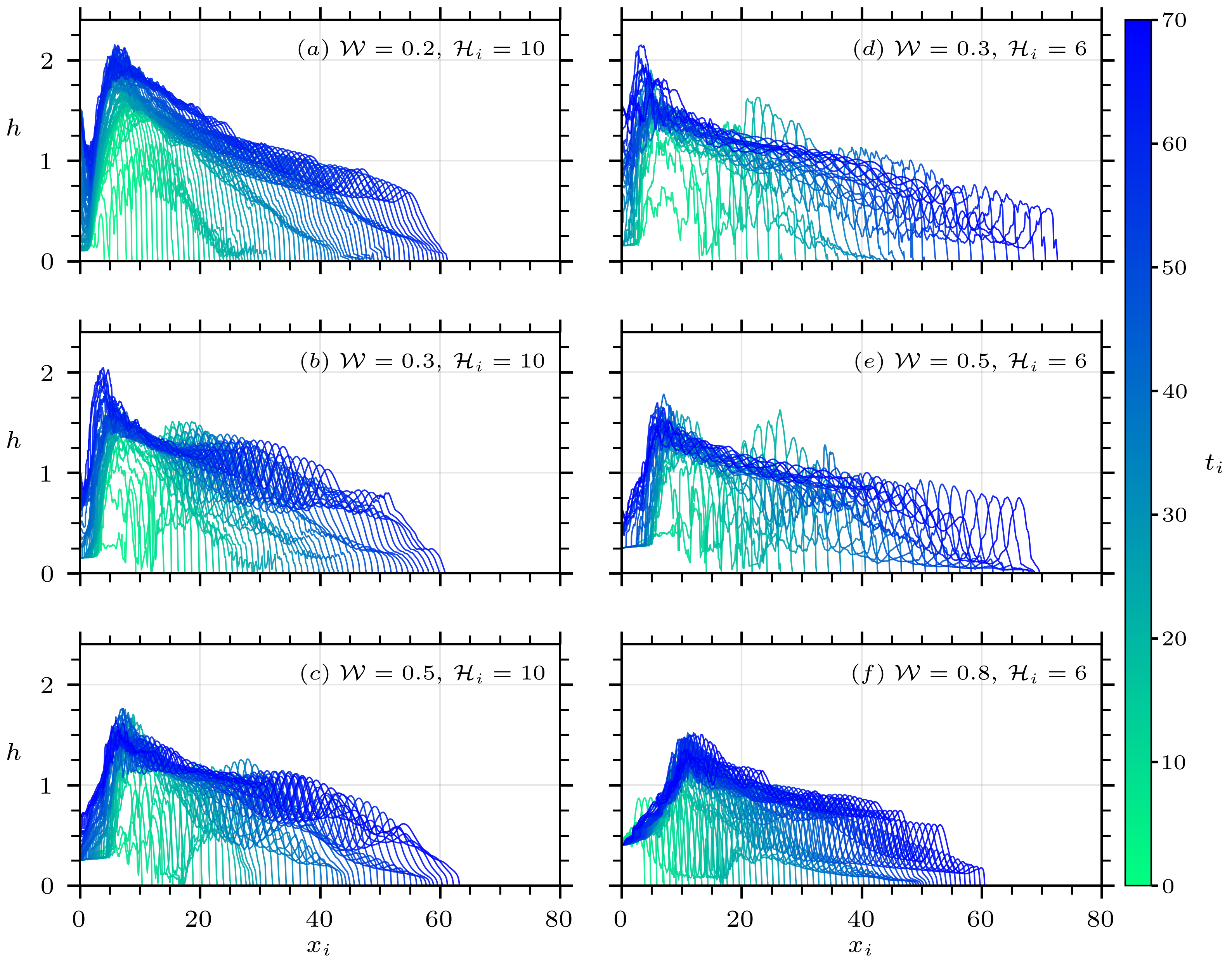}
    \caption{Evolution of the spanwise-averaged thickness profiles ($c_\mathrm{th}=0.1$) for all six simulations: (a--c) $\Hint=10$ with $\Wint=0.2$, $0.3$, $0.5$; (d--f) $\Hint=6$ with $\Wint=0.3$, $0.5$, $0.8$.  Profiles are colored by time up to $\tint = 62.8$, $75$, $84$, $100$, $122$ and $80$ respectively. {We plot $h$ as the maximum vertical extent of the contour $c = 0.1$.}}
    \label{fig:3D_thickness_all}
\end{figure}

The near-source region can be approximated as an initial laminar jet region that transitions to a turbulent expanding jet at $\xint \approx 2\text{--}3$ across all times (FIG.~\ref{fig:3D_late}b, \ref{fig:3D_velocity}a).  The laminar jet portion is initially horizontal but at late times is deflected upwards (in this instance).  A diffuse cloud of intrusion fluid surrounds it.  This causes the region to appear broader at longer times; however, the jet itself is still localized (see FIG.~\ref{fig:3D_velocity}a,b).  Beyond the laminar jet, the flow appears turbulent up to $\xint \approx 5$--$15$ depending on the measurement considered.  The streamlines within it show strong local recirculation (FIG.~\ref{fig:3D_velocity}a) and significant portions have gradient Richardson numbers below a quarter or even that are negative (FIG.~\ref{fig:3D_Ri}). Both suggest enhanced local mixing and entrainment. The overall profile of this region evolves only slowly (FIG.~\ref{fig:3D_thickness_all}) and appears quasi-steady when compared to the remainder of the profile. In this region, the thickness increases away from the source (FIG.~\ref{fig:3D_thickness_all}) and the average horizontal velocity falls (FIG.~\ref{fig:3D_velocity}f).

\begin{figure}
    \centering
    \includegraphics[width=\linewidth]{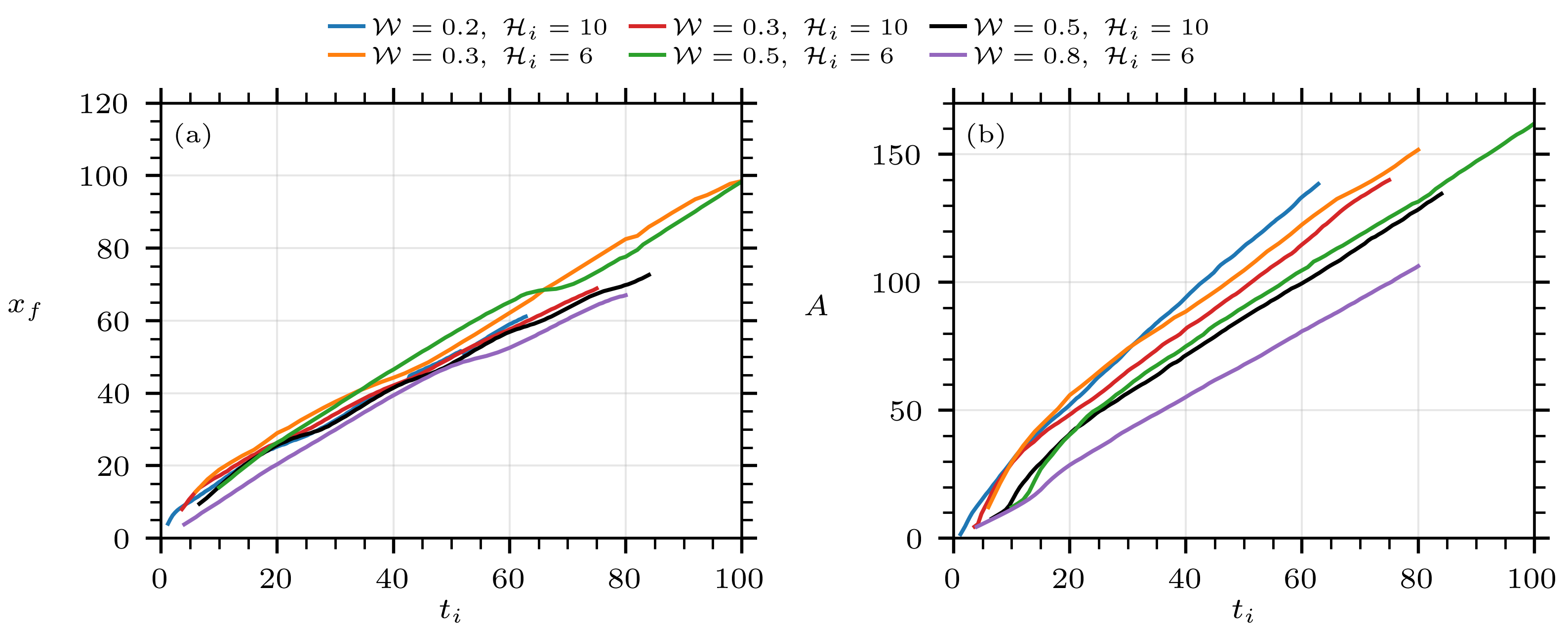}
    \caption{Front position $x_f(\tint)$ and area $A(\tint)$ for the six 3D simulations.}
    \label{fig:3D_tip_and_area}
\end{figure}

Ahead of this near-source region, the intrusion is a well-mixed wedge that thins progressively towards the front (FIG.~\ref{fig:3D_late}g).  This thinning takes the form of a near-linear reduction closest to the source and a more rapid reduction at the front.  The structure is nearly independent of the spanwise direction (FIG.~\ref{fig:3D_late}d-f).  On top of this overall structure, interfacial waves propagate on the upper and lower surfaces of the intrusion towards the tip (FIGs.~\ref{fig:3D_late} and~\ref{fig:3D_waves}).  In the intrusion the flow is fast and the streamlines in the intrusion are nearly parallel (FIG.~\ref{fig:3D_velocity}a).  The vertically averaged horizontal velocity increases with $\xint$ (FIG.~\ref{fig:3D_velocity}f).  The front advances at a nearly constant speed of approximately $0.7$ to $0.85$ (red curve in FIG.~\ref{fig:3D_tip_and_area}a).  The gradient Richardson number is mostly above a quarter (FIG.~\ref{fig:3D_Ri}), suggesting entrainment into the wedge is limited. 

In the ambient, internal waves are clearly evident in the isopycnal displacements (FIG.~\ref{fig:3D_density}).  Mode-$2$ waves propagate ahead of the intrusion.  Immediately above the intrusion there are waves that appear to have a mode-$4$ structure.  However, they propagate forward at a speed closer to that of mode-$2$ waves (FIG.~\ref{fig:3D_waves}).  These internal waves are coupled to the undulations in the intrusion profiles (FIG.~\ref{fig:3D_density}).  The undulations catch the front and plausibly reflect from it (gaps in the crests in FIG.~\ref{fig:3D_waves}).  This interaction between the waves and the front causes a light pulsing in the front propagation.

Overall, this structure is strongly reminiscent of that observed for intrusions from a wide source \cite{Vu25}.

The effective area of the entire intrusion as a function of time is shown by the red curve in FIG.~\ref{fig:3D_tip_and_area}b.  The trend is similar to that for the 2D intrusions: initially the area briefly increases linearly, before increasing rapidly and then returning to a linear growth with a higher coefficient.  The return to a linear growth suggests that overall entrainment into the intrusion has reached a steady rate and we take the slope of the area curve to give the effective flux $Q_e$.  For the baseline 3D simulation this has value $1.64$.

The structure described above is repeated across the simulations we performed.  The combined thickness profiles are shown in FIG.~\ref{fig:3D_thickness_all} and the front positions and areas in FIG.~\ref{fig:3D_tip_and_area}. The effective fluxes are summarized in table~\ref{tab:3D_runs}. The front speed is comparable across all the simulations while the effective flux increases as $\width$ decreases. %except for $\width=0.8$ where the flow has not fully transitioned to Phase~III by the end of the simulation.

%\AS{There are 2 area/tip plots here. Caption not done.}
%\AS{The rescaled panels of these two figures currently use $Q_{new}$ from the model, not the data fit $\Qe$; to be regenerated (or dropped).}

%% AS: please check the panel order against the figure; the old caption listed only the
%%     maximum times (a) 62.8, (b) 64.2, (c) 84, (d) 98, (e) 100, (f) 80.

\subsection{Wedge modelling}\label{sec:3D_wedge}

Given that the near-source expanding jet region appears quasi-steady at late times, it is supplying the wedge region ahead of it with an effectively constant areal flux.  This flux is given by $\Qe$.  The wedge region itself appears well-mixed and non-entraining and so behaves effectively as a wide-source intrusion.  Thus the scalings of \cite{Vu25} should collapse the data across the different simulations and their similarity solutions should provide a good approximation of the profiles.  Testing this requires a spatial shift by $\xs$, the extent of the expanding jet region, and a rescaling of all lengths and velocities by $\sqrt{\Qe}$. We obtain $\xs$ by finding the location of the maximum thickness in the intrusion profile and taking a median of this over late times.
%% AS: Anja's text verbatim (email of Sep 2026); \cite{Vu} -> \cite{Vu25} (key in ref.bib).
%%     Implementation: median over the second half of the run of the position of the
%%     thickness maximum with x < 30 (values in table~\ref{tab:3D_runs}).

\begin{figure}
    \centering
    \includegraphics[width=0.8\linewidth]{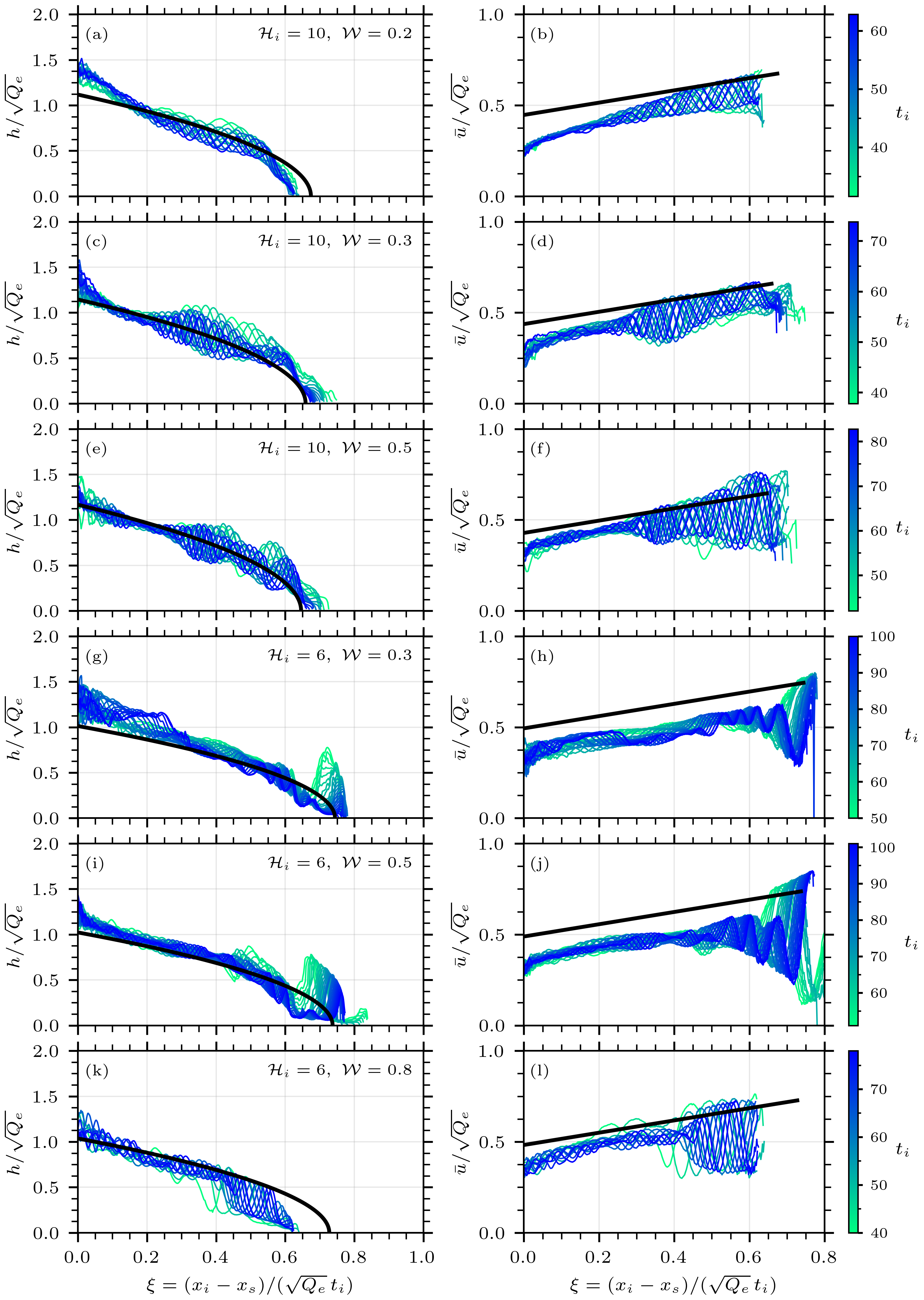}
    \caption{Rescaled half-thickness $h/\sqrt{\Qe}$ (left) and depth-averaged velocity $\bar{u}/\sqrt{\Qe}$ (right) against $\xi = (\xint-\xs)/(\sqrt{\Qe}\,\tint)$, one row per simulation, for profiles over the second half of each run (colored by time). Black: similarity solution of the hybrid shallow-water/DJL model of \cite{Vu25} for the same $\Hint$. $\Qe$ and $x_s$ are given in table~\ref{tab:3D_runs}: $Q_e$ is the effective flux to the intrusion given by the slope of $A(\tint)$ in FIG.~\ref{fig:3D_tip_and_area}b and $\xs$ is the position of the thickness maximum. {We plot $h$ as the maximum vertical extent of the contour $c = 0.1$.}}
    \label{fig:3D_collapse}
\end{figure}

Figure~\ref{fig:3D_collapse} shows the rescaled half-thickness and velocity profiles across all six simulations against the similarity variable $\xi = (\xint-\xs)/\sqrt{\Qe}\,\tint$, together with the similarity solution of \cite{Vu25} using their hybrid shallow-water/DJL model.  The scalings appear to collapse the data well across the suite of simulations.  The similarity solutions provide a comparably good description of the data as for the wide source case of Vu \etal~\cite{Vu25}.  However, as in that case, agreement is not uniform across the entire intrusion. In particular, the solution deteriorates in the vicinity of the advancing front and it cannot capture oscillatory features in the structure.  %The collapse also fails for the critical case $\Hint=6$, $\Wint=0.3$, whose front travels at the mode-2 wave speed and which is therefore outside the subcritical regime assumed by the similarity solution.
%% AS: "all four simulations" in Anja's text -> "six"; if H=6, W=0.3 stays out of the
%%     figure, make it "five" and keep the sentence about the critical case.
Nevertheless, the overall agreement suggests that the intrusions do indeed behave as a quasi-steady entraining jet region that connects to a universal wide-source profile.  This also illustrates the further universality of the wide-source structure.

%We have also attempted to predict $\Qe$ and $\xs$ from the source parameters with an entraining-jet model of the near-source region, in the spirit of Morton \etal~\cite{MortonTaylorTurner1956} and Ellison and Turner \cite{EllisonTurner1959}, in which the momentum and hydrostatic pressure integral $q^2/4h + h^3/3$ is conserved and entrainment follows the Ellison--Turner law until the bulk Richardson number reaches a quarter.  The model produces the right qualitative picture, a thin supercritical jet that entrains and jumps to the thick subcritical wedge state, and $\xs \approx 6$ in agreement with the simulations, but overpredicts $\Qe$ ($1.75$ against $1.6$ for $\Wint=0.3$) because, as shown in FIG.~\ref{fig:3D_near_source}e, the momentum flux is far from conserved in the near-source region.  We therefore use the measured $\Qe$ throughout.
%% AS: this paragraph replaces the "Modelling" subsection of the previous draft
%%     (eqs. 18-29 and model_vs_dns_h_clean.png); delete it if the model is dropped
%%     entirely.  Bib keys checked against ref.bib.

\subsection{Comparison with experimental studies}

The overall structure that we observed in the simulations, a near-source expanding jet connecting to a more laminar far-field wedge, is consistent with the experimental observations of Ceddia \etal~\cite{CeddiaKerr25}. At first glance, the fact that the area does not appear to increase linearly in the experiments and the front propagates sublinearly suggests a discrepancy there, but the inlet width of the experiments is significantly smaller than in our simulations ($\width\approx 0.04$ to $0.11$) and so the nonlinear instability growth phase will last longer.  Re-examining the experiments with the phases of the simulations in mind suggests that there could plausibly be a linear regime at long times. However, taking the effective flux from a linear fit for these times and using it to rescale the data gives a thicker intrusion than observed in the simulations. Thus the overall picture is consistent with the experiments but the two are not in quantitative agreement. Further simulations for smaller $\width$ would be useful. The experimental source also consisted of a series of holes rather than a slot and this difference might also contribute to the different behavior, although this would suggest the nature of the far-field solution is less universal than it appears here.

\section{Conclusions} \label{sec:conclusions}

Two- and three-dimensional direct numerical simulations have been presented of planar intrusions originating from a continuously supplied, narrow source in a linearly stratified ambient, over dimensionless inlet widths between $0.2$ and $1$ and Reynolds numbers of $500$, $1000$ and $2000$. In every instance the intrusion evolves through three stages: a dipole-dominated stage, a sinuous instability of the trailing jet, and finally a buoyancy-dominated stage at late times.

In the first phase the intrusion behaves as a starting planar jet: roll-up of the two shear layers forms a coherent leading vortex dipole that moves at roughly constant speed, followed by a trailing jet with a varicose interfacial wave train, consistent with the known dynamics of starting jets and separated shear layers
\citep{didden1979formation,gao2016formation,zheng2025formation}. Stratification greatly restricts the vertical growth of the dipole but has only a minor effect on its vorticity, while the Reynolds number mainly sets the thickness, number and sharpness of the smaller-scale roll-ups. The dipole also displaces the ambient isopycnals and generates a phase-locked, lee-wave-like response similar to that of a moving obstacle in a stratified fluid \citep{BainesHoinka85}. The velocity profile across the core of the trailing jet is nearly a locally rescaled Bickley profile, with some indication of return flow near its interface.

The second phase begins when a localized sinuous disturbance appears in the trailing jet and
spreads both upstream and downstream. Its growth rate and wavelength closely match inviscid
planar-jet theory applied using the local velocity and width of the jet
\citep{tatsumi1958stability,sato1960stability}, and change only slightly with $\Wint$ or $Re$, confirming that it is a shear instability; the instability occurs up to $\Wint \approx 1.6$, beyond which the trailing jet remains laminar.
The Bickley-like trailing-jet profile connects this behavior to the simpler parallel Bickley jet examined by \citet{VuDeguchi2026}, where the sinuous and varicose modes continue into finite-amplitude exact coherent structures that either distort the mean flow (short-wave structures) or radiate momentum-carrying internal waves (long-wave structures). Both signatures appear to occur in our simulations: the sinuous instability produces a strongly asymmetric finite-amplitude flow, and internal waves carry the disturbance into the surrounding stratified fluid.

After the disturbance has spread across the flow, the intrusion enters a third,
buoyancy-dominated phase with substantial mixing and entrainment.  In two dimensions, this
yields a wedge-shaped intrusion, thickest near the source; allowing spanwise structure instead
resolves the near-source region into a quasi-steady, expanding region.  This latter picture is 
consistent with the narrow-source experiments of
\citet{EPA72} and \citet{CeddiaKerr25},
although quantitative agreement with the latter is not yet established.  Entrainment
is confined to this near-source region. Using its extent as a virtual origin
and rescaling the data using inertial-buoyancy scalings based on the effective areal flux $Q_e$ delivered to the wedge and the buoyancy frequency $N$ brings the three-dimensional simulations into good agreement with the wide-source similarity solution of
\citet{Vu25}.

% An entraining-jet model \citep{MortonTaylorTurner1956,EllisonTurner1959} reproduces the
% qualitative transition from thin supercritical inlet jet to thick subcritical outflow and
% gives a reasonable estimate of the near-source region's size, but overpredicts $Q_e$ because it
% assumes a conserved momentum-plus-pressure integral, whereas the simulations show substantial
% momentum loss during the turbulent adjustment. A predictive source-region theory will
% therefore need a closure for both entrainment and momentum loss, not an entrainment law alone.

The central conclusion is that a narrow-source intrusion behaves as a wide-source intrusion fed through an entraining adjustment region: the inlet width sets how the jet destabilizes, how much ambient fluid it entrains, and how much momentum it loses, but the downstream wedge
responds mainly to the effective flux delivered after this adjustment. This near-field/far-field distinction offers a single framework for narrow- and wide-source intrusions alike, and identifies near-source entrainment and momentum budget as the key ingredients still needed for
a predictive model.

\section*{Data availability}
The data that support the findings of this article are available on figshare via a private link for review (\url{https://figshare.com/s/45fb97d70a3861bbd306}); the dataset will be made publicly available with a DOI upon publication.

\begin{acknowledgments}
The authors thank K. Deguchi for helpful discussions. This research was supported by Monash eResearch capabilities and the National Computational Infrastructure (NCI Australia), an NCRIS enabled capability supported by the Australian Government. This research was supported by the Australian Research Council Discovery Project DP220101660.
\end{acknowledgments}

\appendix
\section{Code verification}\label{app:3D}

%TO INCLUDE HERE: : FIG.~\ref{fig:convergence_2D} compares four two-dimensional meshes and FIG.~\ref{fig:convergence_3D} compares three spanwise resolutions, and in both cases the large-scale structure, the downstream profiles and the integral measures are insensitive to further refinement at the production resolutions.

\begin{figure}
\centering
\includegraphics[width=\linewidth]{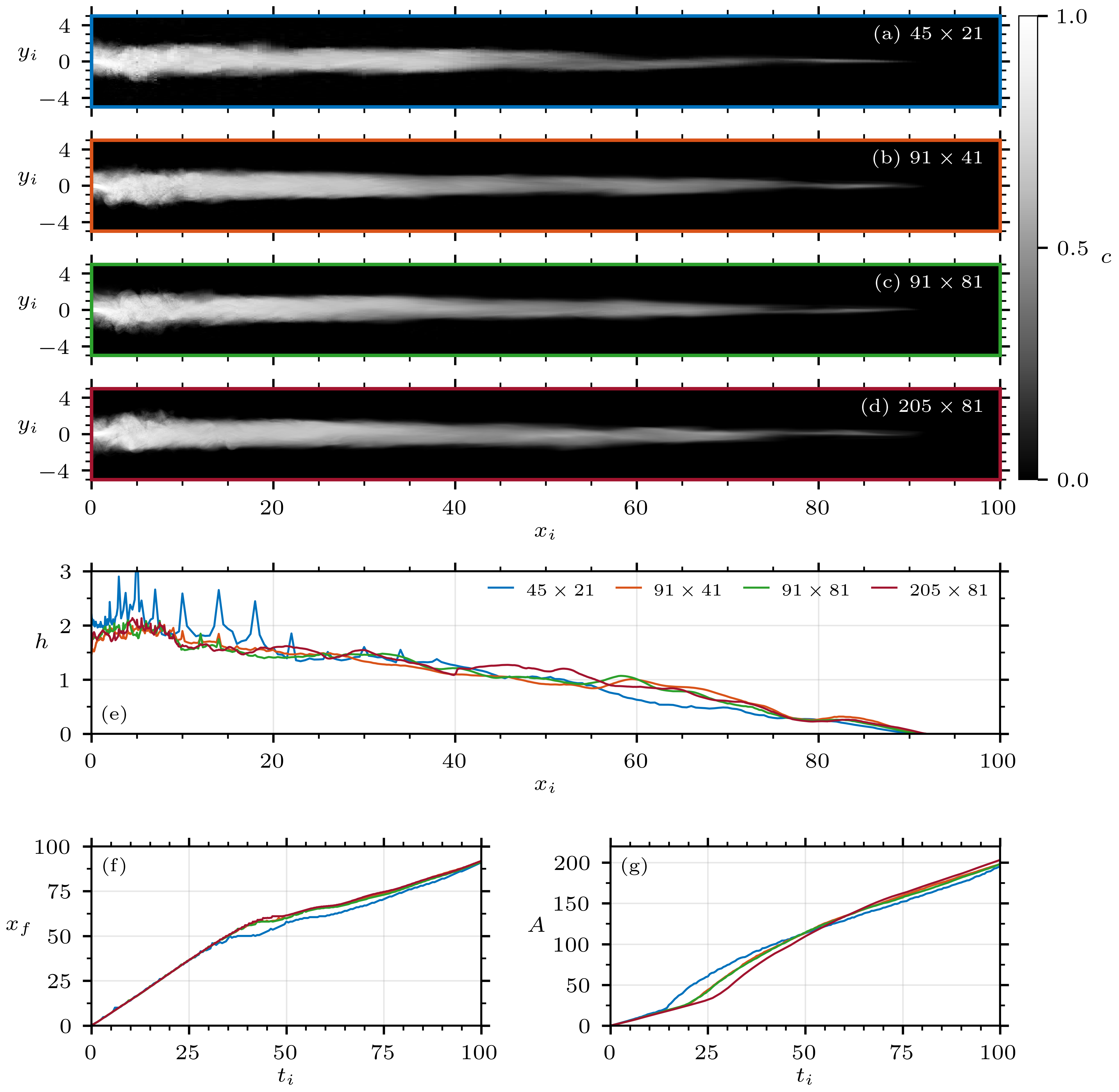}
\caption{Two-dimensional resolution convergence for $\Hint=10$ and $\Wint=0.5$. Panels (a)--(d) show the color function $c$, averaged over $90\le\tint\le100$, using spectral-element meshes containing $45\times21$, $91\times41$, $91\times81$ and $205\times81$ elements, respectively. Panel (e) compares the corresponding time-averaged intrusion thicknesses, while panels (f) and (g) show the evolution of the front position and cross-sectional area. The $205\times81$ mesh is used for the baseline simulations. We plot $h$ as the maximum vertical extent of the contour $c = 0.1$.}
\label{fig:convergence_2D}
\end{figure}

\begin{figure}
\centering
\includegraphics[width=\linewidth]{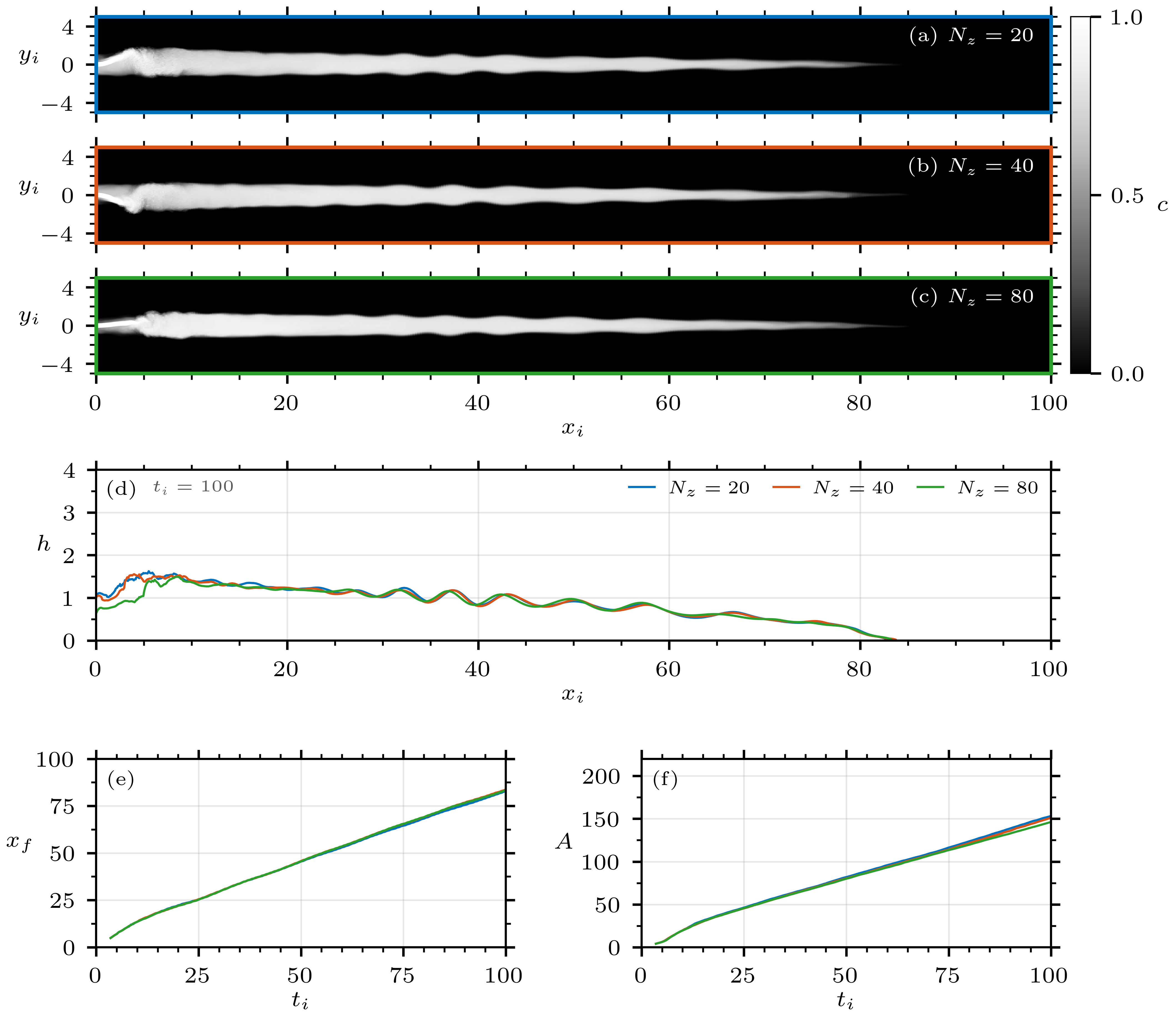}
\caption{Spanwise-resolution convergence for $\Wint=0.5$, $\Hint=10$ and $\Lzint=5$. Panels (a)--(c) show the spanwise-averaged color function $c$, further averaged over $80\le\tint\le100$, using $N_z=20$, $40$ and $80$ Fourier planes, respectively. Panel (d) compares the corresponding intrusion thicknesses at $\tint=100$, while panels (e) and (f) show the evolution of the front position and cross-sectional area. The same three-dimensional perturbation is introduced at $\tint=3$ in each simulation. We plot $h$ as the maximum vertical extent of the contour $c = 0.1$.}
\label{fig:convergence_3D}
\end{figure}

To test mesh convergence, we ran the simulations for our baseline two-dimensional simulation on four different grid resolutions: $45 \times 21$, $91 \times 41$, $91 \times 81$, and $205 \times 81$. As shown in FIG.~\ref{fig:convergence_2D}, the coarsest mesh under-resolves the turbulent near-source region and produces spurious oscillations in the time-averaged thickness for $\xint\lesssim25$. From $91\times41$ onwards, however, the simulations reproduce the same large-scale structure. The downstream thickness profiles and front trajectories also show good agreement, while the front positions and areas obtained with the three finer meshes differ by less than $1\%$ and approximately $2.5\%$, respectively, at $\tint=100$. 
%The remaining differences are localized within the turbulent near-source region, where the positions of individual vortical structures are not expected to converge pointwise.

We also test the influence of the spanwise resolution convergence at fixed $\Lzint = 5$ by running the three-dimensional simulations for $\Hint = 10$, $\mathcal{W} = 0.5$ using $N_z=20$, $40$ and $80$ Fourier planes on the $91\times41$ spectral-element mesh. All are perturbed at the same time $\tint=3$ with a Gaussian-distributed white-noise forcing that has amplitude of $0.1$ applied to all of the variables. All three resolutions reproduce the same overall three-dimensional structure, comprising a short inlet jet, a thickened expanding-jet region and a tapered downstream intrusion (FIG.~\ref{fig:convergence_3D}).  The thickness profiles nearly coincide for $\xint\gtrsim10$, and the front trajectories remain closely aligned. At $\tint=100$, the front positions differ by approximately $1\%$ and the areas by less than $5\%$. The principal resolution sensitivity is confined to the near-source transition, where the finer calculations resolve a slightly thinner laminar jet and a sharper onset of expansion. 
%The three-dimensional simulations use $N_z=160$, a finer spanwise resolution than any of the tests shown here.

Taken together, these tests suggest that the large-scale intrusion structure, downstream profiles and integral measures are insensitive to further refinement at the production resolutions. The remaining resolution dependence is confined primarily to the detailed turbulent structure of the near-source region and does not affect the conclusions drawn from the simulations.

\bibliography{ref}% Produces the bibliography via BibTeX.

\end{document}